\documentclass[ApJ,twocolumn]{aastex631}

\usepackage{graphicx,amsmath,amssymb}
\usepackage{url}
\usepackage{xcolor}
\usepackage[utf8]{inputenc}

\usepackage{ulem}

\newcommand{\todo}{\ifmmode \text{\color{red}{\Huge\textbullet}} \else {\color{red}{\Huge$\bullet$}}\fi}
\newcommand{\tido}{\ifmmode {{\color{red}\bullet}} \else \color{red}{$\bullet$}\fi}

\newcommand{  \mbh      }{\ifmmode M_{\rm BH} \else $M_{\rm BH}$\fi}
\newcommand{  \Lagn     }{\ifmmode L_{\rm AGN} \else $L_{\rm AGN}$\fi}
\newcommand{  \lledd    }{\ifmmode L/L_{\rm Edd} \else $L/L_{\rm Edd}$\fi}
\newcommand{  \RBLR     }{\ifmmode R_{\rm BLR} \else $R_{\rm BLR}$\fi}

\def\e{{\rm E}}

\begin{document}
\title{ULTRASAT: A wide-field time-domain UV space telescope}

\author{Y. Shvartzvald}
\affiliation{Department of Particle Physics and Astrophysics, Weizmann Institute of Science, Rehovot 7610001, Israel}

\author{E. Waxman}
\affiliation{Department of Particle Physics and Astrophysics, Weizmann Institute of Science, Rehovot 7610001, Israel}

\author{A. Gal-Yam}
\affiliation{Department of Particle Physics and Astrophysics, Weizmann Institute of Science, Rehovot 7610001, Israel}

\author{E.~O. Ofek}
\affiliation{Department of Particle Physics and Astrophysics, Weizmann Institute of Science, Rehovot 7610001, Israel}

\author{S. Ben-Ami}
\affiliation{Department of Particle Physics and Astrophysics, Weizmann Institute of Science, Rehovot 7610001, Israel}


\author{D. Berge}
\affiliation{Deutsches Elektronen Synchrotron DESY, Platanenallee 6, 15738 Zeuthen, Germany}
\affiliation{Institut fur Physik, Humboldt-Universität zu Berlin, D-12489 Berlin, Germany}

\author{M. Kowalski}
\affiliation{Deutsches Elektronen Synchrotron DESY, Platanenallee 6, 15738 Zeuthen, Germany}
\affiliation{Institut fur Physik, Humboldt-Universität zu Berlin, D-12489 Berlin, Germany}

\author{R. B\"uhler}
\affiliation{Deutsches Elektronen Synchrotron DESY, Platanenallee 6, 15738 Zeuthen, Germany}

\author{S. Worm}
\affiliation{Deutsches Elektronen Synchrotron DESY, Platanenallee 6, 15738 Zeuthen, Germany}


\author{J.~E. Rhoads}
\affiliation{NASA Goddard Space Flight Center, Greenbelt, MD 20771, USA}


\author[0000-0001-7090-4898]{I. Arcavi}
\affiliation{School of Physics and Astronomy, Tel-Aviv University, Tel-Aviv 69978, Israel}
\affiliation{CIFAR Azrieli Global Scholars program, CIFAR, Toronto, Canada}

\author{D. Maoz}
\affiliation{School of Physics and Astronomy, Tel-Aviv University, Tel-Aviv 69978, Israel}

\author{D. Polishook}
\affiliation{Faculty of Physics, Weizmann Institute of Science, Rehovot 7610001, Israel}

\author{N. Stone}
\affiliation{Racah Institute of Physics, The Hebrew University of Jerusalem, Jerusalem 91904, Israel}

\author[0000-0002-3683-7297]{B. Trakhtenbrot}
\affiliation{School of Physics and Astronomy, Tel-Aviv University, Tel-Aviv 69978, Israel}


\author{M. Ackermann}
\affiliation{Deutsches Elektronen Synchrotron DESY, Platanenallee 6, 15738 Zeuthen, Germany}

\author{O. Aharonson}
\affiliation{Helen Kimmel Center for Planetary Science, Weizmann Institute of Science, Rehovot, Israel}
\affiliation{Planetary Science Institute, Tucson, Arizona, USA}

\author{O. Birnholtz}
\affiliation{Department of Physics, Bar-Ilan University Ramat-Gan 52900, Israel}

\author{D. Chelouche}
\affiliation{Department of Physics, Faculty of Natural Sciences, University of Haifa, Haifa 3498838, Israel}
\affiliation{Haifa Research Center for Theoretical Physics and Astrophysics, University of Haifa, Haifa 3498838, Israel}

\author{D. Guetta}
\affiliation{Department of Physics, Ariel University, Ariel, IL-40700, Israel}

\author{N. Hallakoun}
\affiliation{Department of Particle Physics and Astrophysics, Weizmann Institute of Science, Rehovot 7610001, Israel}

\author{A. Horesh}
\affiliation{Racah Institute of Physics, The Hebrew University of Jerusalem, Jerusalem 91904, Israel}

\author{D. Kushnir}
\affiliation{Department of Particle Physics and Astrophysics, Weizmann Institute of Science, Rehovot 7610001, Israel}

\author{T. Mazeh}
\affiliation{School of Physics and Astronomy, Tel-Aviv University, Tel-Aviv 69978, Israel}

\author{J. Nordin}
\affiliation{Institut fur Physik, Humboldt-Universität zu Berlin, D-12489 Berlin, Germany}

\author{A. Ofir}
\affiliation{Helen Kimmel Center for Planetary Science, Weizmann Institute of Science, Rehovot, Israel}

\author{S. Ohm}
\affiliation{Deutsches Elektronen Synchrotron DESY, Platanenallee 6, 15738 Zeuthen, Germany}

\author{D. Parsons}
\affiliation{Deutsches Elektronen Synchrotron DESY, Platanenallee 6, 15738 Zeuthen, Germany}

\author{A. Pe'er}
\affiliation{Department of Physics, Bar-Ilan University Ramat-Gan 52900, Israel}

\author{H.~B. Perets}
\affiliation{Physics Department, Technion—Israel Institute of Technology, Technion City, Haifa 3200002, Israel}
\affiliation{Department of Natural Sciences, The Open University of Israel, 1 University Road, PO Box 808, Raanana 4353701, Israel}

\author{V. Perdelwitz}
\affiliation{Helen Kimmel Center for Planetary Science, Weizmann Institute of Science, Rehovot, Israel}

\author{D. Poznanski}
\affiliation{School of Physics and Astronomy, Tel-Aviv University, Tel-Aviv 69978, Israel}

\author{I. Sadeh}
\affiliation{Deutsches Elektronen Synchrotron DESY, Platanenallee 6, 15738 Zeuthen, Germany}

\author{I. Sagiv}
\affiliation{ElOp - Elbit Systems Ltd., Rehovot, Israel}

\author{S. Shahaf}
\affiliation{Department of Particle Physics and Astrophysics, Weizmann Institute of Science, Rehovot 7610001, Israel}

\author{M. Soumagnac}
\affiliation{Department of Physics, Bar-Ilan University Ramat-Gan 52900, Israel}

\author{L. Tal-Or}
\affiliation{Department of Physics, Ariel University, Ariel, IL-40700, Israel}
\affiliation{Astrophysics Geophysics And Space Science Research Center, Ariel University, Ariel 40700, Israel}

\author{J. Van Santen}
\affiliation{Deutsches Elektronen Synchrotron DESY, Platanenallee 6, 15738 Zeuthen, Germany}

\author{B. Zackay}
\affiliation{Department of Particle Physics and Astrophysics, Weizmann Institute of Science, Rehovot 7610001, Israel}


\author{O. Guttman}
\affiliation{Department of Particle Physics and Astrophysics, Weizmann Institute of Science, Rehovot 7610001, Israel}

\author{P. Rekhi}
\affiliation{Department of Particle Physics and Astrophysics, Weizmann Institute of Science, Rehovot 7610001, Israel}

\author{A. Townsend}
\affiliation{Institut fur Physik, Humboldt-Universität zu Berlin, D-12489 Berlin, Germany}

\author{A. Weinstein}
\affiliation{Tufts University, Medford, MA 02155, USA}

\author{I. Wold}
\affiliation{NASA Goddard Space Flight Center, Greenbelt, MD 20771, USA}


\correspondingauthor{Yossi Shvartzvald}
\email{yossi.shvartzvald@weizmann.ac.il}

\begin{abstract}

The Ultraviolet Transient Astronomy Satellite (ULTRASAT) is scheduled to be launched to geostationary orbit in 2026. It will carry a telescope with an unprecedentedly large field of view (204\,deg$^2$) and NUV (230-290nm) sensitivity (22.5\,mag, 5$\sigma$, at 900s).
ULTRASAT will conduct the first wide-field survey of transient and variable NUV sources and will revolutionize our ability to study the hot transient universe: It will explore a new parameter space in energy and time-scale (months long light-curves with minutes cadence), with an extra-Galactic volume accessible for the discovery of transient sources that is $>$300 times larger than that of GALEX and comparable to that of LSST. 
ULTRASAT data will be transmitted to the ground in real-time, and transient alerts will be distributed to the community in $<$15 min, enabling a vigorous ground-based follow-up of ULTRASAT sources. ULTRASAT will also provide an all-sky NUV image to $>$23.5 AB mag, over 10 times deeper than the GALEX map.

Two key science goals of ULTRASAT are the study of mergers of binaries involving neutron stars,
and supernovae:
With a large fraction ($>$50\%) of the sky instantaneously accessible, fast (minutes) slewing capability and a field-of-view that covers the error ellipses expected from GW detectors beyond 2025, ULTRASAT will rapidly detect the electromagnetic emission following BNS/NS-BH mergers identified by GW detectors, and will provide continuous NUV light-curves of the events;
ULTRASAT will provide early (hour) detection and continuous high (minutes) cadence NUV light curves for hundreds of core-collapse supernovae, including for rarer supernova progenitor types. 

\end{abstract}

\section{Introduction}
\label{sec:intro}

Time domain surveys are a developing focus of astronomy with an unusual discovery potential. There are three main reasons for this. First, the static sky has already been well imaged in most electromagnetic bands. Second, new technology enables efficient monitoring of large swaths of sky. Finally, some of the most exciting frontiers, particularly those related to cosmic cataclysms, require wide-field synoptic surveys. It is therefore not surprising that large powerful surveys across the electromagnetic spectrum (e.g., SKA, VRO, ZTF, eROSITA, Euclid, Roman), are at the focus of attention by the global community. However, a key piece, the UV, is missing from this synoptic suite. 

The UV band is unique in its ability to probe the physics of hot sources \citep{Sagiv.2014.A,Kulkarni.2021arXiv.A}. Many of the interesting and exotic astrophysical sources shine brightly in the UV, and explosive transients are initially hot, so UV can provide the earliest notification. Thus, even a modest UV time-domain explorer has a vast discovery potential, and can uncover evidence to solve many open questions. 

\begin{figure}
\centerline{\includegraphics[width=0.5\textwidth]{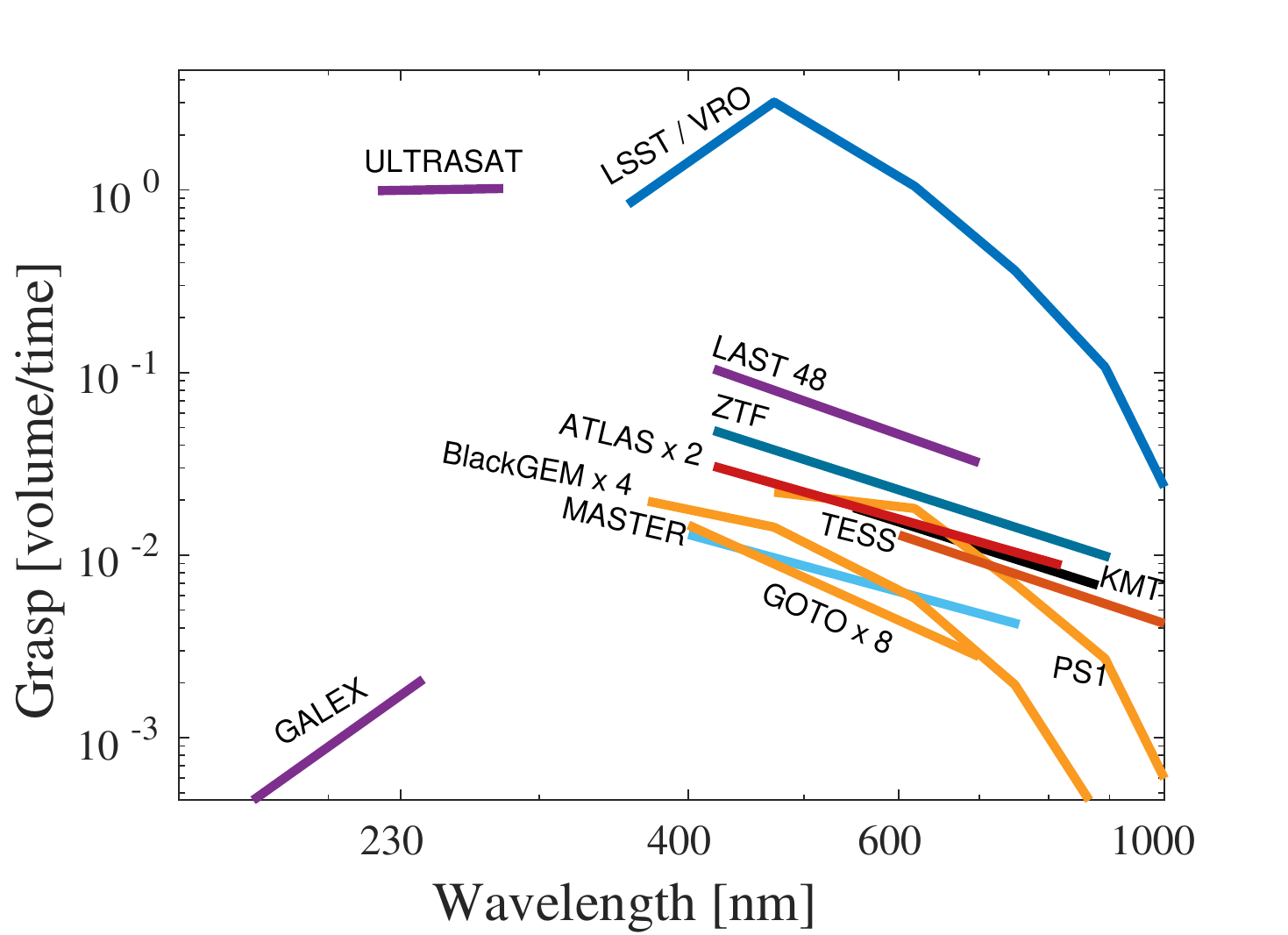}}
\caption{The volume of the universe per unit time (grasp; \citealt{Ofek.2020.A}) accessible to several past, current and future sky surveys (VRO/LSST, \citealt{Ivezic.2019.A}; LAST, Ofek et al. in prep.; ZTF, \citealt{Bellm.2019.A}; ATLAS, \citealt{Heinze.2018.A}; Pan-STARRS, \citealt{Chambers.2016.A}; KMTNet, \citealt{Kim.2016.A}; TESS, \citealt{Ricker.2015.A}; Black-GEM, \citealt{Bloemen.2015.A}; MASTER, \citealt{Gorbovskoy.2013.A}; GOTO, \citealt{Steeghs.2022.A}; GALEX, \citealt{Martin.2005.A}), as a function of wavelength (normalized such that ULTRASAT's grasp is 1). The grasp is given for a 20,000\,K black-body source spectrum (e.g., a hot transient). ULTRASAT's grasp is an order of magnitude larger than that of current surveys, two orders of magnitudes larger than that of GALEX, the largest grasp UV mission to date, and comparable to that of LSST, the largest grasp optical survey under construction.
\label{fig:Grasp}}
\end{figure}

\begin{figure}
\centerline{\includegraphics[width=0.50\textwidth]{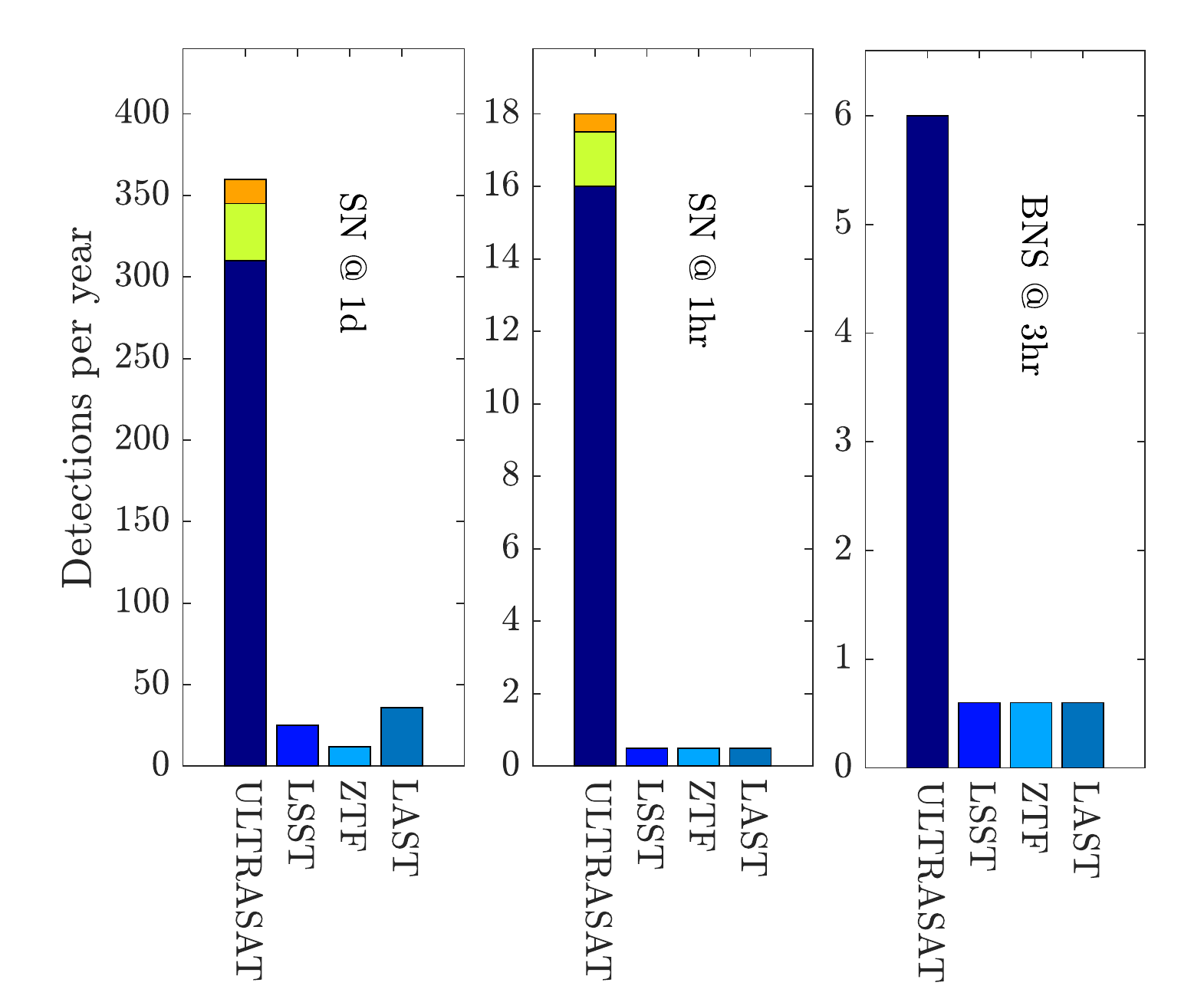}}
\caption{The number per year of supernova (SN) transients and EM transients following binary neutron star (BNS) mergers, expected to be detected at early time (as indicated in the plot) by ULTRASAT (assuming the limiting magnitude for $3\times 300$\,s exposures), the Vera Rubin Observatory (LSST), ZTF \citep{Law_ZPTF09} and LAST \citep{LAST20PASP}. The blue/green/orange parts of the SN bars correspond to RSG/BSG/WR progenitors respectively. The SN rates are taken from \cite{Ganot.2016.A}, based on simulations calibrated to GALEX/PTF observations.
The BNS rates were obtained assuming 12 GW detections per year out to 300Mpc, with UV flux similar to that observed following GW\,170817.}
\label{fig:Rates}
\end{figure}

ULTRASAT is a UV space telescope that will undertake the first wide-field UV time-domain survey, and is planned to be launched to its geostationary (GEO) orbit in 2026. Its main properties are given in Table~\ref{tab:properties}. The celestial volume accessible for transient detection by a survey, and hence the number of extra-galactic objects detected per unit time, are proportional to the survey's {\it grasp} $G\propto\Omega S_m^{-3/2}$, where $\Omega$ is the field of view (FoV) and $S_m$ is the minimum detectable flux \citep{Ofek.2020.A}. Fig.~\ref{fig:Grasp} shows a comparison of the grasp for hot transients of ULTRASAT with that of other surveys. ULTRASAT's grasp is much larger than that of existing optical surveys, comparable to that of the Vera Rubin Observatory (VRO), which is planned to begin operation in 2024, and over 300 times that of GALEX, which had a similar sensitivity but a much smaller FoV. The large grasp, continuous cadence and fast slew rate enable early detection of transients by ULTRASAT at a rate that is much larger than that of existing and planned surveys, as demonstrated in fig.~\ref{fig:Rates}. In addition to its large grasp, ULTRASAT is unique in its energy and time windows. It will provide continuous months-long UV light curves with minutes cadence, as well as early alerts that will enable rapid ground- and space- based followup of transients. Historically, such great leaps in capability have led to marvelous discoveries – a major incentive for ULTRASAT. 

\begin{table*}[ht]
\centering
\caption{ ULTRASAT Key Properties \label{tab:properties}}
\begin{tabular}{|l l|c|l|}
\tableline
\multicolumn{2}{|c|}{Property} & Value & \multicolumn{1}{c|}{Comments}  \\
\tableline
\multicolumn{4}{|l|}{\bf Spacecraft parameters}    \\
\tableline
& Orbit & GEO &    \\
\tableline
& Real-time download of data & Continuous &  \\
\tableline
& Slew rate & $>30^\circ/$min &  \\
\tableline
& Transient alert after observation end & $<15$\,min &  For both survey and ToO modes   \\
\tableline
& Sky accessibility at any given moment & $>50\%$ &  See Figure \ref{fig:Visibility}   \\
\tableline
& Observation start after ToO trigger & $<15$\,min &  At any visible position   \\
\tableline
\multicolumn{4}{|l|}{\bf Payload parameters}    \\
\tableline
& Aperture & 33\,cm &  \\
\tableline
& Total FoV & 204\,deg$^2$ & Covered by four 7.14\,deg x 7.14\,deg  sensors  \\
\tableline
& Pixel scale & 5.4\,\arcsec/pix & Total of 89.9M pixels \\
\tableline
& Operation waveband & 230-290\,nm &  See Figure \ref{fig:Throughput} \\
\tableline
& Mean throughput in operation waveband & 0.25 &  See Figure \ref{fig:Throughput} \\
\tableline
& Out of Band Rejection ($>300$\,nm) & $2.9 \times 10^{-5}$ &  See Figure \ref{fig:Throughput} \\
\tableline
& Mean effective FWHM$^a$  & 8.3\arcsec &  See Figure \ref{fig:Eff_PSF}\\
\tableline
& Mean limiting magnitude$^a$ (in 900\,s, 5$\sigma$)& 22.5\,AB\,mag & See Figure \ref{fig:LimMag_SNR}\\
\tableline
\multicolumn{3}{l}{$^a$ In central 170\,deg$^2$ of FoV, for a $T=20,000$\,K blackbody source.}    \\
\multicolumn{3}{l}{\,\,\,\, Assuming the conservative background values as in Table~\ref{tab:noise}. }    \\
\end{tabular}
\end{table*}

While ULTRASAT is expected to revolutionize the study of a wide range of transients (see \S~\ref{sec:Science} and Table~\ref{tab:highlights}), the mission design is set by two key science goals of fundamental importance -- the study of the mergers of binaries involving neutron stars (binary neutron stars, BNS, or neutron star-black hole, NS-BH, binaries; see \S~\ref{subsec:GW}) and the study of supernovae (SNe; see \S~\ref{subsec:SNe}). Measuring the electromagnetic (EM) emission following BNS/NS-BH mergers will (i) provide direct constraints on the structure and composition of the ejected material, thus providing unique diagnostics of the properties of matter at nuclear density and of the merger dynamics; (ii) enable to determine whether mergers are the sources of r-process elements and gamma-ray bursts; and (iii) allow to determine the location in, and properties of, the host galaxy, thus revealing the stellar antecedents of the binary systems. Measuring the early shock breakout/cooling part of SN light curves will provide unique information on the progenitor star and its pre-explosion evolution, in particular mapping the different types of SNe to the different stellar progenitors, and hence also providing constraints on the explosion mechanisms, which are not fully understood. 

\begin{table*}[ht]
\centering
\caption{ ULTRASAT Science Highlights \label{tab:highlights}}
\begin{tabular}{|l l|c|l|}
\tableline
\multicolumn{2}{|c|}{Source Type} & \# Events per & Science Impact \\
\multicolumn{2}{|c|}{} &3 yr mission &\\
\tableline
\multicolumn{4}{|l|}{Supernovae}\\
\tableline
{    }& Shock break-out and & $>40$ & Understand the explosive death  \\
& Early (shock cooling) of core collapse SNe & $>500$ & of massive stars \\
\tableline
& Superluminous SNe & $>250$ & Early evolution, shock cooling emission  \\
\tableline
& Type Ia SNe & $>1000 $ & Discriminate between SD and DD progenitors, dust reddening   \\
\tableline
\multicolumn{4}{|l|}{Compact Object Transients}\\
\tableline
& Emission from Gravitational Wave events: & $\sim 25$ & Constrain the physics of the sources of  \\
& NS-NS and NS-BH & & gravitational waves \\
\tableline
& Tidal disruption events & $>300$ (high-cadence) & Accretion physics, black hole demographics \\ & & $>4500$ (low-cadence) & \\
\tableline 
\multicolumn{4}{|l|}{Quasars and Active Galactic Nuclei	}\\
\tableline
& Continuous UV lightcurves & $>7500$ & Accretion physics, BLR reverberation mapping, lensed quasars  \\
\tableline
& AGN-related flares \& transients & $>100$ & Accretion physics \\
\tableline
\multicolumn{4}{|l|}{Stars \& Exoplanets}\\
\tableline
& Active \& Flaring stars & $>4\times 10^5$ & Planet habitability, high-energy flare frequency, \\
&  & & stellar magnetic structure, gyrochronology, magnetospheres \\
\tableline
& White dwarfs & $>3\times 10^4$ & Planetary systems, debris accretion, rotation-related variability\\
\tableline
& RR Lyrae & $>1000$ & Pulsation physics  \\
\tableline
& Nonradial hot pulsators, e.g., $\alpha$ Cyg, & $>250$ &  Asteroseismology \\
& $\delta$ Scuti, SX Phe, $\beta$ Cep  types &  &  \\
\tableline
& Eclipsing binaries & $>400$ & Chromosphere and eclipse mapping \\
\tableline
\multicolumn{4}{|l|}{Galaxies and Clusters}\\
\tableline
& All Sky Survey -- galaxies & $>10^8$ & Galaxy Evolution, star formation rate  \\
\tableline 
\multicolumn{4}{|l|}{Gamma Ray Bursts}\\
\tableline
& GRBs occurring in-field & $\sim 30$ & Prompt emission \& afterglow physics, dust reddening \\
\tableline
& Orphan Afterglows & $> 30$ & Fireball $\Gamma$ and opening angle distributions  \\
\tableline
\multicolumn{4}{|l|}{Solar System}\\
\tableline
& Asteroids and other small bodies & $>10^4$ & Asteroid classification, origin  \\
\tableline
\end{tabular}
\end{table*}

Rapid detection and continuous UV measurements of the EM emission following BNS/NS-BH mergers identified by GW detectors set requirements for rapid spacecraft slew capability, instantaneous access to a large fraction of the sky, real-time communication, and a wide FoV. The early (hour) detection and continuous measurements of UV light curves for hundreds of core-collapse supernovae requires a high cadence and a large grasp. A wide FoV, with a correspondingly lower sensitivity for a given grasp, is required in order to obtain a lower characteristic distance, that will enable spectroscopic followup of a large fraction of detected SNe.

In addition to the study of BNS/NS-BH mergers and SNe, ULTRASAT will provide continuous NUV light curves for hundreds of tidal disruptions of stars by super-massive black holes, thousands of active galactic nuclei, and $>10^5$ flaring and variable stars (See Table~\ref{tab:highlights}).

This paper is organized as follows. An overview of ULTRASAT's system design and expected performance is given in \S\ref{sec:design}, a description of the currently planned  modes of operations of ULTRASAT is given in \S\ref{sec:modes}, and the Science Operation Center (SOC) is described in \S\ref{sec:SOC}. An overview of the mission science objectives is given in \S\ref{sec:Science}. A short summary of the expected contributions of ULTRASAT to the key science goals is given in \S\ref{sec:summary}.

\section{Design and performance}
\label{sec:design}

Here we briefly describe the spacecraft, its payload, and performance. ULTRASAT’s key performance characteristics are given in Table \ref{tab:properties}. 
The ULTRASAT spacecraft will be constructed by the Israeli Aerospace Industry (IAI).
It will carry a single instrument - a wide field of view ($\cong204$\,deg$^2$) telescope constructed by Elbit/Elop, with a focal-plane array (hereafter camera) constructed by
Deutsches Elektronen-Synchrotron (DESY),
including a UV-optimized detector produced by TowerJazz and designed by AnalogValue.

ULTRASAT will be launched to a geostationary transfer orbit (GTO) and then self-perform a GTO-GEO transfer to acquire its final Geosynchronous Earth Orbit (GEO) position. It is planned for a minimum 3-year mission operation, with sufficient propellant for a 6-year science mission.
The mission lifetime is limited by propellant availability for station keeping, and final evacuation from its GEO position to the geostationary graveyard belt.

\subsection{Spacecraft}
\label{subsec:S/C}

The ULTRASAT spacecraft bus is constructed around two main components: the ULTRASAT wide-field telescope,
and the bi-propellant propulsion system. The propellant mass contributes about half of ULTRASAT's total launch mass of $\sim$1,100\,kg, and is required to enable the GTO-GEO maneuvers. 
The bus uses high TRL components and is based on the successful IAI heritage series of small satellites, benefiting from that company’s long-term experience in fabricating and operating missions in both LEO and GEO. The spacecraft is designed so that it can be attached as a hosted payload inside the launcher. 

The spacecraft gimbaled high-gain antennas are designed to support ULTRASAT's data rate ($\sim$5Mbps), and together with the GEO orbit allow for continuous real-time data download for all allowed pointings (see pointing restrictions below). 
In case of a communication failure, images will be stored on the ULTRASAT On-Board Recorder (OBR) and will be downloaded once the communication is restored. The OBR can store images of nearly 8 days of continuous observations with the nominal exposure time (300\,s).

In order to enable a quick response to targets of opportunity (ToOs),
ULTRASAT is designed with full-hemispheric telecommand antennas and pointing slew agility ($>0.5$\,deg\,s$^{-1}$), which enable starting observations at any visible position in less than 15 min from an incoming alert trigger (e.g., of a GW event) at the SOC. ULTRASAT's power system (solar panels, battery) supports at least 3 hours of ToO observations with negative power balance (in survey mode the power balance is always positive, see Section \ref{sec:survey_mode}). Assuming randomly distributed ToO's, negative power balance will occur for $\lesssim33\%$ of ToO observations.

\begin{figure}
\centerline{\includegraphics[width=0.5\textwidth]{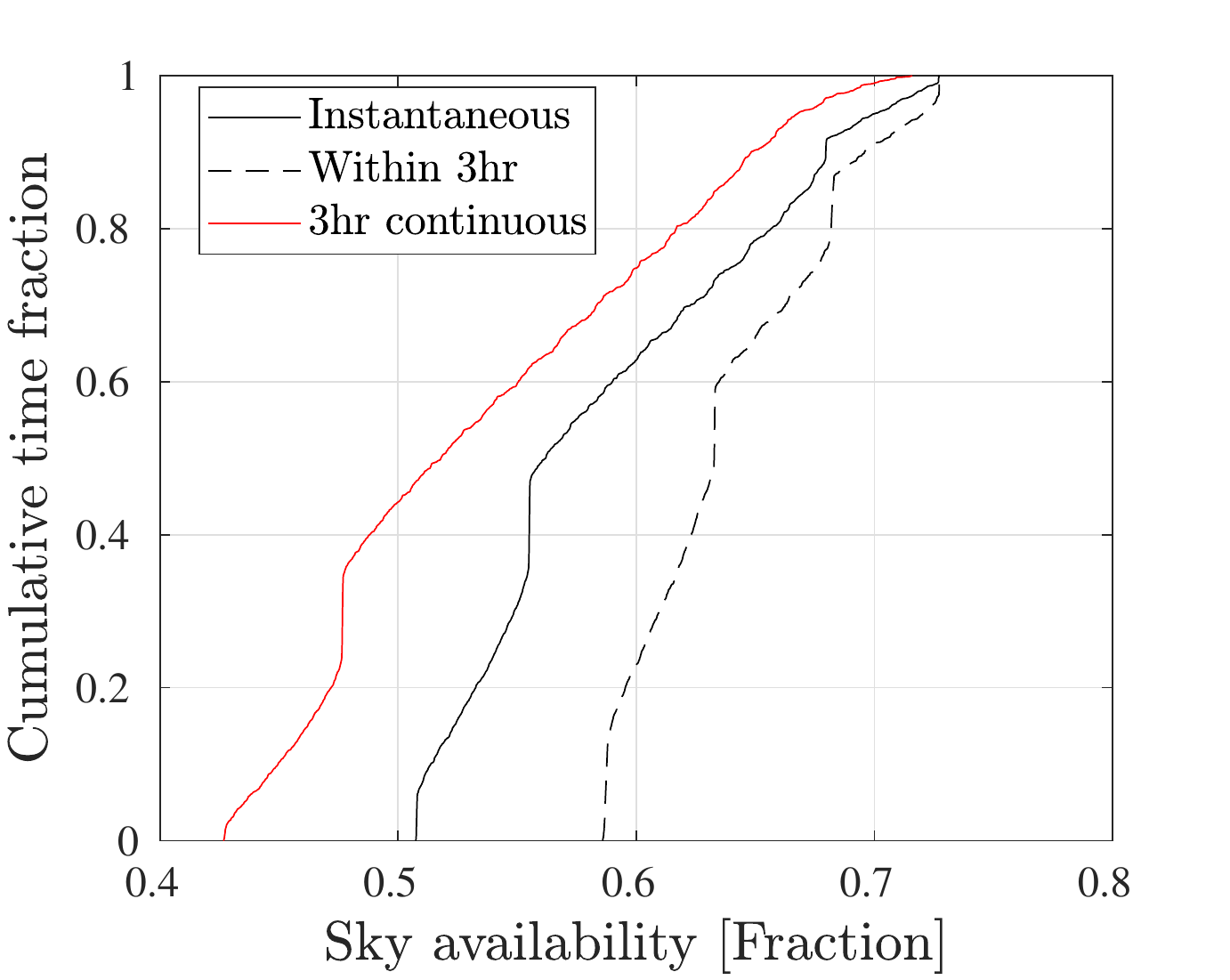}}
\caption{Sky availability for ToO observations under ULTRASAT pointing restrictions. At any given time ULTRASAT can access $>50\%$ of the sky, and $>58\%$ within 3 hrs. The minimal sky fraction for a 3-hr continuous observation (the nominal ToO duration) is 42\%, and the median is 51\%.
\label{fig:Visibility}}
\end{figure}

The (angular sky) directions allowed for ToO observations are limited by stray-light constraints. ULTRASAT is restricted to observe fields for which the optical axis is $>70$\,deg away from the Sun (this restriction also apply during slewing), $>48$\,deg from the Earth limb, and $>35$\,deg from the Moon, in order to limit the stray-light contribution to the background (see \S~\ref{subsec:P/L} for a discussion of stray-light suppression). 
Under these restrictions, ULTRASAT can access instantaneously  (i.e. at any given moment) $>50\%$ of the sky and observe $>42\%$ of the sky continuously for at least 3 hours (see Figure~\ref{fig:Visibility}). 
The fraction of the sky accessible to ULTRASAT within 3 (6) hours is $>58\%$ ($>66\%$).

\begin{figure*}[ht]
\centering
\includegraphics[width=.49\linewidth]{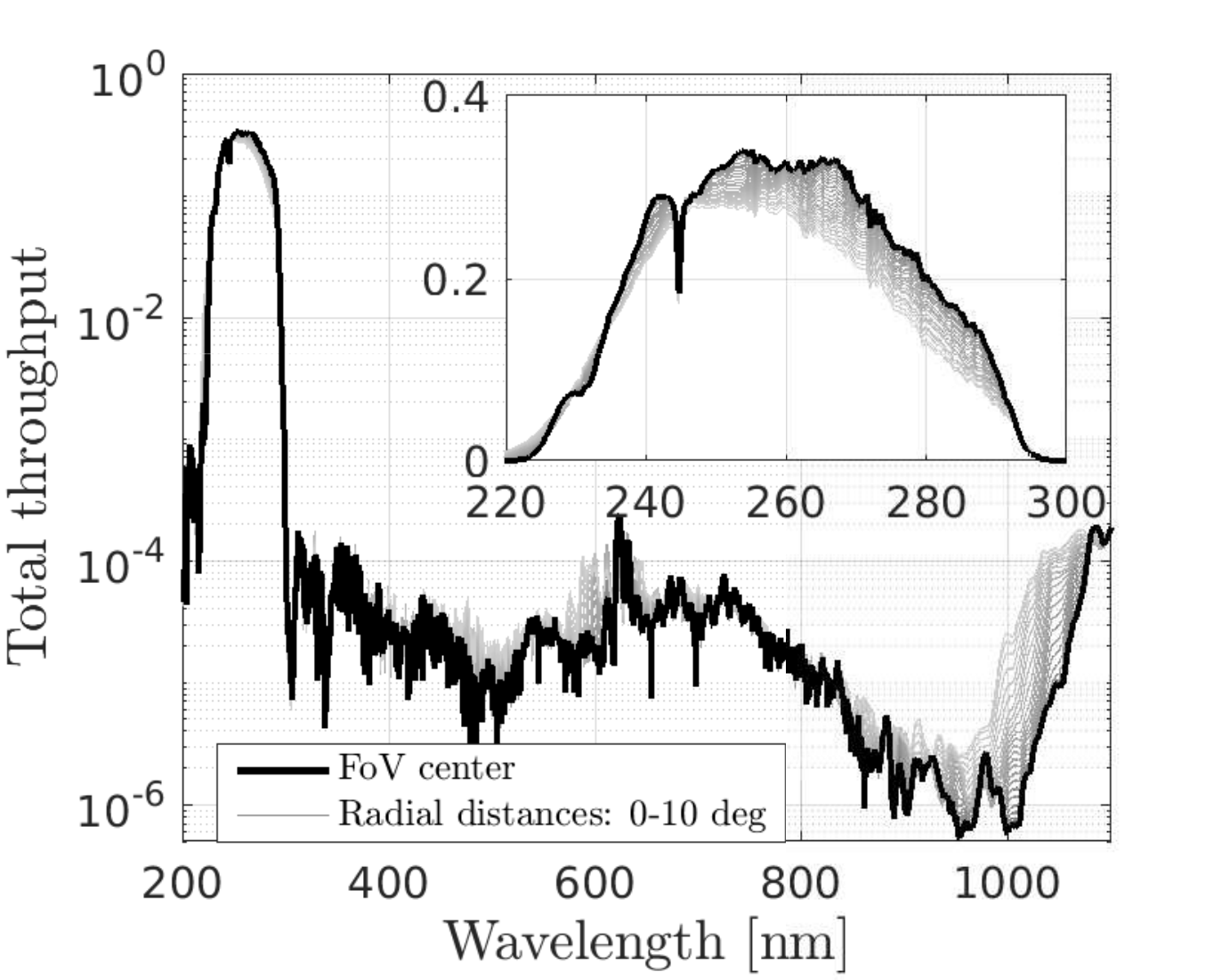}\hfill
\includegraphics[width=.49\linewidth]{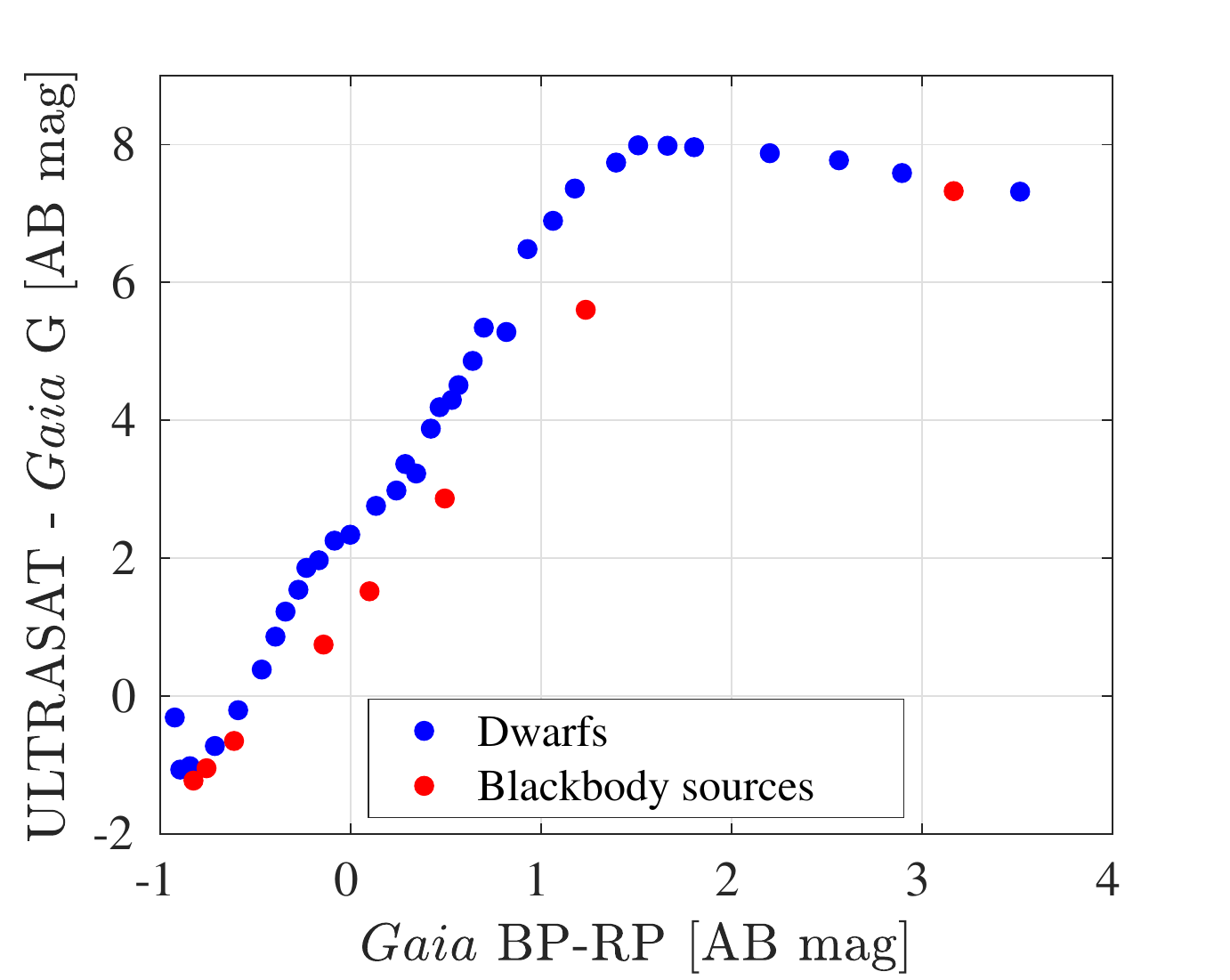}
\caption{{\it Left}: ULTRASAT's total throughput, accounting for all optical elements, the detector QE, and obscuration. The minor variation of the throughput across the field of view (gray curves) is due to the variation of the obscuration and of the angle-of-incidence distribution.
{\it Right}: Color-color relation between ULTRASAT and $Gaia$-bands, for a range of main-sequence dwarfs (blue dots) and blackbody sources (red dots).
}
\label{fig:Throughput}
\end{figure*}

ULTRASAT's pointing stability ("jitter") of $<3\arcsec$ ($3\sigma$) over 300\,s (the nominal exposure time)
is achieved using its attitude and orbit control system (specifically, the star-trackers, reaction wheels, and inertial measurement units). 
For blue sources at radial distance of $\sim5$ deg off the FOV center, the jitter contributes significantly to the image PSF, while for blue sources at other locations in the FoV, and for red sources at any location within the FoV, the jitter contribution is negligible (see \S~\ref{subsec:P/L} and Figure~\ref{fig:Eff_PSF} for more details).

\subsection{Payload}
\label{subsec:P/L}

The ULTRASAT payload has three main components - a baffle, the optical tube assembly (OTA; i.e. telescope) and the focal-plane array (FPA; aka camera). Below we give a brief overview of each of them. Detailed description of the baffle and the OTA is given in \cite{Ben-Ami.2022.A}, and in-depth description of the camera design and characterization is given in  \cite{Asif.2021.A} and \cite{Bastian-Querner.2021.A}, respectively. 

ULTRASAT achieves its wide-field capabilities via a wide-field Schmidt telescope design. A 33\,cm diameter double corrector plate, made of a single fused silica lens and a single CaF2 lens, is mounted at the entrance of the telescope. The 50\,cm diameter Zerodur mirror has a clear aperture of 48\,cm. A field-flattener assembly, made of a fused silica lens, a CaF2 lens and a Sapphire filter, is mounted $\approx$0.55\,mm in front of the focal-plane array. 

Achieving strong out-of-band attenuation, while maintaining high throughput within the operation band, is crucial for meeting the signal-to-noise (S/N) requirements. The payload design addresses the out-of-band attenuation by minimizing the {\it red leak} through a combination of coatings on the sapphire filter, the lenses, the mirror, and the detector. This leads to transmission of only $2.9 \times 10^{-5}$ (band average) of visible light while maintaining $>25\%$ overall in-band throughput. The variation of obscuration and Angle-of-Incidence (AOI) distribution across the FOV results in a minor radial variation of the overall throughput (see Figure~\ref{fig:Throughput}).

\begin{figure}
\centerline{\includegraphics[width=0.5\textwidth]{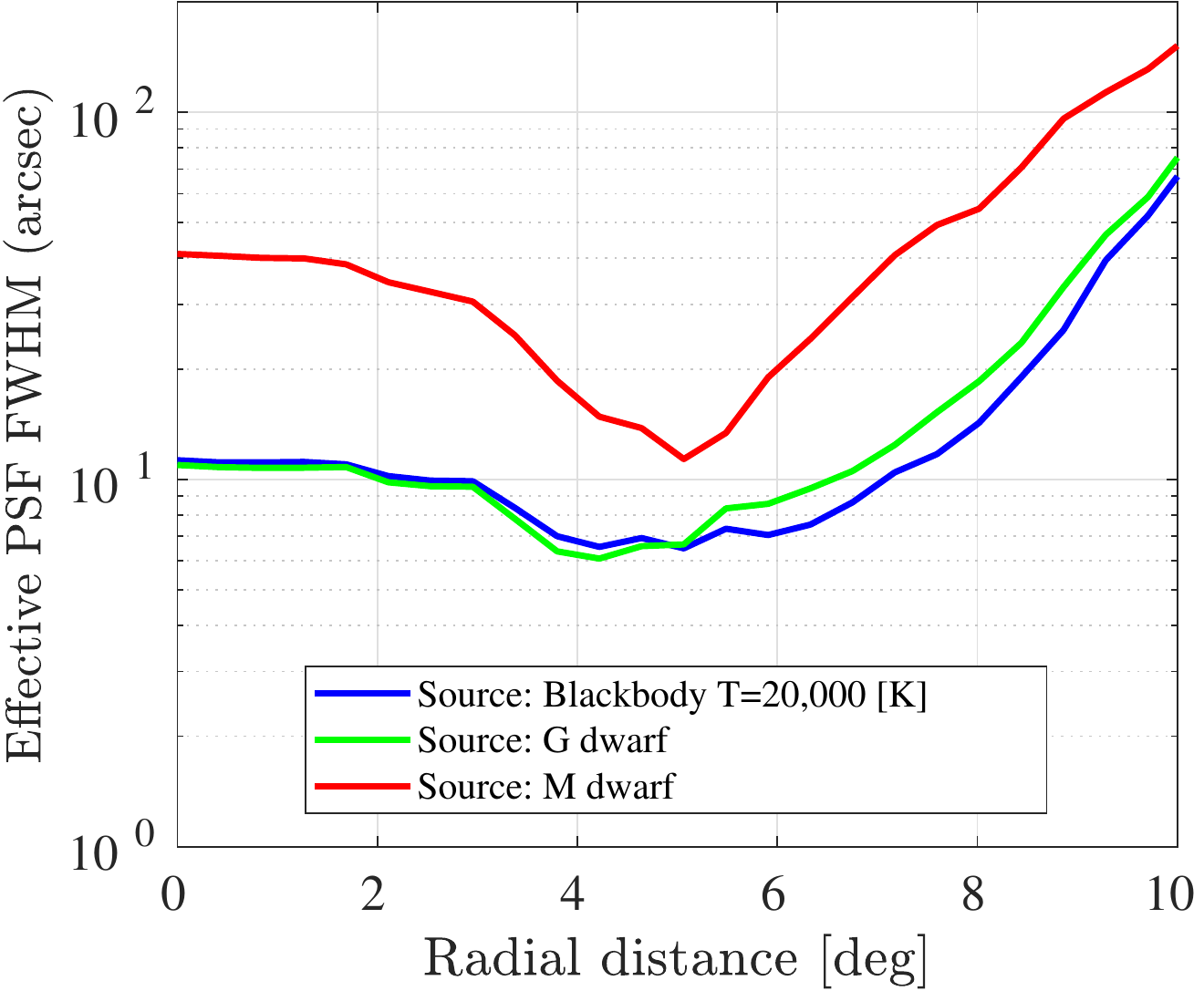}}
\caption{ULTRASAT's radial-dependent effective PSF FWHM for three representative sources -- $G$ and $M$ dwarfs, and a $T=20,000$K blackbody source. 
\label{fig:Eff_PSF}}
\end{figure}

ULTRASAT's optical design maximizes the system grasp via a chromatic radial-dependent PSF optimized to the ULTRASAT band. Figure~\ref{fig:Eff_PSF} gives the PSF full-width at half maximum (FWHM; including spacecraft jitter, thermal variations and assembly errors) for various sources at different radial positions.
The system area-averaged PSF FWHM for a $T=20,000$K blackbody source in the central 170 deg$^2$ of FoV is 8.3’’. 

A 20$^\circ$ slanted baffle is mounted in front of the telescope to prevent stray light from entering the telescope and to reduce Cerenkov radiation. The slanted geometry enables the telescope to point $>70$ deg from the Sun without allowing any Sunlight into the optical system. The baffle includes vanes to suppress the stray light from the Earth and the Moon, achieving, together with the OTA elements (in particular the Sapphire filter and the mirror), better than $2\times 10^{-11}$ stray light suppression\footnote{Stray light suppression is defined as the ratio between the scattered photon flux produced by a source (e.g., Earth) on the detector and the source photon flux at geostationary orbit.}. The second main task of the baffle is to suppress the flux of high energy electrons penetrating the outer corrector lens, reducing by a factor of $\approx 10$ the Cerenkov radiation produced as these electrons propagate through the lens, which is one of the two main noise sources (see \S~\ref{sec:snr}).

The FPA consists of four $45\times45$\,mm$^2$ back-side illuminated (BSI) CMOS detectors, with 9.5\,$\mu$m pixels.
The camera is shutter-less and is read in a rolling-shutter mode, that allows continuous exposures with minimal overhead time ($\sim 2$ milliseconds). The entire pixel array is read in less than 20\,s. High Dynamic Range (HDR) capability is achieved by dual gain 5T pixels. 
The high quantum efficiency (QE) in the ULTRASAT waveband is achieved via a high-K dielectric layer and a UV-optimized anti-reflection coating (ARC).  
See \cite{Liran.2022.A} for more details on the development and characteristics of the ULTRASAT customized sensor.

The FPA is mounted on a spider arm assembly within the telescope tube, thermally isolating the FPA, that operates at 200\,K to minimize dark current, from the telescope structure, that is kept at 293\,K. The instrumental camera noise (e.g., dark current, read noise) contributes less than 30\% to the total background noise variance (see table~\ref{tab:noise}).

\subsection{Sensitivity}
\label{sec:snr}

ULTRASAT's sensitivity (e.g., limiting magnitude and other S/N calculations) is affected by three types of noise sources -- instrumental, external and source related. ULTRASAT's design addresses and minimizes each of these, and specifically ensures that for the nominal observations the contribution of the instrumental noises is sub-dominant. In this section we first describe the contribution of each of the noise sources and then derive the limiting magnitude and S/N for various astrophysical sources. These are used later, in \S~\ref{sec:Science}, to derive their estimated detection rates. The S/N is calculated using publicly available tools (\citealt{Ofek.2014.A}; MAAT\footnote{\url{https://github.com/EranOfek}}).

For the sensitivity calculations we assume conservative observing conditions (see details below for each noise source) and a minimal observing visit per pointing of $3\times300$\,s (the three exposures are required to reliably identify and eliminate cosmic-ray signatures). The resulting contributions to the background variance of the various noise sources for a single $300$\,s exposure are summarized in Table~\ref{tab:noise}. The three external noise sources -- Zodiacal light, Cerenkov radiation, and stray light -- dominate ULTRASAT's background noise. The variation across the FoV of these external noise sources (due to the varying throughput) is negligible ($<0.1\%$). 

{\it Zodiacal light}. ULTRASAT's out-of-band ($\gtrsim300$\,nm) rejection suppresses most of the Zodiacal light, mainly by the Sapphire filter. 
The Zodiacal light varies with Ecliptic latitude and longitude, increasing for observations closer to the Sun with a relatively sharp increase at angles $<90$\,deg from the Sun. ULTRASAT's survey fields and the all-sky map observations (see \S~\ref{sec:survey_mode} and \S~\ref{sec:allsky_map}) are planned to always be at $\geq90$\,deg away from the Sun. In our conservative reference sensitivity we calculate the Zodiacal light for our survey fields close to the Ecliptic poles, when they are in a direction perpendicular to the Sun's direction. We use the Zodiac spectrum from the {\it HST} STIS instrument handbook\footnote{STIS instrument handbook, chapter 6.6, table 6.4. 
}, and adjust for the Ecliptic coordinates using the table from the {\it HST} WFC3 instrument handbook\footnote{WFC3 instrument handbook, chapter 9.7.1, table 9.4.
}. The resulting noise of 27 $e^-$\,pix$^{-1}$ per 300\,s exposure is the highest background noise.

\begin{figure*}[ht]
\centering
\includegraphics[width=.46\linewidth]{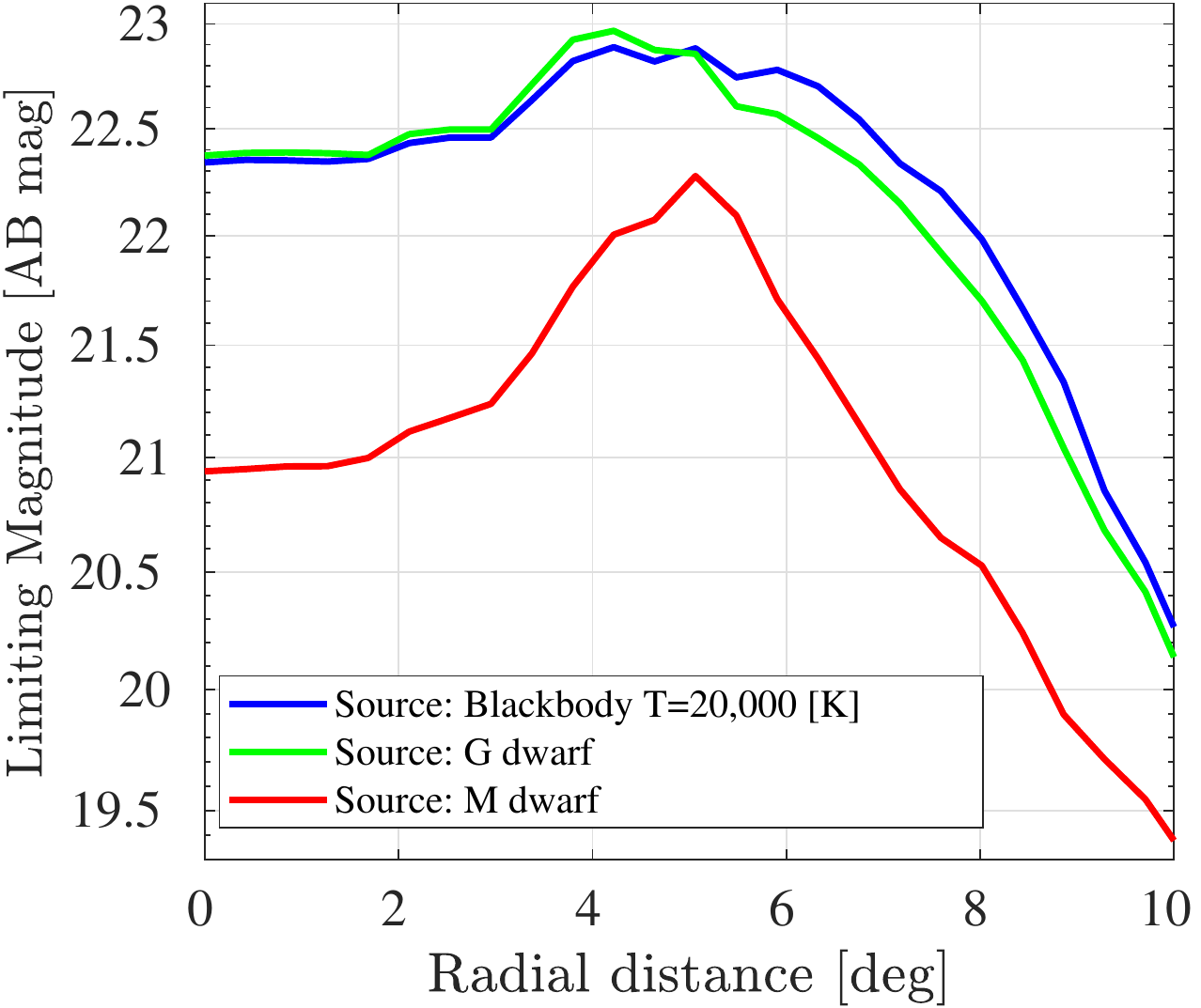}\hfill
\includegraphics[width=.52\linewidth]{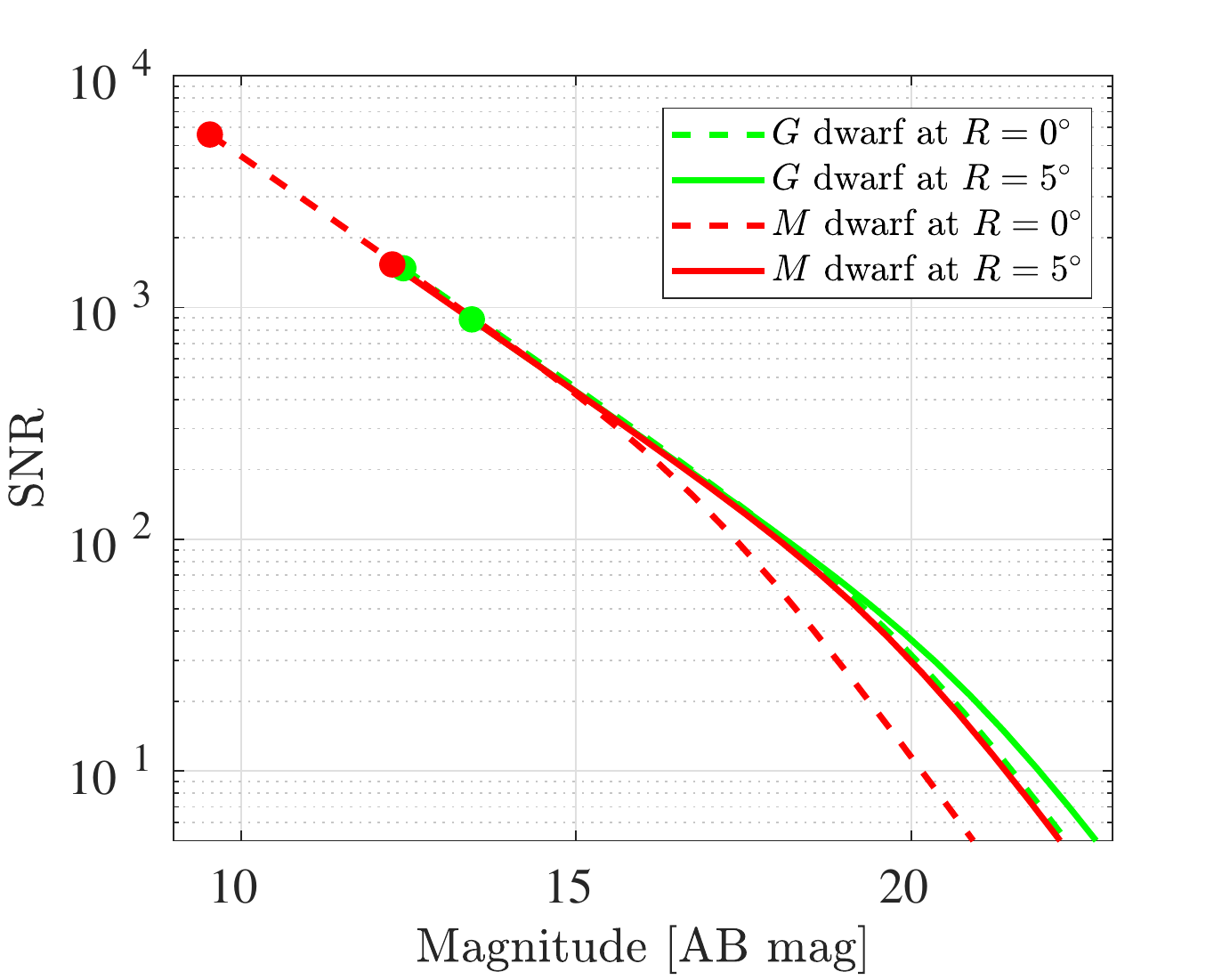}
\caption{{\it Left}: Limiting point source AB magnitude achieved in $3\times 300$\,s ULTRASAT exposures, plotted as a function of radial distance from the field center.
{\it Right}: ULTRASAT's Signal-to-Noise Ratio (SNR) as a function of AB magnitude for $3\times 300$\,s exposures, for $G$ and $M$ dwarfs at the FoV center and at $R=5^\circ$. The dots at the end of each curve mark the saturation limit.  
}
\label{fig:LimMag_SNR}
\end{figure*}

\begin{table}[]
    \centering
    \caption{ULTRASAT estimated background noise in a single $300$\,s exposure}
    \begin{tabular}{|l|c|}
         \tableline
         \multicolumn{1}{|c|}{Source} & Variance\\
         \multicolumn{1}{|c|}{} & [$e^-$/pix]\\ 
         \tableline
         \tableline
         Zodiac (perpendicular to Sun's direction) & 27\\
         \tableline
         Cerenkov (75 percentile high flux) & 15\\
         \tableline
         Stray light (max) & 12\\
         \tableline
         Dark current & 12\\
         \tableline
         Readout noise [squared] & 6\\
         \tableline
         Electronic cross-talk & 2\\
         \tableline
         Gain & 1\\
         \tableline
         Quantum yield & $<1$\\
         \tableline
         \tableline
         Total & 75\\
         \tableline
    \end{tabular}
    \label{tab:noise}
\end{table}

{\it Cerenkov emission}. Trapped energetic electrons hitting the external fused-silica corrector lens are the main source for the Cerenkov radiation illuminating the detector. ULTRASAT's baffle significantly reduces the flux of electrons reaching the outer lens (dominated by the flux entering through the baffle opening angle), and hence the resulting Cerenkov background.
The Cerenkov noise contribution was calculated using a detailed Monte-Carlo ray-shooting simulation to estimate the energy dependent electron intensity incident at various depths and radial positions across the outer lens. The electron intensity distribution was then used to derive the resulting Cerenkov emission at relevant angles (for which propagation through the optical system leads to illumination of the detector). The numeric results are consistent with those we obtained using a simplified analytic calculation (confirming in particular that the contribution to the Cerenkov light from optical elements other than the outer lens is negligible).
The trapped electron flux at GEO orbit is highly variable on timescales of days, with variations of 3-4 orders of magnitude. For our conservative noise estimations we used the 75 percentile of electron flux based on the AE9 model \citep{Ginet.2013.A}. The resulting Cerenkov noise contribution is about half of the Zodiacal.
For the 95 percentile electron flux, the Cerenkov noise will be 2.8 times larger, while for the median electron flux it will be 2.5 times smaller (thus already below the instrumental noise). Our estimates of the Cerenkov contribution are consistent with the results of \cite{Kruk.2016.A}, who studied the radiation-induced background at GEO orbit and found that for exterior glasses (e.g., un-shielded external lens) Zodiacal light dominates at $\gtrsim300$\,nm, while Cerenkov radiation dominates at shorter wavelengths (see their Fig.~8).

{\it Scattered light}. Under our pointing restrictions (described above), scattered light from Earth is the dominant stray light source. The baffle design (e.g., vanes and Acktar vacuum black coating) was optimized to suppress this contribution. The Earth's flux at GEO orbit was estimated based both on the observations of the Sentinel 5P satellite \citep{Veefkind.2012.A}, and on calculations using the Solar irradiance and Earth's albedo, with consistent results between the two methods. The stray-light flux reaching the detector was calculated by detailed Monte-Carlo simulations of light propagation through the optical system. For our conservative noise estimations we assumed full Earth at the closest allowed angle from the lowest side of the baffle. The resulting stray light noise contribution is similar to (but slightly smaller than) the Cerenkov noise.

{\it Instrumental noise}. ULTRASAT's sensor was designed to minimize the instrumental noise \citep{Liran.2022.A}. The nominal sensor design performance was already measured and verified. The sensors are kept at 200\,K, yielding a low dark current of $<0.05e^{-}$\,s$^{-1}$\,pix$^{-1}$. Long-term radiation effects (TID) will lead to some increase in dark current, but the sensors are sufficiently shielded so that the worst-case dark-current increase at the end of the mission is expected to be less than a factor of 2. The read noise was measured to be $<2.5e^-$\,pix$^{-1}$ (averaged over the PSF, taking into account also the tail of high read noise pixels). The remaining instrumental noise sources -- electronic cross-talk, gain, and quantum yield -- do not contribute significantly to the overall noise.

{\it Resulting sensitivity}. ULTRASAT's chromatic and radial-dependent PSF yields a radial and source-dependent sensitivity. Figure~\ref{fig:LimMag_SNR} shows ULTRASAT's radial-dependent limiting AB magnitude (5$\sigma$ detection in 3 co-added $300$\,s images) for various point sources, and 
the S/N achieved with 3 co-added $300$\,s images as a function of AB magnitude. For these calculations, we use either the \cite{Pickles.1998.A} spectral flux library for stellar sources or a black-body spectrum. The measured source flux is derived using ULTRASAT's throughput curve and by conservatively assuming 80$\%$ PSF photometry efficiency (i.e., 20$\%$ source flux losses due to imperfect PSF representation and fitting). The noise is calculated within an effective PSF, which is calculated by applying the ULTRASAT chromatic and radial-dependent PSF to the source spectrum propagated through ULTRASAT's throughput. A faint 24 AB mag host galaxy is assumed for the calculations, thus not dominating the background noise in these estimates. 

{\it Confusing limit}. 
ULTRASAT will observe some regions of the sky repeatedly for a total of more than 100--$10^4$ hours (e.g., the high- and low-cadence survey fields. See \S~\ref{sec:survey_mode}). Crowding may become a major issue in deep co-added images of these regions.
In \S~\ref{subsec:galaxies} below, we carry a detailed analysis of the confusion limits as a function of ULTRASAT PSF FWHM, as part of the estimates for expected galaxy counts.
The radial-dependent confusion limit in the central 170\,deg$^2$ of FoV is in the range 24.0-25.5 AB mag (achieved by co-addition of 200-400 images of $300$\,s exposures) . See Figure~\ref{fig:confusion} and \S~\ref{subsec:galaxies} for more details.

\section{Modes of Operation}
\label{sec:modes}

ULTRASAT will have two main modes of operation: (i) {\it Survey observations} during which the telescope stares at the same field (or cycle through a few adjacent fields) for a long period of time, and (ii) {\it Target of Opportunity} (ToO) observations, a triggered mode where ULTRASAT stops its survey observations and slews to a new position. In addition, at the beginning of the mission an all-sky UV map will be constructed. 

During the survey mode, the extra-galactic survey volume will be maximized by pointing ULTRASAT at fields near the ecliptic poles, minimizing Galactic extinction, and switching between North and South in summer/winter, minimizing Zodiacal background. Below we describe in some detail our current observing plan, which is used in \S~\ref{sec:Science} for the calculations related to observations designed for meeting the science goals. This plan will be re-evaluated, and potentially revised, before the beginning of the mission and annually during the mission.

\subsection{Survey modes}
\label{sec:survey_mode}

ULTRASAT will devote $\approx90\%$ of its time during the first year to a high cadence survey, and $\approx10\%$ to a low cadence survey. In both survey modes, the images will be transmitted to the ground in real-time. ULTRASAT's SOC will issue alerts of new transients detected by ULTRASAT within 15 min from image capture. Every 6 months, as the Sun switches hemispheres, ULTRASAT will slew to the region in the opposite hemisphere. During each semester ULTRASAT rotates about the center of the FoV  by $\approx$1 deg every $\approx$1 day to keep the Sun within $\approx$10 deg of the solar panels' normal.

{\it High cadence survey (21.25 hours/day):} 
ULTRASAT will be pointed towards one predetermined field near the ecliptic poles. The fields selection is based on minimizing both the Zodiacal background and Galactic extinction, such that the extra-galactic survey volume is maximized. Each field will be observed for six months, with continuous 300\,s exposures.

{\it Low cadence survey (2.75 hours/day):} 
During each seasonal (north/south) dwell ULTRASAT will cycle through 40 fields covering $\sim$8000 deg$^2$, observing 10 fields each day (i.e., a 4-day cadence per field). In this mode, ULTRASAT will stare at a given field for 15min (3 consecutive 300\,$s$ exposures), and then move to observe a nearby field. The angular distance between adjacent fields will be $<40$\,deg and the slew time between adjacent fields will be $<1.5$\,min. The survey fields will meet the Sun and Moon minimum angle limits during the entire 6-month observing window.

\subsection{Target of Opportunity (ToO)}
\label{sec:too}
Upon reception of a ToO trigger command, ULTRASAT will interrupt observations and slew to the designated pointing. At any given moment, $>$50\% of the sky will be available for ToO observations (limited by the pointing restrictions with respect to the Sun, Earth and Moon). ULTRASAT will slew to any observable point on the sky within $<15$ min from the time a trigger was received at ULTRASAT's SOC. The interruption time per ToO is nominally capped at $<3$\,hr. ULTRASAT will observe the ToO field continuously with a nominal exposure time of 300\,s.

\subsection{All-sky UV map}
\label{sec:allsky_map}
During the first six months of the mission, ULTRASAT will undertake an all-sky survey of total integration of $6\times300$\,s (1800\,s) at low Galactic latitudes ($|b|<30$\,deg) and $50\times300$\,s (15,000\,s) at high Galactic latitude ($|b|>30\deg$). The high latitude survey with an AB limiting magnitude of 23.5 to 24 mag will be $>$10 times deeper than the GALEX all-sky survey. Furthermore, it will provide an all-sky UV variability survey.

The All-sky UV map is required as a deep reference image
for transient detection using image subtraction. Specifically, following triggers of gravitational wave events (see more details in \S~\ref{subsec:GW}). The scheduling of the all-sky map observations will be optimized with respect to the LIGO/VIRGO O5 run.

\section{Science Operation Center}
\label{sec:SOC}

The ULTRASAT Science Operations Center (SOC) will be located at the Weizmann Institute of Science (WIS) in Israel, and will support all scientific aspects of the ULTRASAT mission, including observation planning, data reduction and alerts distribution. The SOC interfaces with the Ground Control Segment (GCS) at IAI, which is responsible for all direct communications with the ULTRASAT spacecraft. The GCS will acquire the spacecraft after separation from the host vehicle and maintain continuous communication through a ground antenna. 
 
The SOC will plan and schedule in advance the survey and all-sky UV imaging observations (see \S~\ref{sec:modes}). In addition, the SOC will receive real-time alerts from several sources (mainly GW events; GRB, TDE, and neutrino triggers are also under consideration). The SOC will send a new immediate ToO observing plan to the GCS in cases where a ToO observation is both relevant (meeting predefined criteria) and feasible (considering, e.g., target visibility and spacecraft power condition). Observations will start within no more than 15 minutes from the time the incoming alert was received at the SOC.
 
The GCS will distribute the science and telemetry data to the SOC for real time processing, calibration, and archiving. 
The ULTRASAT pipeline will generate a large number of data products, the goal of which is to minimize the time required to perform scientific research using the data. Focus is therefore given to quality, diversity, and ease of use. The pipeline is based on the software package described in \cite{Ofek.2014.A} and Ofek et al. (in prep).

The SOC will issue public alerts of new transients, for both survey and ToO observations, within less than 15 minutes of image capture. All other ULTRASAT data products will be made available via periodic data releases (DRs) following full calibration and verification. The proprietary period will be 12 months. Members of the ULTRASAT collaboration and the science working groups will have immediate access to all ULTRASAT data products.

\section{Science Objectives}
\label{sec:Science}

ULTRASAT will detect transient events in an unprecedented large volume of the universe, and will explore a new parameter space in energy (NUV) and time-scale (minutes to months). It is thus expected to have a major impact on a wide range of astrophysics topics, with a vast space for serendipitous discoveries.
In this section we discuss ULTRASAT's broad scientific impact across the fields of GW sources (\S~\ref{subsec:GW}), deaths of massive stars (\S~\ref{subsec:SNe}), cosmology (\S~\ref{subsec:cosmology}), gamma-ray bursts (\S~\ref{subsec:GRBs}), stars and stellar remnants (\S~\ref{subsec:stars}), exoplanets and star-planet connection (\S~\ref{subsec:exoplanets}), tidal disruption events (\S~\ref{subsec:TDEs}), active galactic nuclei (\S~\ref{subsec:AGN}), galaxies (\S~\ref{subsec:galaxies}), and the Solar system (\S~\ref{subsec:solar_system}). Table~\ref{tab:highlights} summarizes ULTRASAT's science highlights.


\subsection{Search for UV Emission from Gravitational Wave (GW) Sources}
\label{subsec:GW}

The era of GW astrophysics has begun with the detection of GW from black-hole (BH) and Neutron Star (NS) mergers by the laser interferometer observatories LIGO and Virgo  \citep{1stBHGW,Abbott17PhRvL}. The possibilities are many and exciting \citep{2019GW_roadmap,Margutti21Rev}: New tests of general relativity, a new probe of stellar death in binary evolution, and determination of the demographics of stellar remnants. Coalescences involving NS provide unique insights into the physics of our Universe: They are the most likely sites of production of r-process elements (e.g., Pt, Au)  \citep{Fernandez.2016.A,2021RvMPr-proc} and provide unique diagnostics of the physics of the densest matter in the Universe  \citep{2009PhinneyDecadal}.

The detection of the associated electro-magnetic (EM) emission will be the key to using these events for addressing fundamental physics and astrophysics questions: It will provide direct constraints on the structure of the ejected material, which will in turn provide unique diagnostics of the properties of matter at nuclear density and of the merger dynamics; It will enable us to determine whether and which r-process elements are produced and whether highly relativistic gamma-ray burst jets are produced; It will allow us to precisely localize the mergers. Determining the location in, and properties of, the host galaxy will reveal the stellar antecedents of NS binaries, and the host redshift distribution can be used to measure cosmological parameters, in particular H$_{0}$.

By 2026, GW interferometer networks are expected to provide a few to a few tens of detections of mergers involving NS per year within $\approx 300$~Mpc, with angular localization of about $\sim100$\,deg$^2$ \citep{Abbott.2016.A,Abbott.2018.A,Abbott.2020.A}. In this era, issues of GW alerts with rough parameter estimates (NS-NS, NS-BH or BH-BH; rough sky position and distance) are expected within minutes and with refined parameters within hours.

With a large fraction ($>50\%$) of the sky instantaneously accessible, fast (minutes) slewing capability and a field-of-view that covers the error ellipses expected from GW detectors beyond 2025, ULTRASAT will rapidly detect, and provide continuous UV light-curves of, the EM emission following BNS mergers identified by GW detectors. The alerts provided by ULTRASAT will further enable early ground/space-based follow-up spectroscopy and monitoring at other wavelengths. 

In \S~\ref{subsubsec:GW-UV-detection} we show that the distance out to which the EM emission will be detectable by ULTRASAT is expected to exceed the $\approx300$~Mpc horizon of the GW detectors. In \S~\ref{subsubsec:ULTARASAT-GW_advantage} we discuss ULTRASAT's advantages, compared to the capabilities of other surveys, in detecting EM emission following GW events. In \S~\ref{subsubsec:GW-UV-importance} we discuss the unique constraints that will be provided by early (hour time scale) light curves, in particular in the UV, on the structure and composition of the merger ejecta.

\subsubsection{Detecting UV emission following BNS mergers}
\label{subsubsec:GW-UV-detection}

Direct calculations of the structure and composition of the ejecta produced in NS-NS/BH mergers are highly complicated and computationally demanding since they involve a density range of over 30 decades, the entire periodic table, sub-relativistic to relativistic flows, and require consideration of neutrino transport, GR and magnetic field effects \citep{2019ShibataRev,2020RadiceRev,2022RosswogRev}. Modeling the EM output is thus a challenging but vibrant field \citep{Fernandez.2016.A,2020PhR-Nakar-EMGW,Margutti21Rev}. 

Many models predict an UV signal, with a $10^{41}-10^{42}$\,erg\,s$^{-1}$ luminosity lasting for hours to a day, that may be produced by different ejecta components (see \cite{Fernandez.2016.A,2020PhR-Nakar-EMGW,Margutti21Rev} for reviews, and recent analyses by \cite{2020TanakaLowYe,Dean_2021,2022IokaCocoon,2022TanakaEarlyHot,2022CombiUVdetectability}): free neutron beta decay in high-velocity matter, high $Y_{e}$ ejecta with lower r-process element content and opacity, hot low $Y_{e}$ ejecta, and boosted relativistic material and/or shock cooling. These predicted signals will be detectable by ULTRASAT out to $\approx300$\,Mpc.

The detection of EM emission following the nearby (40\,Mpc) NS merger event GW\,170817 lends strong support to the estimates of the expected UV signal, and highlights the advantages of ULTRASAT as an EM counterpart detection machine.  The EM counterpart of GW 170817 (AT\,2017gfo) was detected 10 hours following the merger, with a luminosity of $10^{42}$\,erg\,s$^{-1}$ and a temperature exceeding 10,000\,K.
The absolute NUV magnitude of AT\,2017gfo was about $-14.5$ at about 0.6\,days
after the merger. Furthermore, the NUV light curve was possibly decaying between
the two first epochs, indicating that the NUV peak luminosity
was earlier and brighter.
Given these parameters, ULTRASAT can detect AT\,2017gfo-like events to a distance
of about 240 (400)\,Mpc, in 15\,min (2\,hr) integrations.
However, it is likely that the AT\,2017gfo was brighter in NUV in earlier times.
In this case, ULTRASAT may detect these events to larger distances.

\subsubsection{ULTRASAT's advantages in detecting EM emission following GW events}
\label{subsubsec:ULTARASAT-GW_advantage}

Major ground-based optical/IR (OIR) facilities will also react to GW events. These facilities have the sensitivity to detect the optical emission,
but have much lower real-time sky accessibility ($\sim6\%$ per location, given day/night, horizon, moon, and weather effects). 

The ground-based detection of EM emission 10\,hours following GW\,170817, by targeting known galaxies in the GW source error volume, is an impressive result. However, earlier detection would be crucial for obtaining spectra and early light curves, which are essential for determining the properties of the ejecta (see more below). Furthermore, although the GW angular localization is expected to improve to 10-100\,deg$^{2}$ by 2026, EM transient identification will remain challenging as the GW detection horizon is increased to $\approx 300$\,Mpc, implying a larger error volume containing hundreds to thousands of galaxies and fainter signals with a larger number of background transients that would need to be classified.

Before the launch of ULTRASAT, several more GW events with exceptional properties (dense ISM, a rare on-axis orientation) might have been localized to their host galaxies. If so, the EM/GW frontier will be in early localization and in constructing larger, unbiased samples and detailed study of BNS mergers, for which ULTRASAT is ideally suited. 

Finally, let us comment about EM searches in other bands. It is generally accepted that GW events involving NS can produce short hard gamma-ray bursts. So X-ray and $\gamma$-ray searches are clearly useful. However, there is sound empirical evidence that the prompt high energy emission is strongly “beamed” (conical emission). Only $\sim$1 in 100 events will be aimed towards us and detectable at high energies, to distances larger than tens of Mpc. This is well supported by the very weak X/$\gamma$-ray counterpart of GW\,170817, that would not have been detectable beyond 60\,Mpc. The explosive energy of the coalescence will also drive a strong shock in the ambient medium, from which radio emission is expected to be detectable after months to years. The radio channel is attractive because it is isotropic. However, the long delay precludes key observational diagnostics for r-process elements – specifically OIR spectroscopy on timescales of days to a week. This conclusion is also well supported by the radio emission from GW\,170817, that was detected with a two-week time delay despite its proximity.

\subsubsection{The importance of early UV measurements}
\label{subsubsec:GW-UV-importance}

Despite the extensive observations and study of the EM emission following the nearby BNS merger GW\,170817, large uncertainties remain regarding both the structure and the composition of the ejecta.

\noindent[1] The observed UV-IR emission is consistent with the emission of radiation from a mildly relativistic expanding ejecta, which is being continuously heated by radioactive energy release at a rate corresponding to radioactive elements heavier than the Iron group (e.g. \cite{Kasen17Nat,2017DroutKN-LC,Rosswog17,CowperthwaiteBerger173KN,Tanaka17,PeregoRadice173KN,2018WOKG}). This is remarkably consistent with the "kilonova" emission predicted to follow NS mergers \cite{Fernandez.2016.A}.
However, the detailed properties of the ejecta, which are inferred from observations, are inconsistent with those obtained in merger simulations (see \citet{nedora_2021b} for a recent detailed discussion). In particular, the mass of the ejecta is larger than obtained in simulations, and it is difficult to explain (e.g. \cite{2018WOKG,2018MetzgerQuataertMagnetarKN,2018FujibayashiViscousEjecta}) the existence of a fairly massive, $\simeq0.05M_\odot$, fast, $v\sim0.3c$, component with low opacity corresponding to a large initial electron fraction $Y_e$ and a low Lanthanide mass fraction $X_{\rm Ln}$, that is inferred from the early UV/blue emission \citep{2018WOKG}.
Alternative models have thus been proposed, in which the blue emission is produced by boosted relativistic material and/or shock cooling of an expanding mildly relativistic shell (e.g. \cite{2017Kasliwal,Piro17,Gottlieb18Cocoon}). 

\noindent[2] The abundances of elements produced in GW\,170817 is not yet clear. The blue to red evolution of the emission may be explained by the existence of several ejecta components characterized by largely differing compositions, with higher opacity components corresponding to $X_{\rm Ln}\approx10^{-1.5}$ dominating at later times \citep{Kasen17Nat,2017DroutKN-LC,Rosswog17,CowperthwaiteBerger173KN,Tanaka17,PeregoRadice173KN}. We have
shown \citep{2018WOKG}  (see also \cite{2017Natur.551...75S}) that an alternative explanation is possible, in which the entire ejecta is composed of low opacity material corresponding to $X_{\rm Ln}\approx10^{-3}$, spanning a wider velocity distribution than previously assumed. 
Inferences of composition based on spectral analyses (e.g. \cite{Jerkstrand22NLTE,Hotokezaka21NLTE,Hotokezaka22Spitzer,Watson19Sr,Perego22Sr,Gillanders22Sr,Domoto22Ln}) are challenged by the partial atomic data (opacities, excitation/ionization cross sections) available for heavy elements, and by the large density of lines of such elements combined with relativistic expansion velocity. \cite{Watson19Sr,Perego22Sr,Gillanders22Sr} find that features in the early ($\sim1$\,d) spectra may be explained as due to first peak $r$-process elements (Sr \& Zr) with low $X_{\rm Ln}$ ($<5\times10^{-3}$ \cite{Gillanders22Sr}), while $X_{\rm Ln}$ values inferred from (the large IR opacity implied by) later spectra differ widely between different analyses \citep{2018WOKG,Gillanders22Sr,Domoto22Ln}.

Different models for the structure and composition of the ejecta differ in their predictions for the early, hour time scale emission, in particular in the UV. Earlier detection of the EM transients and early UV light curves will therefore be highly useful for discriminating between these models (e.g. \cite{Arcavi.2018.A,2020TanakaLowYe,Dean_2021,2022IokaCocoon,2022TanakaEarlyHot,2022CombiUVdetectability,2022Dorado,2022QUVIK}).


\subsection{Characterizing the Death of Massive Stars}
\label{subsec:SNe}

Most stars of $>8M_\odot$ end their lives in supernova (SN) explosions, which create and distribute the majority of the heavy elements \citep[e.g.,][]{woosley2002}. The explosion mechanisms are a subject of vigorous research. Theoretical models begin with initial conditions – the assumed structure of a model star – and attempt to predict the properties of the resulting explosion. Progress thus requires assembly of a data set connecting observations of pre-explosion stars with measured explosion properties. Such observational constraints are critical, but exceedingly scarce – few SNe have an identified progenitor visible in pre-explosion imaging \citep[e.g.,][]{smartt2015}, and few more are likely to be found in the foreseeable future, due to the low rate of such events in very nearby galaxies where individual stars can be observed. Charting the fate of the diverse populations of massive stars to specific explosive outcomes is important in order to elucidate the feedback of massive stars on their gaseous environments, and their role in galaxy evolution.

Detection of SNe shortly after explosion and observations following the early emission from such events will enable  dramatic advances in our understanding of the way massive stars evolve shortly prior to explosion, the connection between the properties of the progenitor stars and the resulting explosions, and the physics of the explosive process.  

The earliest emission of radiation from a SN explosion is associated with ``shock breakout'' \citep[e.g.,][]{waxman2017,Levinson2020}. As the radiation mediated shock (RMS) that drives the ejection of the SN envelope expands outwards, the optical depth of the material lying ahead of it decreases. When the optical depth drops to $\sim c/v_{shock}$ where $v_{shock}$ is the shock velocity, radiation escapes and the shock dissolves. Such breakout may take place once the shock reaches the edge of the star, producing a bright, $10^{44}-10^{45}$\,erg\,s$^{-1}$, X-ray/UV flash lasting typically over the progenitor light crossing time $R_*/c$ (seconds to a fraction of an hour, depending on the radius of the exploding star $R_*$), forming the first electromagnetic signal that can reach an external observer. 

Alternatively, the breakout may take place at larger radii, within the circum-stellar material (CSM) ejected from the progenitor star prior to the SN explosion, e.g. by a steady stellar ``wind'' or by an episodic ejection of an outer envelope shell, provided that the CSM optical depth is larger than $c/v_{shock}$ \citep[e.g.,][]{Ofek2010}. In this case, the breakout time scale may be extended from $\sim {\rm hour}$ to many days \citep[e.g.,][]{Ofek2010}.

Following the shock breakout flare, the hot ejecta cool and expand, emitting UV/optical emission from the expanding envelope (the “shock cooling” phase), with L$\sim10^{43}$\,erg\,s$^{-1}$ on a day time scale.

\subsubsection{Simple shock breakout observations}

Due to the short duration of the breakout pulse, only a handful of cases have been observed with an indication for a breakout signal \citep[e.g.,][]{Campana2006,Soderberg2008,Gezari2008,Gezari2015,Schawinski2008}. ULTRASAT will change that.

For spherical stars with standard density profiles, and lacking dense CSM, the duration of the shock breakout flare provides a direct measurement of the stellar radius at explosion ($R_*/c$).
This duration ranges from minutes to about an hour for supergiant stars, making the ULTRASAT cadence uniquely powerful to obtain such measurements. 

However, the observed duration may be dominated by the intrinsic pulse duration rather than by light travel time, $R_*/c$, as well as by differences in the arrival time of the shock to the stellar surface at different (angular) positions in strongly a-spherical explosions or progenitors \citep{Katz2012,Afsariardchi2018}. In such cases, the breakout flare temporal structure provides instead a unique probe of the pre-explosion inner structure of the progenitor star, that is mostly inaccessible to other probes  \footnote{We note though that the $>R_*/c$ breakout duration, obtained in the 3D calculation of a breakout from a convective red supergiant envelope in \citep{Goldberg2022}, is dominated by the intrinsic duration of the pulse (which is significantly longer than $R_*/c$ for the chosen progenitor parameters), rather than by the convective density inhomogeneities \citep{Morag2022}}.

Though theoretical details of the shock breakout flares at very early times ($<1$ h) are still debated (see. e.g., Figure 1 of \citealt{Ganot.2016.A}), the UV shock-breakout emission must match smoothly to the well-understood, and observed subsequent shock-cooling phase. The high temperature of the breakout and shock-cooling implies that the UV and optical bands are in the Rayleigh-Jeans regime. The decrease in flux due to falling temperature is nearly offset by radius expansion. Thus, for large (supergiant) progenitors the ratio of breakout to shock cooling flux is close to unity at a few hours. Using these robust flux estimate and observationally-calibrated event rates we can determine that ULTRASAT is expected to detect about 15 flares per year (\citealt{Ganot.2016.A}; see Figure \ref{fig:Rates}).

\subsubsection{CSM shock breakout observations}

Consideration of CSM breakouts is motivated e.g., by the indication that pre-SN “precursors” are common for SNe IIn \citep[e.g.,][]{Ofek2014, Strotjohann2021} and occur also in other SN types \citep{Foley2007,Jacobson-Galan2022}. These month-long precursor events occurring within the final years prior to explosion provide direct evidence for intense mass loss episodes in many SN progenitors shortly before the explosion. This challenges the canonical picture \citep[e.g.,][]{Langer2012} of a rapidly evolving (through nuclear burning) core surrounded by a nearly time independent envelope. Various mechanisms have been proposed to generate rapid mass loss, including pair instability pulsations \citep{Woosley2007}, convection and radiation driven instabilities \citep{Suarez-Madrigal2013,Smith2014}, waves excited by core convection \citep{Quataert2012,Fuller2018}, and common envelope interaction \citep{Chevalier2012,Soker2013}.

Observations of CSM breakout flares provide information about the properties of the surrounding CSM, and, most interestingly, on the rate and duration of the mass loss episode \citep[e.g.,][]{Ofek2010}, which map the violent pre-explosion evolution of massive SN progenitors, which is observationally hard to observe and theoretically poorly understood. 

Furthermore, non-relativistic CSM breakouts are interesting because they may be the sources of several classes of powerful transients. Non-relativistic CSM breakouts are considered as possible explanations of (at least part of) the super-luminous SN (SLSN) class \citep{Ofek2010,Chevalier2011,Balberg2011,Ginzburg2012,Moriya2013}, of “double peak” SNe, of some rapidly evolving transients \citep[e.g.,][]{Margalit.2022} and of the early part of the emission of SNe of type IIn \citep{Ofek2014b}. 
Specifically, early UV observations were found to be essential to accurately estimate the temperature, radius, and bolometric luminosity of interacting SNe, because they provide a better handle on the blackbody spectrum shape compared to visible light alone. Such observations have recently shown that SNe IIn are hotter and brighter than previously thought, with their radius growing faster \citep{Soumagnac.2020.B}. Additionally, early UV observations can help determine the geometrical distribution of the CSM surrounding these events, shedding light on the mass-loss processes which occur before the explosion and on the nature of the progenitors. Recent studies using early UV observations have found that at least a third of SNe IIn have non-spherical CSM \citep{Soumagnac.2019.A,Soumagnac.2020.B}, challenging the usual assumption of spherically symmetric models. ULTRASAT will provide UV light curves for about a thousand of SNe IIn and enable to perform such studies on a much larger scale. This could open a new chapter in the study of SNe IIn, allowing for a stronger constraint on the number of aspherical CSM cocoons and bringing us closer to understand their progenitors and explosion physics.

Relativistic breakouts are also very interesting, especially for high-energy signals \citep[e.g., GRBs and XRFs][]{WaxmanCampana07,Campana06,Calzavara04,Tan01,Budnik10,Katz10,NakarSari12}. ULTRASAT detection and prompt alerts of such rare phenomena could provide highly interesting information.

\subsubsection{Shock cooling observations}

Following the breakout flare, the shock cooling emission phase provides additional unique signatures of the structure of the progenitor star (including radius, envelope mass and surface composition) and of its mass-loss history close to the explosion (and possibly also of CSM density inhomogeneities, e.g. \citealt{Fryer2020, Goldberg2022}). 

For non-relativistic ($v/c<0.1$) breakouts from stellar surfaces, existing theoretical analyses \citep[e.g.,][]{Rabinak.2011.A,Sapir2017,Katz2012,Sapir2013,Nakar2010,Kozyreva2020,Piro2021,Morag2022}, provide a good understanding, and tools for accurate description of the radiation emitted during and following breakout \citep[see][for reviews]{waxman2017,Levinson2020}. 

Early SN observations were used to set, utilizing these theoretical analyses, important constraints on the progenitors of SNe of Types Ia, Ib/c and II \citep{Maoz2014,waxman2017,Levinson2020,Soumagnac.2020.A,Irani2022}. A major challenge facing utilizing the early emission for a systematic study of SN progenitors and explosion parameters is obtaining the required early, $<1$\,d, high cadence, $\sim1$\,hr, multi-band observations: for most shock cooling observations, early multi-band observations, including in particular at short, UV, wavelengths, are not available at the high cadence and accuracy required for an accurate determination of model parameters \citep[see, e.g.,][for examples and discussion] {Rubin2016,Soumagnac.2020.A, Ganot2022}. UV measurements are essential for an accurate determination of the high color temperature \citep{Rabinak.2011.A,Sapir2017,Rubin2016}; high cadence multi-band observations are required for a determination of the relative extinction \citep{Rabinak.2011.A}, which strongly affects the inferred luminosity and color temperature. ULTRASAT will revolutionize this field, with both the quantity and quality of data expected to improve significantly. This will enable a systematic accurate determination of progenitor and explosion parameters based on shock cooling, and possibly shock breakout, observations.

In summary, studying the early SN emission provides unique information on the SN progenitor and its pre-explosion evolution, which cannot be directly inferred from later time observations. This information is highly instructive for the study of the supernova explosion mechanisms, which are not fully understood despite many years of research \citep[e.g.,][]{Janka2016,Burrows2021,Maoz2014}.


\subsection{Cosmology  \& Type Ia SNe }
\label{subsec:cosmology}

Standard candles such as Type Ia supernovae (SNe Ia), plateau core-collapse supernovae (SNe IIP) and kilonovae (KNe) can be used to derive distances in the Universe. Together with redshifts these constrain the physical properties of the Universe at large scales. Distant SNe Ia were fundamental in the discovery of the accelerated expansion of the Universe and the existence of dark energy \citep{perlmutter99,riess98}. More recently, standard candles in our local neighborhood indicate that the local expansion velocity (Hubble constant) is in tension with values predicted based on CMB observations \citep{2019ApJ...876...85R,2020MNRAS.498.1420W} (however, see \cite{Rigault:2014kaa, Mortsell:2021tcx}).
Looking towards the future, large field-of-view surveys such as ZTF and LSST will provide tens of thousands of supernovae at intermediate distances \citep{2019JCAP...10..005F,2018arXiv180901669T}. These samples can be used to construct peculiar velocity maps of the full distribution of matter, i.e. also including the otherwise "invisible" dark matter \citep{2019BAAS...51c.140K}. Standard candles, observed at different redshifts, thus uniquely allow us to probe all aspects of cosmology, from the nature of dark energy and dark matter through tests of the early Universe to tests of general relativity. Today, the discovery and classification of large numbers of transients have been largely streamlined \citep{2019A&A...631A.147N,2020ApJ...895...32F} and the fundamental limitation has instead shifted to the possible impact of systematic effects \citep{2019ApJ...874..150B}. Even rare objects, such as Kilonovae and strongly lensed SNe, where any discovery provides immediately new insights \citep{2017Sci...356..291G,2017Natur.551...85A}, are facing  systematic uncertainties.
In the following, we discuss how ULTRASAT observations can have significant impact on these cosmology related questions.

\subsubsection{SN Ia: progenitors, dust and cosmology}

Thousands of SNe Ia have been observed with high precision during the last decade, but these observations, typically occuring around peak light, have still not answered the fundamental question of what triggers the explosion. A SN Ia is driven by the thermonuclear explosion of a white dwarf, but the exact nature of the precursor system and how the explosion is ignited is still not known. Two main progenitor scenarios have been identified, either involving the merger of two double degenerate white dwarfs (DD) or mass transfer from a companion onto a Single Degenerate (SD) white dwarf \citep{2016IJMPD..2530024M, 2018PhR...736....1L}. For each of these, further unknowns create additional uncertainty: Is the companion a main sequence, giant or helium star and what is the role of circumstellar medium (SD)? How violent and clumpy is the merger, and is the detonation caused by an initial explosion in an outer layer of material (DD)? It is likely that several of these explosion scenarios are realized in nature, which could map to the intrinsic diversity among SNe Ia \citep{2013ApJ...770L...8P,2019NewAR..8701535S}. 
The question about the progenitors, besides being of fundamental interest, also implies a systematic uncertainty for cosmology as the potential progenitor scenarios can be expected to evolve differently with time or occur more frequently in some galactic environments. As an example, DD scenarios involving two degenerate objects can be expected to be found also long after star formation occurred while SD models typically require a younger companion stars \citep{2012PASA...29..447M}. Observational evidence does suggest cosmological differences between SNe Ia in different environments exist \citep{2013A&A...560A..66R}. 
Any unaccounted difference in luminosity between these progenitor classes could quickly bias cosmological constraints. 

Lightcurves of SNe Ia are famously standard close to peak light, as the strength of the detonation erases the signatures of the progenitor setup. Distinguishing these instead requires to observe at very early phases (or very late). Comparisons between theoretical predictions yield significant differences during the first hours/days \citep{noebauer,2016ApJ...826...96P, 2020A&A...642A.189M}. This is particularly true for the SD scenario, where the expanding ejecta is expected to strongly interact with the companion star for certain viewing angles \citep{kasen10, 2017MNRAS.465.2060B}. 

Extensive searches during the last decade have yielded claims of individual SNe both with and without an early flux excess \citep[e.g.][]{2011Natur.480..344N, 2015Natur.521..328C, 2016MNRAS.459.4428K, 2017ApJ...845L..11H, 2017Natur.550...80J, 2018ApJ...852..100M, 2020ApJ...898...56M, 2020ApJ...900L..27S, 2023MNRAS.521.1162D, 2023arXiv230305051L}, but consistent studies of larger samples are limited and hard to interpret \citep{2011ApJ...741...20B, 2013ApJ...779...23M,2015Natur.521..332O,2018A&A...614A..71N, 2021ApJ...908...51F, 
2022arXiv220707681B}. In particular, it has been seen that both SD and DD scenarios are expected to produce early variability through effects such as Ni mixing or heating from outer Ni shells / clumps \citep{2021MNRAS.502.3533M}.

UV observations have the power to finally solve the progenitor puzzle, as the additional information allows to disentangle the currently discussed models: Companion interaction is expected to produce a shock-like signature at early times where the flux increase is even larger at short wavelengths while radioactive material distributed to the outer ejecta will at least partially absorb UV emission \citep{2018ApJ...861...78M}. 
As it is now likely that multiple explosion channels exist, and that each could be associated with stochastic variations (e.g. due to viewing angle), a significant statistical sample of O($100$) SNe Ia with early ($<3$ days) UV observations will be needed.   
ULTRASAT is uniquely positioned for obtaining this sample, and thus for solving the long-lasting progenitor debate.

An ULTRASAT SNIa sample could also be used to directly improve SNIa distances and thus cosmological parameter estimates. The standardized SN peak magnitude for large samples currently displays an intrinsic scatter around $0.13$\,mag \citep{2014AA...568A..22B}, but several extended standardization methods shows that the underlying true intrinsic dispersion is likely at least $30\%$ lower \citep{2015ApJ...815...58F}. U-band features have been shown to be one possible way to achieve this reduction \citep{2018A&A...614A..71N}.
An improved intrinsic dispersion directly translates into improved cosmological constraints, in a way which for example an increased sample does not. Interestingly, a low redshift SNeIa sample observed in the ULTRASAT band (in addition to the optical bands accessible with many surveys) maps well to the $g$-band of the LSST survey for SNeIa detected at redshifts $z\sim0.9\pm 0.1$, allowing for a sensitive test for SNeIa evolution over more than half the Universe's age.   
So far, only a handful of SNIa have good UV coverage, but already this small sample shows interesting signs of variation not easily captured in the standard lightcurve analysis \citep{Brown_2018}. It is plausible but untested that the standardization potential extends, or even increases, into the UV. If this would be the case, ULTRASAT observations could be used to immediately improve cosmological constraints. 

Photon absorption and/or scatter by dust along the line of sight constitutes a fundamental uncertainty for luminosity distances to both SNe Ia and SNe~IIp \citep{2011ARNPS..61..251G}. This effect causes a systematic bias whereby objects appear fainter, and thus more distant, than they really are. Extinction is routinely corrected for as dust absorption is stronger at bluer wavelength, assuming  color measurements are obtained \citep{1989ApJ...345..245C,1998ApJ...500..525S}. However, for cosmology this corrections needs to be accurate at the 1\% level, far below what standard methods provide \citep{2003ApJ...599..408B}. Reaching this level  requires knowing both the intrinsic transient color as well as the wavelength dependence of the dust absorption. 
The distribution and property variations of dust is only well studied in the most nearby galaxies, while  supernovae are more likely to explode in particular (star forming) galactic environments. An extreme example of this is the potential existence of dust in the immediate supernova environment, possibly previously ejected, which might evolve in time through the interaction with the explosion itself. 

In fact, attempts to model reddening of SNe~Ia empirically during the recent decades have yielded models that vary significantly from the expected properties of any kind of dust~\citep{2014A&A...566A.102S, 2015MNRAS.453.3300A}. The potential cause of this difference was at the core of the different approaches taken by the two teams who detected the expansion acceleration, and it has not yet been resolved. The underlying problem is that supernovae are typically observed at optical wavelengths where it has turned out to be difficult to separate variations due to dust absorption with intrinsic variations in the SN SED. Both can reasonably be expected to vary e.g. with the surrounding environment. 

The ULTRASAT observations have the opportunity to fully reform our understanding of dust absorption of SNe through directly probing the kind of dust individual SNe have encountered, as well as look for any variation in time. This is because different dust models show significant variations in the UV, even for similar amounts of extinction in the optical. In addition, extinction in the UV is very pronounced, making this an easy observation. So far, such studies are confined to only a small set of SNe Ia observed in the UV by HST \cite{2015MNRAS.453.3300A}.  

\subsubsection{SN Ia: rates}

Estimating the expected rate of Type-Ia SNe (SNeIa) observable by ULTRASAT is difficult, since the established emission parameterizations do not extend into the UV (e.g.\ \cite{salt2,salt3}), and theoretical models show a large variation. 
In order to nevertheless obtain a robust estimate of the expected rate, we have used the fact that the rest frame UV band of high-redshift SNeIa falls into the well observed optical bands. We focused on data from the SNLS survey \citep{2014AA...568A..22B}.
The redshifted $g$-band light curves from SNLS can be approximated as rest-frame ULTRASAT UV light curves for a redshift $z = \lambda_g / \lambda_{\rm UV} - 1 \approx 0.92$,
where $\lambda_g \approx 5000$ \AA~and $\lambda_{\rm UV} = 2600$ \AA. The difference in limiting magnitudes between the SNLS $g$-band and ULTRASAT translates to a redshift of $\sim 0.2$, up to which ULTRASAT would still have similar sensitivity. The SNeIa of the SNLS/JLA sample within the range $z = 0.92 \pm 0.1$ were analysed for emission in the SNLS $g$-band. 
The light curves of the 46 objects in the high redshift bin were evaluated both by eye, and by using statistical information. Seventeen objects where selected that have at least one UV detection, i.e.\ 37\% of the initial sample. Given the high-redshift of the SNeIa sample, we must expect a significant Malmquist bias. Indeed, in comparison to a large sample of local SNeIa from ZTF, it became evident that the SNLS SNeIa are missing redder objects ($c<0.1$). The redder objects make up about 50\% of the ZTF sample, and accordingly, and it is unknown if these would show UV emission, if probed with a more sensitive instrument (if the reddening is due to dust one would expect to observe a large fraction of the nearby objects). Accordingly, a conservative assumption is that the 50\% missing red objects are not detectable in the UV.  

Combined, we can make a robust prediction that $\ge 18\%$ of SNeIa below 0.2 are detectable in the UV by ULTRASAT. For a volumetric rate of SNe Ia of $3\times 10^{-5}\, {\rm Mpc}^{-3}\,{\rm yr}^{-1}$ \citep{SDSS:2010ays} and a survey area of 8000 deg$^2$, the rate will be $\ge 300\,\,(2400)$ SNeIa per year up to a redshift of 0.1 (0.2). This is sufficient statistics to make the above mentioned studies promising.

\subsubsection{Hubble constant from Kilonovae, time delay of lensed SNe and other exotic transients}

Gravitational waves can function as cosmological sirens and in principle allow sub-percent measurements of the Hubble parameter, $H_0$ \citep{Schutz:1986gp,LIGOScientific:2017adf}. The major uncertainties are caused by the distance measurement and inclination of the GW source, both almost fully degenerate with $H_0$. The redshift can most easily be determined from the detection of the optical counterpart (kilonova) caused by a neutron star merger. The most promising methods for estimating inclination focus either on measuring the GRB afterglow time-delay, the color evolution of the early kilonova lightcurve or through polarimitry \citep{2022Univ....8..289B}.
Both the afterglow as well as the early kilonova evolution are prime targets for high-cadence ULTRASAT observations. Current radiative transfer kilonova studies do not focus on the UV region, but it is clear that the dominant inclination effect is in the u-band (the bluest band included), where inclination cause a four magnitude variability during the first day after explosion \citep{2023MNRAS.520.2558B}. The opacity difference between "blue" and "red" kilonova components are likely even stronger at UV wavelengths, making colors based on ULTRASAT observations an excellent inclination indicator.

An alternative path to measure the Hubble constant is the time-delay between multiply (strongly) lensed transient sources. This technique has now reached maturity for lensed quasars \citep{2020MNRAS.498.1420W} and is expected to become of use for supernovae within the next years \citep{2018ApJ...855...22G}. As gravitational lensing is achromatic, ULTRASAT will not in itself increase the sensitivity for strongly lensed sources. However, as the cosmological constraint directly depends on the precision with which the time-delay can be measured, blue and short-lived events such as the shock-breakout of core collapse SNe could allow ULTRASAT to measure the time-delay between strongly lensed events even when these cannot be spatially resolved. 
Such events are rare, but the shorter signature will allow to probe less massive lenses, which can increase the rate compared to that of current surveys. A sample characteristic signature would be a shock breakout with multiple peaks which evolves into an over-luminous SN (during which the time-difference cannot be seen).


\subsection{Gamma Ray Bursts (GRBs)}
\label{subsec:GRBs}

ULTRASAT will enable GRB studies both through target-of-opportunity followup and through direct detection of their ultraviolet counterparts in time domain fields.  

While GRB afterglows are effectively panchromatic, and near-UV observations probe roughly the same phenomena as optical observations, ULTRASAT has significant strengths for GRB astrophysics.   First, its combination of sensitivity, field of view, duty cycle, and response time stands out among present and near-future facilities.   Second, the near-UV sky has a lower density of confounding sources (both steady and transient) than the optical sky.  Third, because near-UV observations are more sensitive to dust extinction than optical or near-IR observations, ULTRASAT data may effectively probe dust columns along GRB sight lines.  

Approximately 10 cosmological GRBs per year are expected to occur within ULTRASAT's field of view.  UV afterglow emission from these GRBs can be detected to cosmological distances, given ULTRASAT's sensitivity.  Moreover, prompt emission may also be detected, provided the fluence within a 300s ULTRASAT exposure exceeds the detection limit.   A typical prompt UV luminosity of $> 10^{45}\,  \hbox{erg}\,\hbox{s}^{-1}$ and a prompt emission duration $\sim 10\,\hbox{s}$ is sufficient up to a luminosity distance $d_L \approx 5 \,\hbox{Gpc}$, corresponding to redshift $z \sim 0.8$.

ULTRASAT may be particularly valuable in constraining the existence of orphan GRB afterglows and of dirty fireballs.  

GRBs may be part of a larger class of cosmological fireballs.  
Such fireballs are generically expected when a sufficiently large amount of energy is released into a sufficiently small volume along with a (relatively small) quantity of ordinary baryonic matter.  This results in optically thick conditions, so that a large part of the energy is converted into kinetic energy of ejecta.  The smaller the baryon loading is, the more relativistic the ejecta become.  For bulk Lorentz factors $\Gamma \ga 100$, internal shocks within the ejecta and/or external shocks between ejecta and ambient medium can produce gamma rays.   However, there is no {\it a priori} reason that ``dirty fireballs,'' which are physically similar apart from larger baryon loading and lower peak Lorentz factors, could not exist.  Such explosions would not produce much gamma ray emission, but could resemble GRBs in most other ways.  In particular, they could still produce bright transients with properties similar to GRB afterglows at longer wavelengths, including the near-UV.

Afterglows without accompanying gamma rays may also be expected as a consequence of GRB collimation.   Relativistic beaming of photons has a characteristic half-angle $\sim 1/\Gamma$.  When ejecta from a fireball that is collimated into angle $\theta_{jet}$ decelerate to $\Gamma < 1/\theta_{jet}$, the fireball becomes visible at viewing angles $\theta_{obs} \sim 1/\Gamma > \theta_{jet}$ \citep{Rhoads.1997.A,Rhoads.1999.a,Sari.1999.a}.  Photon energies from the expanding remnant decrease as $\Gamma$ decreases, with the net effect that some off-axis events should be seen as ``orphan afterglows'' without accompanying gamma rays.

In principle, event rates of both dirty fireballs and off-axis orphans could greatly exceed the rate of observable GRBs.   In practice, preliminary indications from ZTF do not show large numbers of orphans \citep{Ho.2022.A}.  

Ultimately, we would like to study the overall population of relativistic fireballs.  Suppose we describe each fireball with its total energy $E$, collimation angle $\theta_{jet}$, and peak Lorentz factor $\Gamma$.  We would like to measure the event rate in this multidimensional parameter space $R(E,\theta_{jet},\Gamma,z)$.

GRB experiments constrain this distribution for the $\Gamma \ga 100$, subject to thresholds in fluence or flux.   Multiwavelength monitoring of GRB afterglows yields constraints on the distribution of $\theta_{jet}$.   ULTRASAT offers the wavelength coverage to probe down to $\Gamma$ of a few, and the survey efficiency (grasp) to measure values of $R$ that are appreciably smaller than the intrinsic GRB rate.

Full interpretation of the fireball population observed by ULTRASAT will doubtless require more sophisticated physical models than the arguments presented above.   Relativistic hydrodynamical simulations \citep{Granot.2018.a} can be used to examine the light curves expected as a function of wavelength, jet angle, and viewing angle.


\subsection{Stars and stellar remnants}
\label{subsec:stars}

In the UV range, as in the optical and in the IR, the majority of sources brighter than $\sim20$~mag in an astronomical image, particularly at low Galactic latitudes, are stars and their white-dwarf (WD) remnants.
ULTRASAT's unique combination of a large field of view, a long, continuous, dwell time with a high cadence, and all this in the UV, will open up largely uncharted territories in the study of stars and WDs.

\subsubsection{Stellar rotation and chromospheric activity}
The link between rotation and magnetic activity in stars can appear in all layers of the stellar atmosphere, be
it in the photosphere (e.g. \citealt{SuarezMascareno.2016.A}),
the chromosphere (e.g. \citealt{Olmedo.2013.A,AstudilloDefru.2017.A}) or the corona (e.g. \citealt{Wright.2011.A,Pizzocaro.2019.A}).
The manifestations of stellar spots and flares and their connections
to magnetic field, stellar age, and rotation, have been probed only scantily in the UV, let alone have they been well characterized.  
Apart from the poorly understood physics underlying these phenomena, observing stellar activity 
has gained also practical importance for the discovery and characterization of exoplanets.
The radial velocity (RV) accuracy achievable
using advanced methods and stabilized spectrographs (e.g. \citealt{AngladaEscude.2016.A}) is ultimately limited  by the magnetic activity of the host stars, which can cause RV jitter
with amplitudes of $\gtrsim 10$\,m\,s$^{-1}$ \citep{TalOr.2018.A}, and
can mask or mimic the signal caused by substellar companions.

\begin{figure}
\centerline{\includegraphics[width=.55\textwidth]{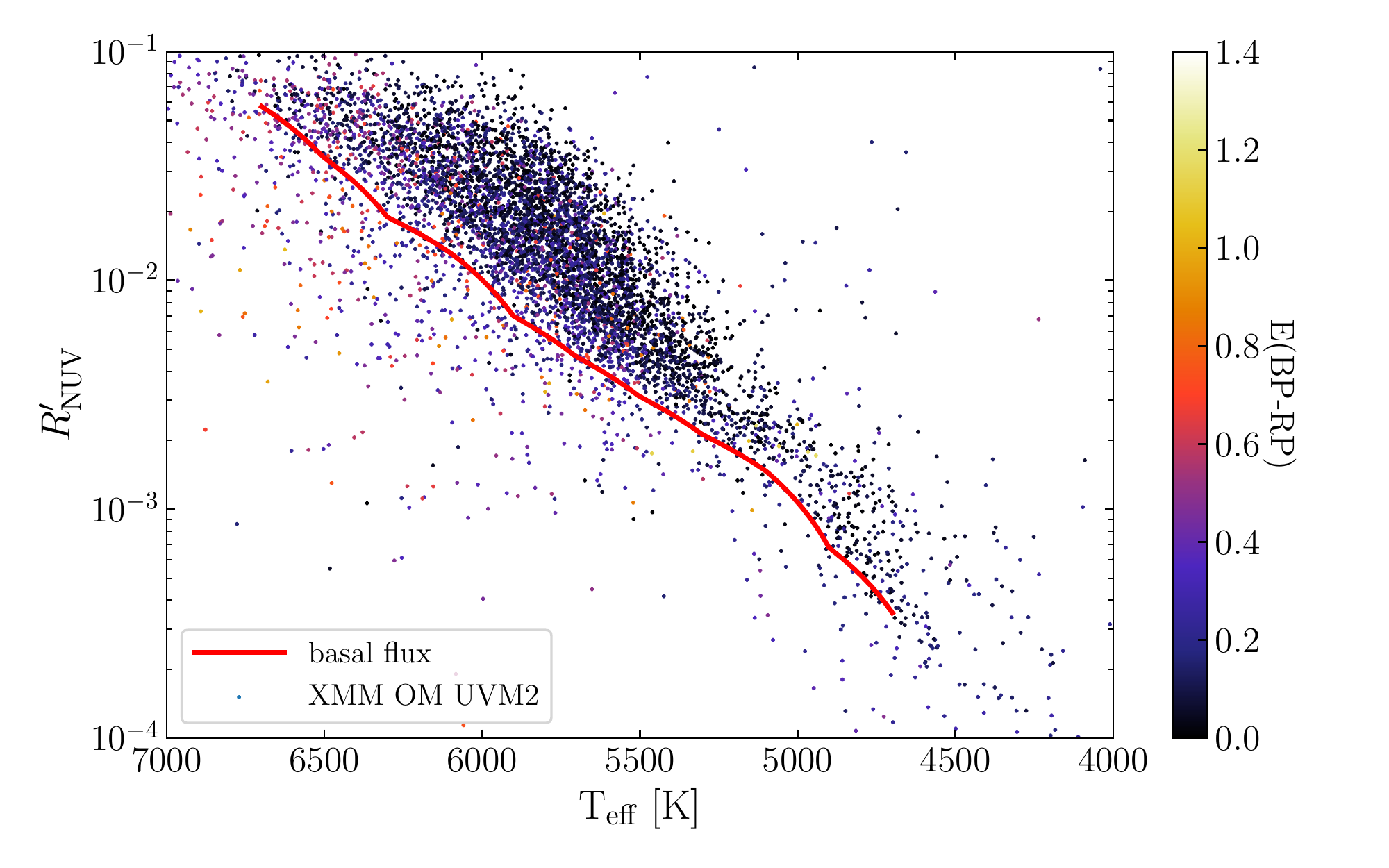}}
\caption{Relative chromospheric flux excess over the model photospheric flux, versus effective temperature, for 8859 stellar sources observed
in the UV with the XMM-Newton Optical Monitor, in the Gaia
DR2 source catalog, and in the LAMOST DR5 catalog. The color coding of the points reflects the reddening. The red line denotes the
the flux excess of the bottom 5th percentile of stars having reddening coefficient
$E(BP-RP)\le 0.05$.\label{fig:UVexcess_stars}}
\end{figure}

To assess quantitatively the amount of magnetic-activity-related emission from normal stars that ULTRASAT will see,   
 we have analyzed archival data from 
the XMM-Newton Optical Monitor (OM; \citealt{Mason.2001.A}),
whose
UVM2 filter is similar in
width and central wavelength to the ULTRASAT bandpass.
Point sources from the XMM-SUSS 4.1 catalog \citep{Page.2012.A} with a signal-to-noise ratio $\ge5$ in
the UVM2 filter were cross-matched with the GAIA DR2
source catalog \citep{GaiaCollaboration.2018.A} to obtain parallaxes, and with the LAMOST DR5 catalog \citep{Luo.2019.Cat.A} to
obtain stellar parameters (log g, [Fe/H], $T_{\rm eff}$) for a total of
8859 stars.
The UV flux measured by the XMM OM consists of 
photospheric and chromospheric contributions. 
To estimate the photospheric contribution, we folded each spectrum from the latest
PHOENIX stellar atmosphere grid \citep{Husser.2013.A} through
the UVM2 filter transmission function, yielding a grid of theoretical UV fluxes
for the full range of surface
gravities, effective temperatures and metallicities. For each star
in the XMM OM catalog we then interpolated, on this grid,
its theoretical photospheric UV flux. In analogy to the stellar activity indicator $R^{'}_{HK}$,
we define a UV excess index,
\begin{equation}
R^{'}_{\rm NUV}=\frac{f_{\rm UVM2} - f_{\rm phot}}{f_{\rm bol}}
\end{equation}
where $f_{\rm UVM2}$
is the flux measured by the XMM OM, $f_{\rm phot}$
is the theoretical photospheric flux in the UVM2 band from PHOENIX,
and $f_{\rm bol}$ is the bolometric flux based on Gaia.

Figure ~\ref{fig:UVexcess_stars} shows the UV non-photospheric flux excess as a function of stellar 
effective temperature, where the color coding of the points denotes
the Gaia/Apsis reddening coefficient $E(BP-RP)$. The red
curve shows the $R^{'}_{\rm UVM2}$ of the bottom 5th percentile of all
sources with a reddening coefficient $E(BP-RP)\le 0.05$.
We see that, in a large majority of all stars, chromospheric emission in the UVM2 band constitutes a non-negligible fraction, 0.1\% to 5\%, of the bolometric luminosity of the star, with typical values of $\sim 1-3$\%, and a clear rising trend with effective temperature. Note that this excess UV flux is accumulated over generally long XMM integrations, and likely consists of both variable and steady-state components. ULTRASAT will thus be ideally positioned to measure and characterize this important component of stars, and its behavior (particularly in the temporal regime).   

The magnetic activity of stars manifests itself observationally as variability in two forms: stochastic variability from stellar flares that arise from magnetic instabilities in the chromosphere and the corona; and periodic variations resulting from photospheric stellar spots, combined with stellar rotation. The mission-long UV light curves of all stars observed with ULTRASAT will be an unparalleled resource for characterizing
both of these types of stellar variability in the UV.

Recent studies have revealed, in growing detail, large scale trends with stellar properties of the optical variability periods that trace the rotation periods (e.g. \citealt{McQuillan.2014.A,Davenport.2018.A,CantoMartins.2020.A,Gordon.2021.A,Briegal.2022.A}). Apart from an empirical trend between stellar color and rotation period, which permits deducing a star's age ("gyrochronology"--stars lose angular momentum and magnetic-field strength as they age), several of these studies and others have shown an intriguing bi-modal structure in the parameter space of rotation periods of main-sequence stars with respect to color, with a dearth of stars in a period gap between $\sim15$ and 25 days. These results have emerged largely from Kepler and TESS data with their continuous, high-cadence, temporal sampling, similar to that planned for ULTRASAT.
The ULTRASAT data will permit extending such studies, for the first time, into the UV, with this band's sensitive, independent, indicators of magnetic activity. Systematic large surveys of stellar activity are in progress also in X-rays (e-ROSITA; \citealt{Merloni.2012.arXiv.A}). In radio, stellar activity studies require observations
based on pointing and dwelling on individual stars, but a large body of data has already been collected (e.g. \citealt{Crosley.2018.A,Villadsen.2019.A,Vedantham.2020.A}; and references therein). ULTRASAT data will thus be instrumental for the inter-comparison of these phenomena across the electromagnetic spectrum.  
Active short-period binaries and planet hosts are another promising category for ULTRASAT studies. The topology of the stellar magnetic field may be modified by the presence of a close companion. The UV emission from the
corona, which is in interaction with the magnetic field, should be modulated with
the stellar binary period, and with the rotation period, assuming the system has reached synchronization via tidal dissipation. If
synchronization has not been achieved, the stellar magnetic topology can be complex,
with possible induced instabilities that will be reflected in the UV variability
observed by ULTRASAT.

In summary, the still-poorly understood question of how magnetic activity operates in normal stars will receive a major resource from the characterization by ULTRASAT of UV variability, potentially for a large fraction of all the stars that it will observe. Planned ULTRASAT studies of the stochastic flaring activity (rather than periodic variability) of normal stars are further discussed in Section~\ref{subsec:exoplanets}, below.

\subsubsection{White dwarf accretion and rotation}
Distinct from pulsating WDs (which are on the WD instability regions in the HR diagram, with periods of minutes), 
photometrically variable WDs with amplitudes of order $\sim 10\%$ and periods of hours to months are known, but rare.
Except in a few individual cases that may be explained by effects connected to an orbit with a companion mass (beaming, reflection, ellipsoidal tidal distortion), the variability has been generally associated with WD rotation (which has the same typical range of periods), combined with non-uniform surface emissivity. Except for very cool WDs, most WDs have fully radiative atmospheres, devoid of the convective cells that can produce star spots. The photospheric inhomogeneity  implied in variable WDs has therefore  been ascribed, instead, to extreme magnetic field strengths that produce magnetic dichroism \citep{Angel.1981.A}. 

However, \cite{Maoz.2015.B}, analyzing the Kepler time series of 14 normal WDs, found periodic photometric modulations in seven of them. The variation periods were of order hours to 10 days, with amplitudes of order $10^{-4}$ to $10^{-3}$, much lower than could have been detected with pre-Kepler technology. This discovery raised the possibility that most or all WDs have low-level variability associated with rotation, but also    
exacerbated the problem of how to produce, as a rule in WDs, an inhomogenous WD photosphere. \cite{Maoz.2015.B} and \cite{Hallakoun.2018.A} hypothesized that the optical-band low-level photometric modulation seen in many WDs could be associated with
photospheric metal pollution, which is likewise seen in a large fraction of WDs \citep{Koester.2014.A}. 

Over the past couple of decades it has become widely accepted that the optical and UV metal absorption lines seen in WD atmospheres are the result of ongoing accretion of planetary debris onto the WD \citep{Jura.2003.A,Zuckerman.2003.A,Zuckerman.2010.A,Gansicke.2012.A,Koester.2014.A}. Slightly inhomogeneous surface coverage
of the accreted material (e.g. due to moderate magnetic fields) could lead to inhomogeneous
UV absorption. Optical fluorescence of the absorbed UV photons \citep{Pinto.2000.A}, combined with the WD rotation, could then potentially produce the observed levels and periods of optical modulation.

We estimate that $\sim 400$~WDs will be detected above the limiting magnitude in every ULTRASAT field. Although the $10^{-4} - 10^{-3}$ variation amplitudes seen with Kepler will be near the limit detectable with ULTRASAT (depending on the level of hard-to-foresee systematics),
many WDs show modulations with larger amplitudes, mostly in optical-band
photometric surveys (such as Kepler’s K2 continuation mission). One example of a
high-amplitude UV-variable WD is GD 394. This hot, metal-polluted, WD was observed in the past by the Extreme
Ultraviolet Explorer (EUVE) and
displayed 25\% variations in the extreme UV (70-380 \AA) with a period of 1.15 d \citep{Christian.1999.A,Dupuis.2000.A}.
Follow-up far UV spectroscopy (1144-1710 \AA) taken in 2015 with $HST$ failed to detect any variability, down to the 1\% level
\citep{Wilson.2019.A}. However, recent TESS data have revealed
12\%-amplitude optical variations, with a similar period \citep{Wilson.2020.A}. GD 394 has some striking similarities
to WD\,J1855+4207 \citep{Maoz.2015.B,Hallakoun.2018.A}. Both are hot
($\gtrsim 30,000$\,K) WDs with indications of highly ionized circumstellar metals, in addition to the
detected photospheric metals, and both have relatively large periodic optical modulations. The monitoring of the numerous WDs 
in the ULTRASAT-monitored fields can potentially detect additional such large-amplitude or changing-amplitude WDs.

Furthermore, in the case of a WD  with significant metal pollution in its photosphere, UV
variability over short (minutes to days) or long (months to years) timescales, could indicate a
change in the debris accretion rate, or in the structure of the circumstellar debris around the WD.
Combined with simultaneous photometric monitoring in multiple bands (from the UV to the near infrared) using
ground-based facilities
(e.g. \citealt{Hallakoun.2017.A,Xu.2019.A}), ULTRASAT may thus finally provide answers to the question of how periodic variability is produced in a large fraction of WDs.


\subsection{Exoplanets \& Star-Planet Connection}
\label{subsec:exoplanets}

The discovery of thousands of extrasolar planets ranks among the most exciting scientific developments of the past decades. ULTRASAT capabilities allow us to explore unique regions of exoplanet and star-planet connection parameter space. In this section we focus on two research topics that ULTRASAT will significantly advance - host activity (Section~\ref{subsec:host_activity}) and the search for planets around WDs (Section~\ref{subsec:WD_planets}). We also discuss ULTRASAT potential to study exoplanet atmospheres (Section~\ref{subsec:exoplanet_atmospheres}), albeit this will require modifying the survey modes.

\subsubsection{Planet's host activity}
\label{subsec:host_activity}
The UV radiation emitted from a host star on an orbiting planet can dictate whether a planet retains its atmosphere, and govern the photochemistry in it. 
In addition, the UV flux will have both positive and negative contribution to the likelihood of life developing on it \citep{Buccino.2006.A,Ranjan.2016.A}. As a stressor, high fluxes of UV radiation, in either steady state emission or through flares, can destroy nascent biomolecules through photolysis. As a eustressor, recent studies indicate UV radiation can play a key role in prebiotic chemistry, and might have been the most abundant energy source on the young earth. In the context of the search for biologic activity, UV emission can result in abiotic generation of bona-fide biosignatures, such as the disassociation of water molecules in exoplanets orbiting M-dwarf (dM) stars that will in turn result in high concentration of molecular oxygen  \citep{Meadows.2017.A}. Therefore, one must know the UV radiation field around a planet to understand its atmosphere evolution. Defining a UV habitable zone (UV HZ) – the range of orbits in which the UV radiation from the host star will allow (and contribute) to the evolution of life as we know it, is critical for putting the evolution of life on our own planet on a galactic scale, and for the search of bioactivity on other worlds. 

Despite the impact of UV radiation on exoplanet atmospheres evolution and the likelihood of bioactivity, our understanding of the UV emission from stellar objects is limited, and the commonly used photospheric models routinely underestimate it, as the UV emisison for {\it e.g.,} dM stars, originates in the chromosphere and transition region \citep{France.2013.A}. Our current capability to characterize it for various stellar types is hindered by the small number of UV observatories to date, the lack of large UV datasets, and the large variance in the UV radiation in both short and long time scales, \textit{i.e.,} flares for the former and variability for the latter. 

Initial studies performed with the limited data available are portraying a complicated picture, specifically for dM stars, the most common stars in the solar neighborhood and preferred targets for many exoplanet transit searches due to favorable planet-to-host radius ratio, and the potential for atmosphere studies. Some suggest that the HZ and the UV HZ do not necessarily coincide for various stellar types \citep{Buccino.2006.A}. Another study using HST data indicates that some dM stars that seem to be in a quiescent state when observed in the visible band are active in the UV band \citep{France.2016.A}. These results demonstrate clearly the need for a detailed study of UV radiation from a large sample of stars of various stellar types. Additional data, albeit for a limited sample, is expected from two CubeSats missions that have recently been launched: The Colorado UV Transit Explorer \citep[CUTE;][]{Fleming.2018.A}, and The Star-Planet Activity Research CubeSat \citep[SPARCS;][]{Ramiaramanantsoa.2022.A}. 

\begin{figure}
\centerline{\includegraphics[width=0.49\textwidth]{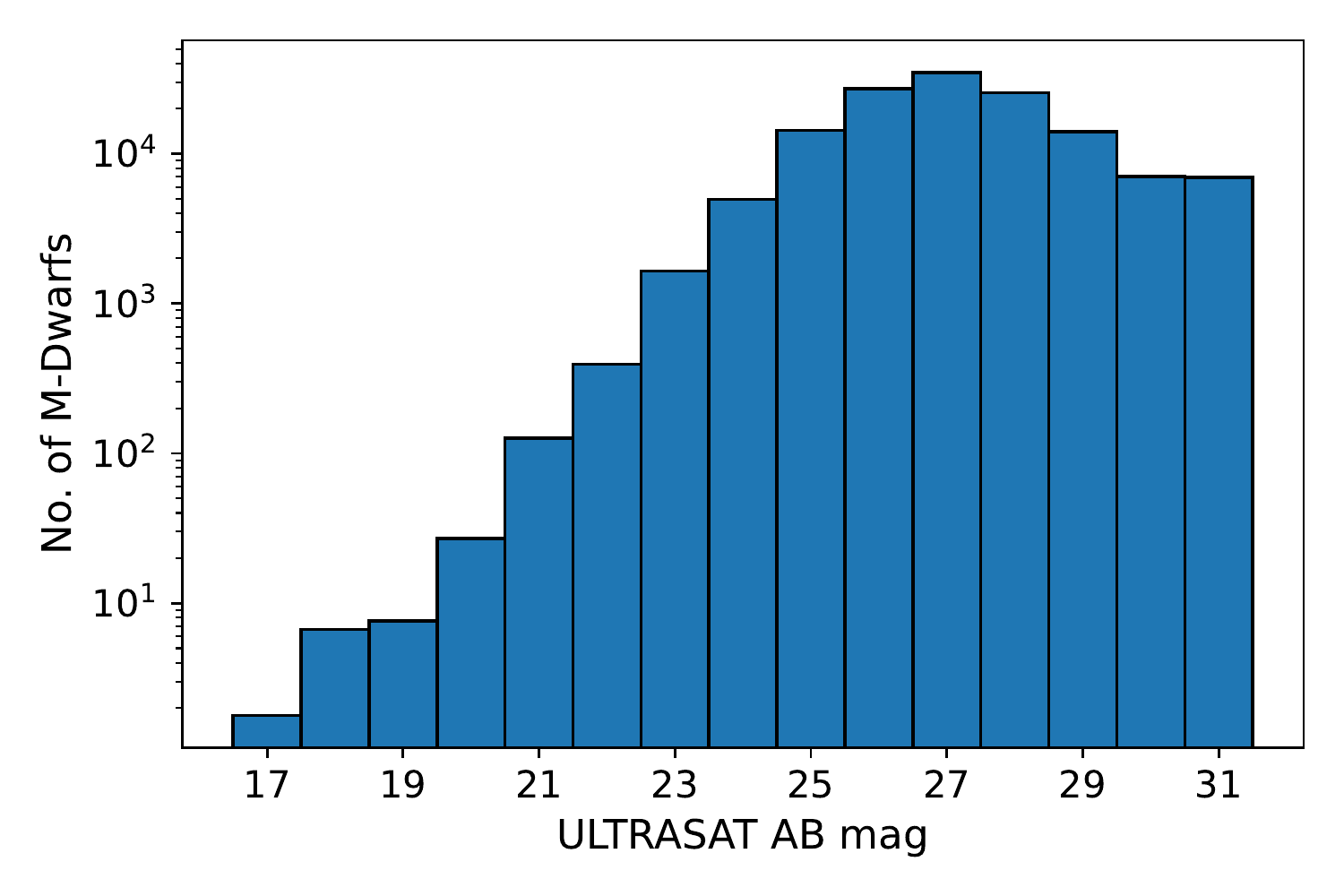}}
\caption{Magnitude distribution of dM stars in quiescence up to 800\,pc in an exemplar ULTRASAT FoV of $200\,$deg$^{2}$. While flaring they can get up to $\sim$5 mag brighter (Rekhi et al. 2023, submitted). Input catalogues used include the TESS Input Catalogue \citep{Stassun.2019.A}.
Limiting magnitude for a 3 image co-add for dM stars is $\approx21.4$. 
\label{fig:dM}}
\end{figure}

\begin{figure}
\centerline{\includegraphics[width=0.49\textwidth]{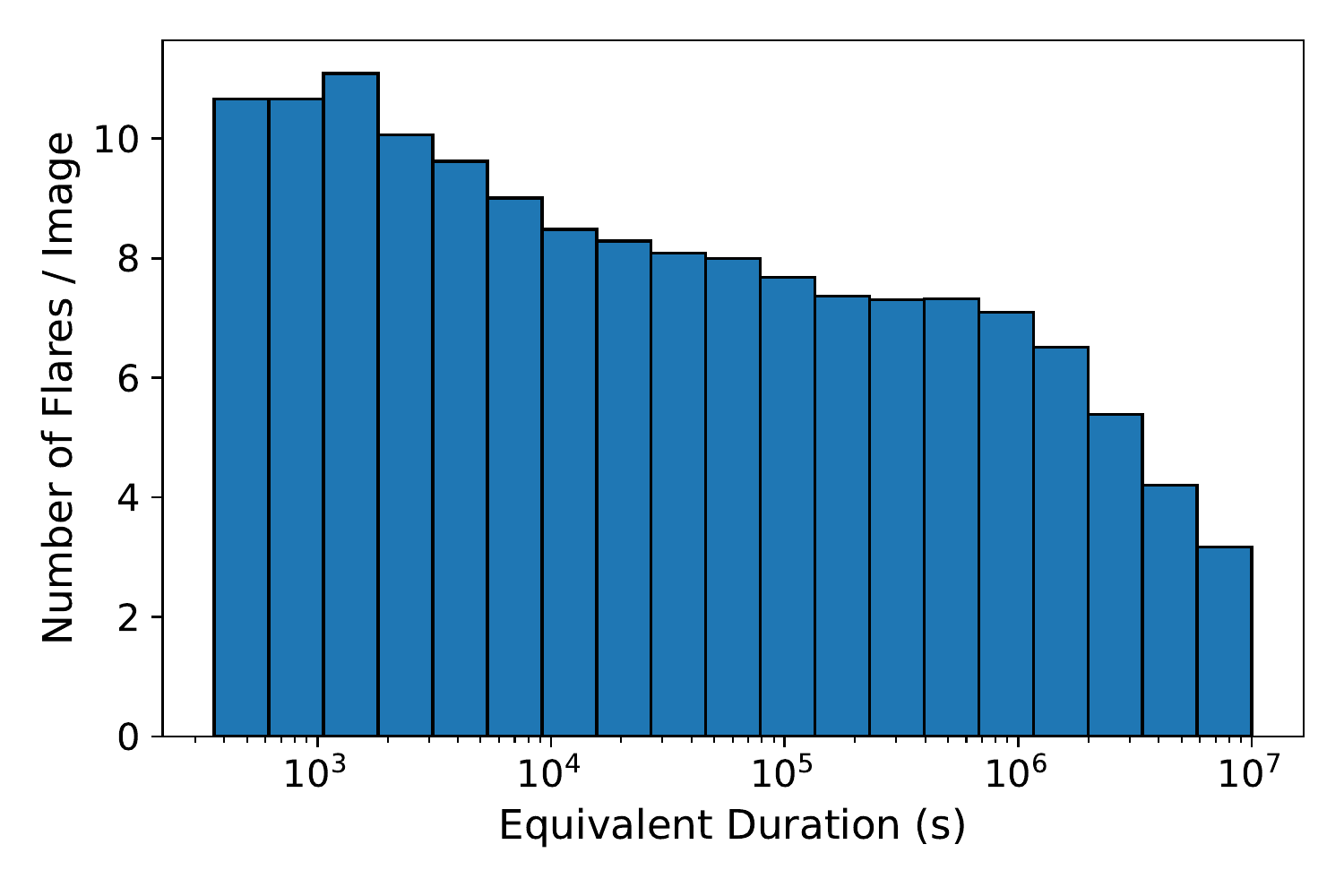}}
\caption{The expected distribution of flares detected in each 300\,s ULTRASAT image, for the the dM population shown in Figure \ref{fig:dM}. The high energy tail of the distribution, with equivalent duration $\geq10^4\,$s, is poorly constrained. 
A Monte-Carlo simulation in which the $NUV$ FFD was empirically derived from $GALEX$ observations suggests each ULTRASAT image will capture $\sim150$ flares from dM stars (Rekhi et al. 2023, submitted).
\label{fig:dM_Flares}}
\end{figure}

The total number of stars within ULTRASAT $200\,$deg$^2$ FoV and up to $\sim800$\,pc is estimated to be $\approx200,000$ stars, with roughly half of them being dM stars, see Figure \ref{fig:dM}.
For the latter in quiescence, we expect to detect only $O(1-10)$ stars for co-adds of three images, and $O(10^3)$ stars for co-adds reaching an estimated confusion limit of 24 magnitude in the ULTRASAT waveband. 
Assuming the NUV flare frequency distribution (FFD) derived by  
Rekhi et al. (submitted)
using archival $GALEX$ data, each 300\,s ULTRASAT image will capture $\sim150$ flaring dM stars, with equivalent duration distribution shown in Figure \ref{fig:dM_Flares}, for which $\approx50\%$ of the flares are in the poorly constrained high energy regime.
The distributions shown in Figures \ref{fig:dM} and \ref{fig:dM_Flares} demonstrate how the long baseline observations of ULTRASAT will allow us to monitor the high energy tail of the FFD ({\it i.e.,} equivalent duration $\geq10^4\,$s), which is poorly constrained to date. Flares in this energy range are expected to have a significant effect on an exoplanet atmosphere. The data will provide us a more complete picture of the high energy radiation environment around stellar hosts, and will allow us to determine the best candidates for expensive spectroscopic studies of exoplanet atmospheres by \textit{e.g.,} JWST and the upcoming ELTs.

\subsubsection{Planets around white dwarfs}
\label{subsec:WD_planets}

Occurrence rate studies show that most main sequence stars host exoplanets \citep{Winn.2015.A,Cassan.2012.A}. The vast majority of these hosts will evolve into WDs. Planets in close orbits are unlikely to survive this transition, as the host star climbs the red giant branch, and its envelope expands. Despite the expected fate of close-in planets, both direct and indirect evidence have been accumulated for the presence of sub-stellar objects and minor bodies in close orbit around WDs \citep{Veras.2021.A}. 
Several explanations for the presence of these objects have been suggested. Planets can be captured and/or migrate from wide orbits into close orbits after the host transition into a WD is completed \citep{Debes.2002.A}. Planets can also form out of gas near the WD, e.g., via the interaction or merger of binary stars (i.e., second-generation planets; \citealt{Livio.2005.A}). Studying the occurrence rates of planets around WDs can shed light on the evolution of these systems. Moreover, given the slow change in the flux emitted by a WD as it ages, the continuous habitable zone (CHZ) around a WD can have a lifetime of $\gtrsim3$\,Gyr for orbital periods $\approx4-32$hr \citep{Agol.2011.A}. Planets detected in the CHZ, will be optimal for atmosphere studies and search for biomarker in their transmission spectra \citep{Loeb.2013.A}.

The search for planets around WDs is a rapidly growing field, with a few detections of both disintegrating planets or planetesimals, and a transiting giant planet candidate \citep[\textit{e.g.,}][; See review by \citealt{Veras.2021.A}]{Vanderburg.2020.A}.  Due to the small radii of WDs, the transit depth induced by a planet is large and can be detected with standards photometric precision telescopes (e.g., ground-based, ULTRASAT), making transit detection possible despite WDs being faint compared to typical targets of exoplanet surveys.  Currently, all occurrence rate constraints are based on null detections (thus giving only upper limits). \cite{Fulton.2014.A} used a sample of $\sim$1700 WDs observed by the PanSTARRS survey to constrain planets occurrence rates in the HZ of WDs. They find an occurrence upper limit of 3.5\% for $R_{\rm pl}=2-5R_\oplus$, but cannot constrain the occupancy rate for planets with smaller radii. \cite{van_Sluijs.2018.A} investigated the light curves of 1148 WDs from {\it Kepler}'s K2 campaign, and conclude that the occurrence of habitable Earth-sized planets ($R_{\rm pl}=1-2R_\oplus$) around WDs is $<28\%$, approximately equal or less than their main-sequence occurrence. A key to move forward is to significantly increase the number of monitored WDs. ULTRASAT, with a large FoV that can image hundreds of WDs in a single exposure, is ideal for this task.

The number of WD transiting planets is given by
\begin{equation}
N_{\rm WD,pl} = N_{\rm WD} f_{\rm WD,pl} P_{\rm transit} P_{\rm detection},
\end{equation}
where $N_{\rm WD}$ is the total number of monitored WDs, $f_{\rm WD,pl}$ is the fraction of WDs hosting planets with orbital period $<P$,  $P_{\rm transit}$ is the transit probability, $(R_{\rm pl} + R_{\rm WD})/a$, and $P_{\rm detection}$ is the probability to detect at least $N_{\rm tr}$ transits. ULTRASAT low-cadence survey will monitor $N_{\rm WD}\approx32,000$ WDs (40 fields in each hemisphere, 400 WDs per field), an order of magnitude larger than previous studies (see \citealt{Ben-Ami.2023arXiv.A} for a future planned survey). In addition, $\sim800$ WDs will be monitored by ULTRASAT high-cadence survey each year.\footnote{We assume that the high-cadence fields will be different each year, though this is still to be decided.}
The detection probability, assuming the transit can be detected in a single visit, is:
\begin{equation}
P_{\rm tr,det} = 1-\sum^{N_{\rm tr}-1}_{k=0} \binom{n_{\rm vis}}{k}\left(\frac{t_{\rm vis}}{P}\right)^k \left(1-\frac{t_{\rm vis}}{P}\right)^{n-k}, 
\end{equation}
where $n_{\rm vis}$ is the number of visits and $t_{\rm vis}$ is the duration of each visit. The 40 fields of the low-cadence survey (8000 deg$^2$)  cover almost all of the low-extinction ($A_{\rm ULTRASAT}<1$) sky in each hemisphere. Thus, for our current calculations we assume these will be the same fields for the entire 3-year mission. Under this assumption, the low-cadence survey will have $\sim$135 visits per field, each of $t_{\rm vis}=15$\,min. The high-cadence survey fields will have over 45,000 images each year.

\begin{figure}
\includegraphics[width=0.53\textwidth]{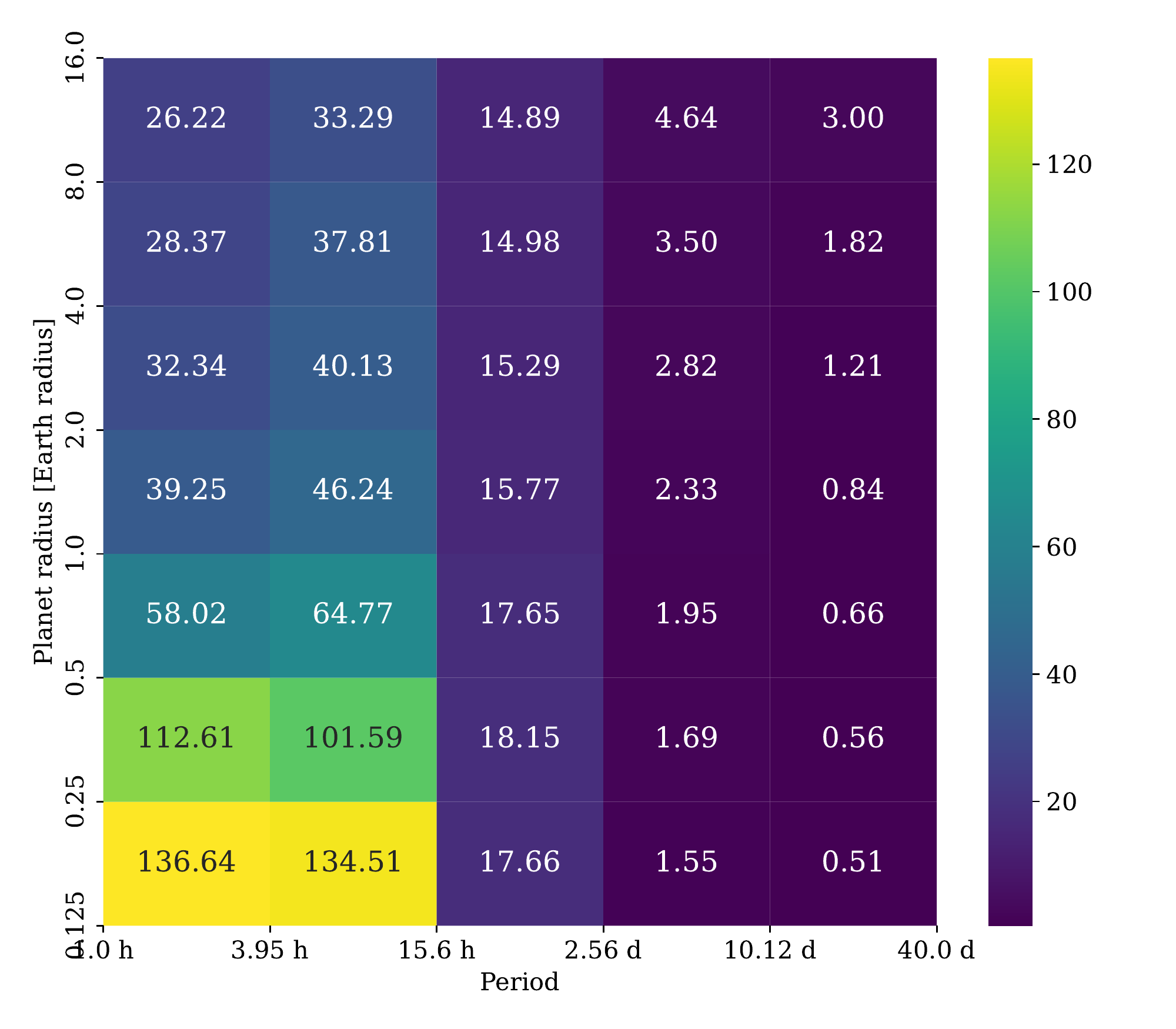}
\caption{
Maximum number of WD transiting planets detected by ULTRASAT, by both the low cadence survey and high cadence survey. Based on the current occurrence rate upper limits from \cite{van_Sluijs.2018.A}.}
\label{fig:WD_planets}
\end{figure}

We use the 68\% upper limits on $f_{\rm WD,pl}$ from \cite{van_Sluijs.2018.A}, to estimate the maximal number of WD-transiting planets with at least $N_{\rm tr}=3$ detected transits by ULTRASAT, for a range of planet's radii and orbital period (Figure \ref{fig:WD_planets}). We find that ULTRASAT is highly sensitive to planets in the entire CHZ, with maximal number of detected planets well over 100 (assuming occupancy rates identical to the upper limits derived by \citealt{van_Sluijs.2018.A}). The high number of detections is a result of the large sample of WDs monitored by the low-cadence survey. ULTRASAT will thus be able to deliver significantly tighter upper limits to those available by any other existing survey, or better yet, will deliver a unique sample of planets and minor bodies in short orbits around WDs. Planets at wider orbits than the CHZ can also be detected (mainly via the high-cadence survey), but with a smaller statistical impact.

\subsubsection{Exoplanet atmospheres}
\label{subsec:exoplanet_atmospheres}
The near UV bandpass holds unique prospects for the study of exoplanet atmospheres \citep[\textit{e.g.,}][]{Christian.1999.A}. Several studies show the potential of $NUV$ observations to differentiate between atmosphere models, as this wavelength range contains telltales for the presence of clouds and haze \citep{Goyal.2018.A}. In addition, an escaping atmosphere will result in deeper transits in the $NUV$, as the optical depth of the escaping atmosphere is higher at shorter wavelengths \citep[\textit{e.g.,}][]{Vidaz-Madjar.2013.A,Salz.2019.A}.
While the cases above are typically targeted by FUV measurements, as the signature is expected to be intrinsically stronger at this band, the $NUV$ bandpass hold several unique advantages \citep[\textit{e.g.,}][]{Christian.1999.A}, mainly the uniform distribution of $NUV$ emission across the stellar disk when compared to the FUV – which makes the scientific derivation more secure, and the reduced stellar attenuation in the $NUV$ bandpass relative to the FUV. 
It is not clear whether ULTRASAT will deliver the precision needed for such studies. As a starting point and capability demonstration, ULTRASAT will attempt to detect the signature of escaping atmosphere from several bona-fide examples, such as WASP-12b. The signature of the escaping atmosphere in these cases is several milimag \citep{Fossati.2010.A}, well within the instrument capabilities, see Figure \ref{fig:LimMag_SNR}. These cases will allow us to establish the precision limits of the instrument and will allow us to plan a dedicated survey for \textit{e.g.,} the search for escaping atmospheres from exoplanets in short orbits. We emphasize that the science case described in this section is not a part of the standard survey, and will require several changes to the standard observing strategy, such as changes in the exposure time.


\subsection{TDEs}
\label{subsec:TDEs}

Tidal disruption events (TDEs) offer a novel probe of massive black hole (MBH) demographics, super-Eddington accretion physics, and stellar dynamics in galactic nuclei.  Theoretical dynamics calculations indicate that the volumetric TDE rate should be dominated by the smallest galaxies with a high MBH occupation fraction \citep{Wang.2004.A, Stone.2016.A}, indicating that statistical samples of TDEs can resolve open questions about the poorly-understood low end of the MBH mass function \citep{Greene.2020.A}.  As probes of MBH demographics, TDEs have the unique ability to sample {\it quiescent} galactic nuclei to cosmological distances \citep{Bloom.2011.A, Chornock.2014.A}, and perhaps to measure MBH spins \citep{Stone.2012.A, Guillochon.2015.A, Wen.2020.A} as well as masses \citep{Rees.1988.A, Mockler.2019.A, Ryu.2020.A}. There are indications that TDEs may be multimessenger sources, as two different high-energy IceCube neutrinos have been temporally and spatially coincident with TDE flares \citep{Stein.2021.A, Reusch.2022.A}.  More speculatively, TDEs may be linked to {\it LISA}-band gravitational wave sources, either preceding \citep{Seto.2011.A} or following \citep{Stone.2011.A} the GW signal from a merging MBH binary, and may even lead to novel tests of fundamental physics \citep{Lu.2017.A, Wen.2021.A}.

A TDE is set in motion when a star is scattered onto a highly radial orbit \citep{Frank.1976.A} around a MBH of mass $M_\bullet$.  Once its pericenter $R_{\rm p}$ passes within the tidal radius $R_{\rm t}\approx R_\star(M_\bullet / M_\star)^{1/3}$, the star is disrupted \citep{Hills.1975.A, Rees.1988.A}.  Half of the stellar debris eventually returns on highly eccentric ($e \gtrsim 0.99$) but bound orbits, powering a highly luminous flare.  While the hydrodynamic evolution and radiative emission processes in TDEs remain highly uncertain and hotly debated (see \citealt{Roth.2020.A, Bonnerot.2021.A} for recent reviews), the last decade of detections has shown that typical TDEs are blue, quasi-thermal transients of high luminosity \citep{Gezari.2008.A, van_Velzen.2011.A, Gezari.2012.A, Arcavi.2014.A}.  Typical peak bolometric luminosities range from $L_{\rm BB}\sim 10^{43-44.5}\,{\rm erg\,s}^{-1}$; typical blackbody temperatures range from $T_{\rm BB} \approx 2-4 \times 10^4$\,K; there is little color evolution over timescales of weeks-months \citep{Hung.2017.A, van_Velzen.2021.A}.  Despite their high luminosities, TDEs are challenging to discover because of their low event rate, which is likely $\dot{N}_{\rm TDE} \sim 10^{-4}\,{\rm galaxy}^{-1}\,{\rm yr}^{-1}$ \citep{Stone.2016.A, van_Velzen.2018.A}, about $1\%$ of the core-collapse supernova (CCSN) rate.

The high effective temperatures and low volumetric rates of TDEs make them ideal targets for {\it ULTRASAT}, given the mission's large grasp and UV sensitivity.  In contrast to other transients of interest for {\it ULTRASAT} (e.g. shock breakout), TDEs evolve slowly, over timescales of days to weeks \citep{Gezari.2012.A, Arcavi.2014.A, Hung.2017.A}, and therefore will benefit from the low-cadence survey mode. 

We now quantify the number of TDEs that {\it ULTRASAT} will detect and identify every year.  We emphasize in advance that these estimates are quite approximate and incorporate significant assumptions on TDE flare populations as well as unavoidable ambiguities in what counts as a secure identification.

\begin{figure}
\centering
\includegraphics[width=8.0cm]{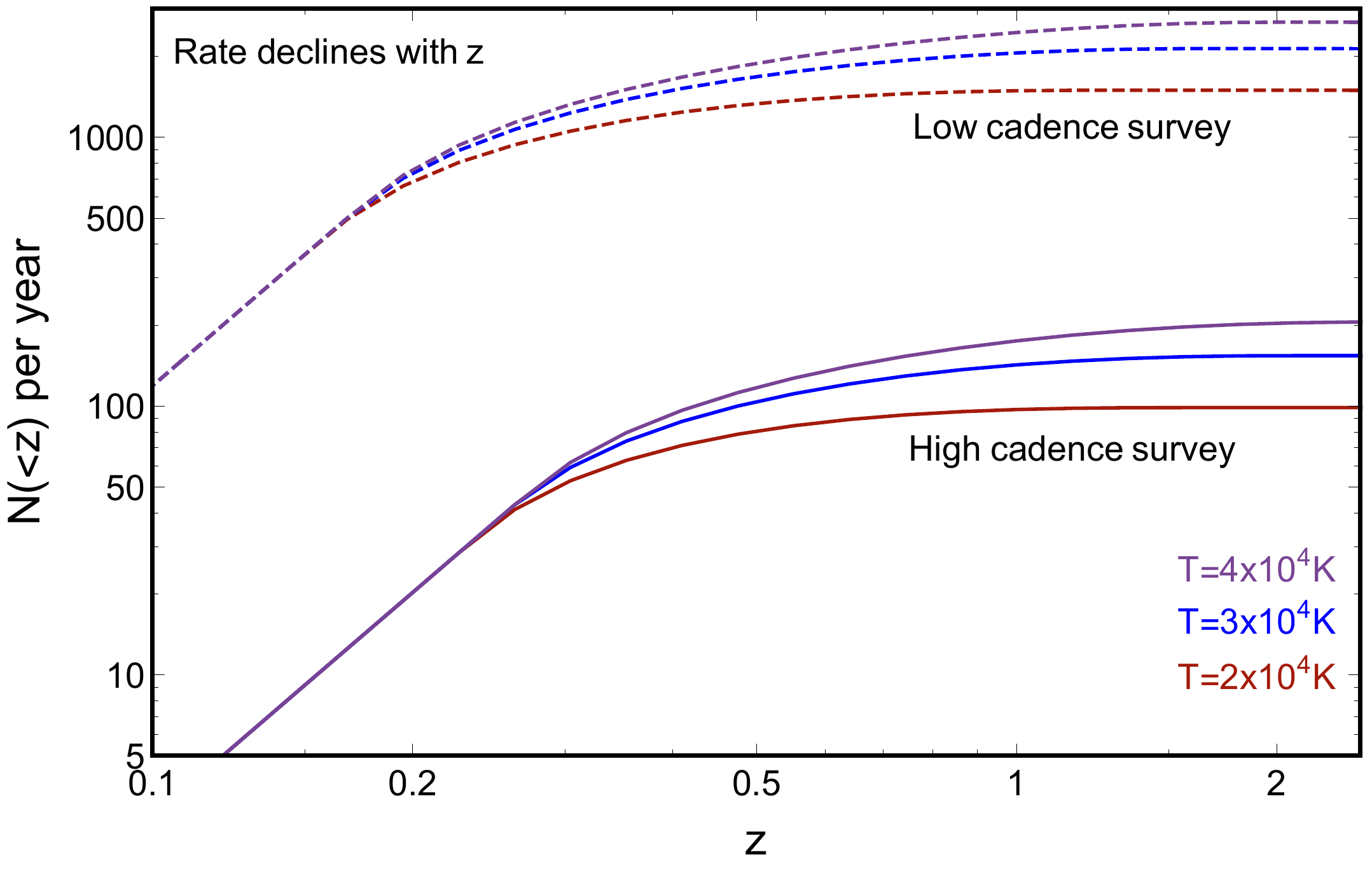}
\includegraphics[width=8.0cm]{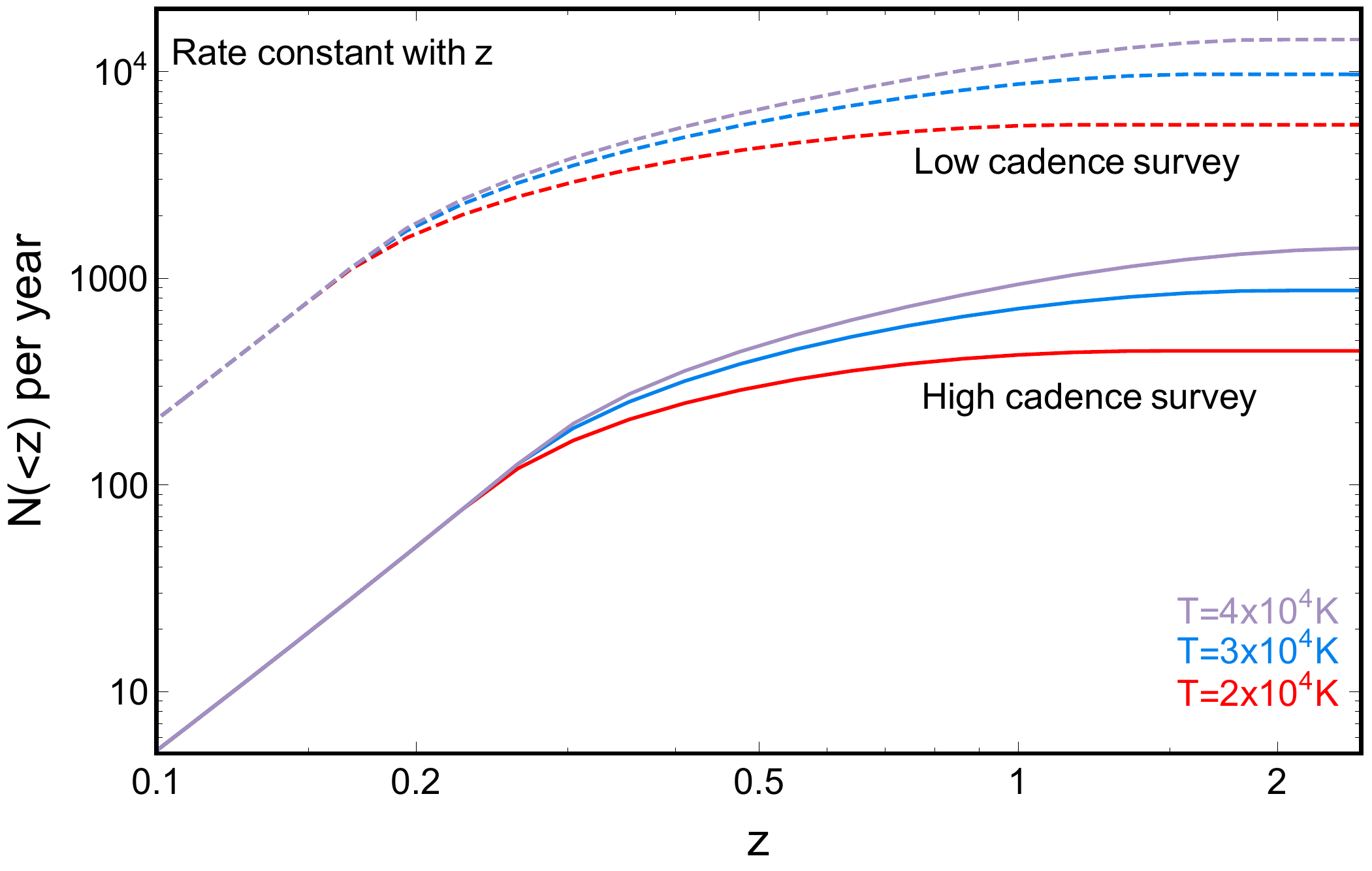}
\caption{The cumulative number of TDEs detected per year within a given redshift, $N(<z)$.  Colors show the effective blackbody temperatures $T$ of TDE spectra at peak; we consider the idealized cases of $T=2\times 10^4~{\rm K}$ (red), $T=3\times 10^4~{\rm K}$ (blue), $T=4\times 10^4~{\rm K}$ (purple). Detection rates are shown assuming a volumetric rate that declines steeply with $z$ (top panel, dark colors) or one that is constant with $z$ (bottom panel, light colors).  In this rate calculation, we consider both the high-cadence (solid lines) and low-cadence (dashed lines) survey modes.  In the low-cadence calculations, we count as a ``detection'' every TDE with a peak magnitude $m \le 22.5$, while in the high-cadence calculations, we assume faint exposures over the course of one day are co-added up to a confusion limit of $m = 23.5$.}
\label{fig:cumulTDE}
\end{figure}

\begin{figure}
\centering
\includegraphics[width=8.0cm]{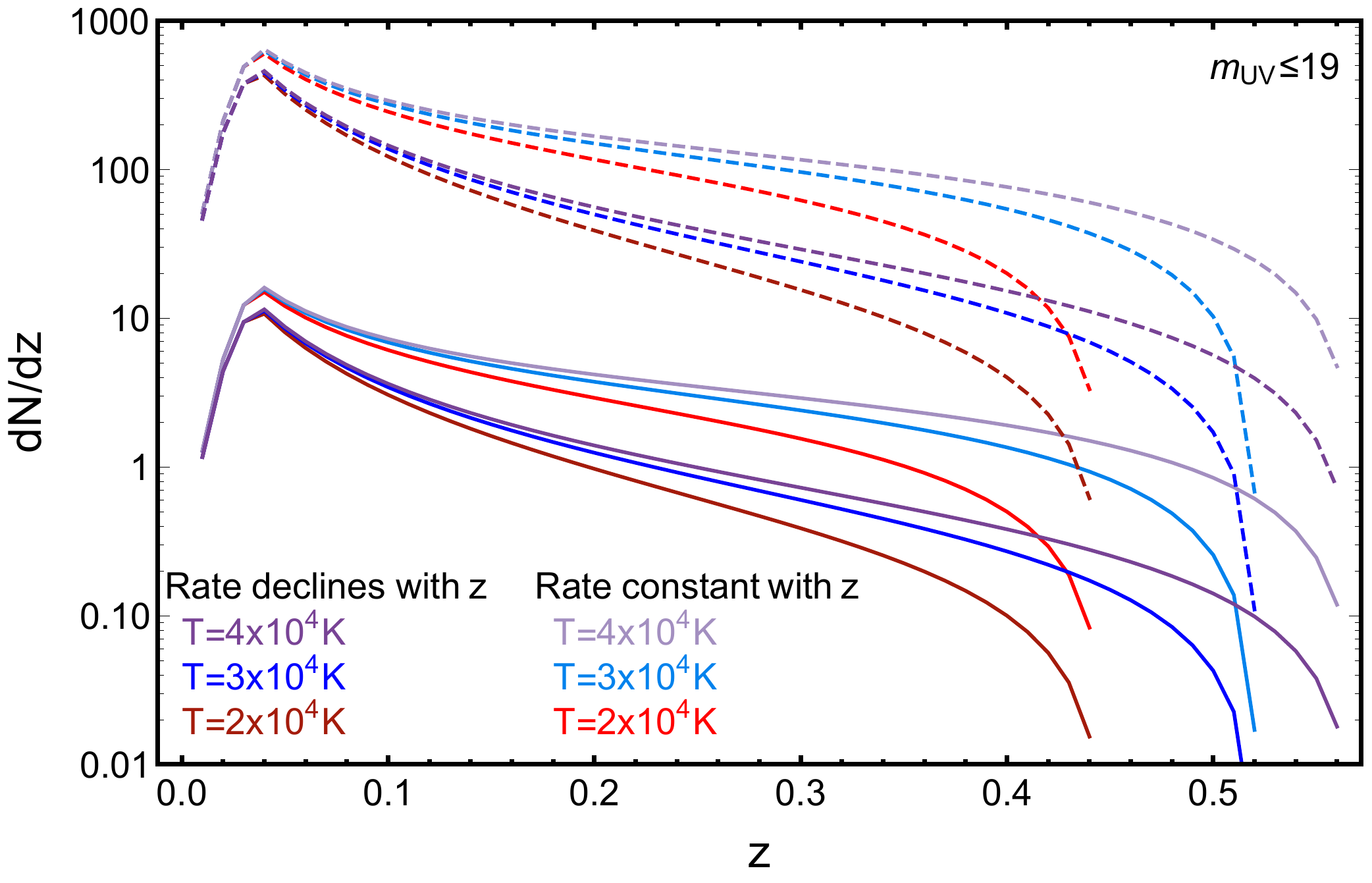}
\includegraphics[width=8.0cm]{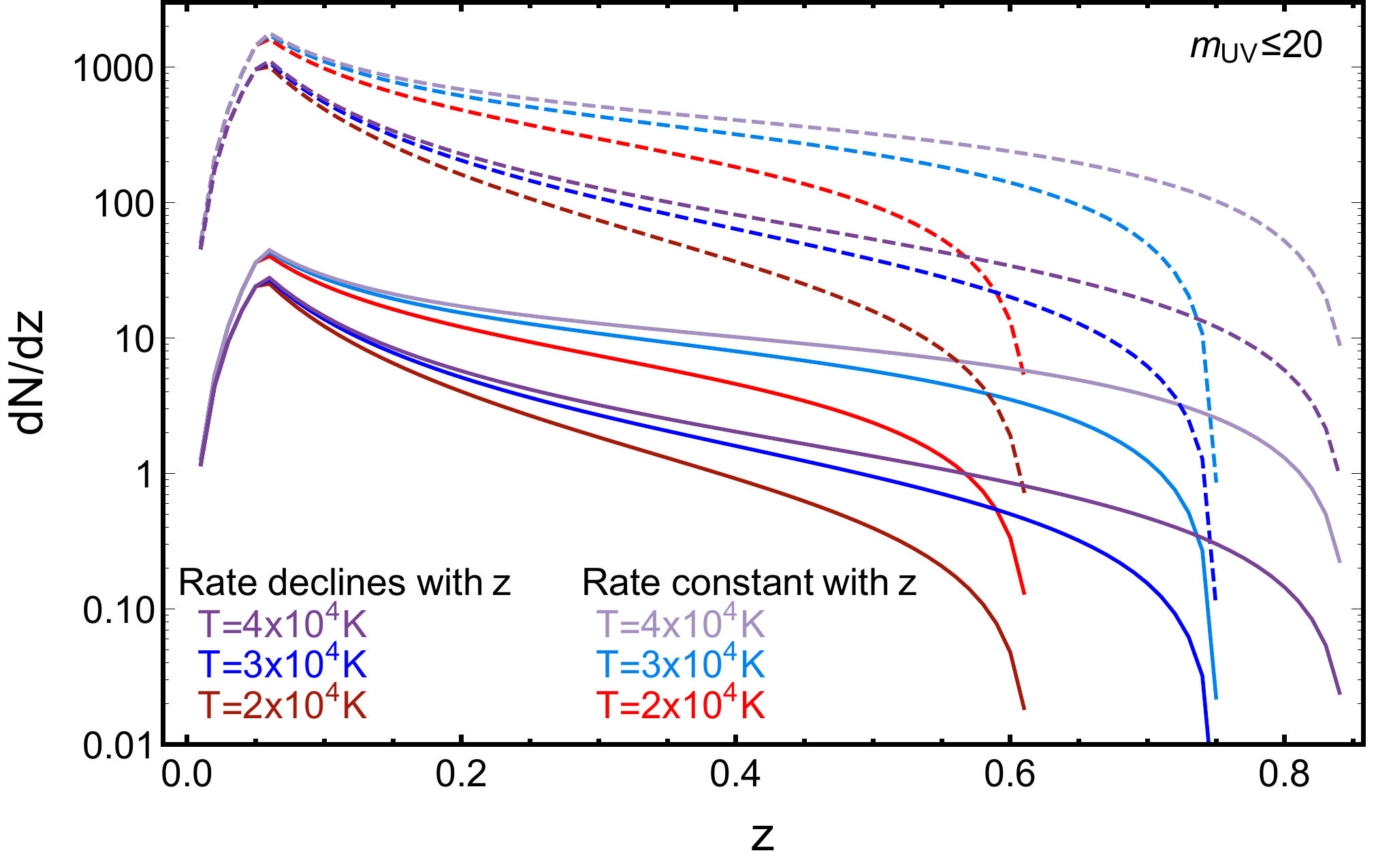}
\caption{The differential number of TDEs {\it suitable for spectroscopic followup} detected per year around a certain redshift, ${\rm d}N/{\rm d}z$.  Line styles and colors are the same as in Fig. \ref{fig:cumulTDE}, except that we now only count TDEs with peak UV magnitude $m\le 19$ (top panel) or $m \le 20$ (bottom panel).  This restriction greatly reduces the maximum redshift out to which {\it ULTRASAT} will find TDEs of interest, and represents a factor of $\sim 50-150$ reduction from the total number of detected TDEs (for the $m\le 19$ case).}
\label{fig:followupTDE}
\end{figure}

We crudely estimate the limiting luminosity distance for detecting a TDE:
\begin{equation}
    D_{\rm L}^{\rm max} = \sqrt{ \frac{ L_{\nu, \rm peak}}{ 4 \pi F_{\nu, \rm lim}}},
\end{equation}
where $L_{\nu, \rm peak}$ is the peak UV luminosity density of a TDE and $F_{\nu, \rm lim}$ is the limiting flux density of {\it ULTRASAT}. If we consider a fiducial TDE with a single-temperature blackbody spectrum, an effective temperature (at peak) $T = 2\times 10^{4}~{\rm K}$ and peak (bolometric) luminosity $L_{\rm peak}=10^{44}~{\rm erg}~{\rm s}^{-1}$, then the peak luminosity density at the center of the {\it ULTRASAT} band (250 nm) is $L_{\nu, \rm peak}=5.2 \times 10^{28}~{\rm erg}~{\rm s}^{-1}~{\rm Hz}^{-1}$.  Considering the low-cadence survey mode with a typical limiting AB magnitude of $m=22.5$, the UV detection horizon will be cosmological in distance, so we include K-corrections, and find that $D_{\rm L}^{\rm max} \approx 3$ Gpc, or $z^{\rm max} \approx 0.6$.

Since TDEs show minimal UV brightness evolution over the course of a few days (see \citealt{vanVelzen2020} and \citealt{Gezari2021}, and references therein), neither survey mode will be cadence-limited, and the number of TDEs found in a single survey of duration $\Delta T$ and sky coverage $\Delta \Omega$ will be
\begin{align}
    N \approx & \frac{\Delta\Omega}{4\pi} \Delta T \int_{L_{\rm min}}^{L_{\rm max}} {\rm d}L_{\rm peak} \int_0^{D_{\rm L}^{\rm max}(L_{\rm peak})}{\rm d}D_{\rm L} \notag \\
    & \times 4\pi D_{\rm L}^2 (1+z)^{-3} \frac{{\rm d}\dot{n}}{{\rm d}L_{\rm peak}}    . 
\end{align}
Here the extra factor of $(1+z)^{-3}$ converts to an integral over comoving distance.  In the following rate calculations, we will estimate the total number of TDEs $N$ found during a survey of $\Delta T = 1$ yr, considering both the low-cadence ($\Delta \Omega = 6800~{\rm deg}^2$) and high-cadence ($\Delta \Omega = 170~{\rm deg}^2$) modes of operation.  For the low-cadence survey, we take a limiting AB magnitude of $m_{\rm low}=22.5$ for detection.  For the high-cadence survey, we assume that individual exposures can be co-added over the course of a single day to increase detection sensitivity.  As $m=22.5$ was the limiting magnitude for a set of exposures with total duration 900 s, we estimate the limiting magnitude for a detection using 21 hours of stacked observations as $m_{\rm high}=m_{\rm low}+(5/4)\log_{10}(75600/900) \approx 24.9$.  However, {\it ULTRASAT} fields will typically be confusion-limited at magnitudes $m > m_{\rm conf} \approx 23.5$, so we conservatively assume any TDEs dimmer than $m_{\rm conf}$ will not be found.  

We will approximate the volumetric (and bolometric) TDE luminosity function as
\begin{equation}
    \frac{{\rm d}\dot{n}}{{\rm d}L_{\rm peak}} = \dot{n} \times \frac{{\rm d}N}{{\rm d}L_{\rm peak}},
\end{equation}
where $\dot{n}$ is the volumetric TDE rate, and 
\begin{equation}
     \frac{{\rm d}N}{{\rm d}L_{\rm peak}} = 1.5 \frac{L_{\rm peak}^{-2.5}}{L_{\rm min}^{-1.5}-L_{\rm max}^{-1.5}}.
\end{equation}
This is the empirical {\it bolometric} optical/NUV luminosity function fitted by \citet{van_Velzen.2018.A}, where $L_{\rm min}$ is the unknown bottom end of the luminosity function and $L_{\rm max}\approx 10^{45}~{\rm erg}~{\rm s}^{-1}$ \citep{Leloudas.2016.A, van_Velzen.2021.A, Reusch.2022.A}.  Based on our current flux-limited sample, $L_{\rm min}\lesssim 10^{42.5}~{\rm erg\,s}^{-1}$; as these faint TDEs will dominate the total volumetric event rate, we set $L_{\rm min}=10^{42.5}~{\rm erg\,s}^{-1}$ as a conservative choice.  For simplicity, we will convert bolometric luminosities into UV magnitudes by assuming that every TDE is a blackbody of constant temperature $T_{\rm BB}$, although we consider three different cases: $T_{\rm BB}=\{2, 3, 4\} \times 10^4~{\rm K}$.

A final assumption that enters our calculation concerns the volumetric TDE rate $\dot{n}(z)$.  At low redshift, we take $\dot{n}(0)=3\times 10^{-6}~{\rm Mpc}^{-3}~{\rm yr}^{-1}$ (the most conservative choice from the rate calculations of \citealt{Stone.2016.A}, and a result in reasonable agreement with the volume-corrected empirical analysis of \citealt{van_Velzen.2018.A}).  However, the redshift evolution of this rate is almost completely unknown.  Theoretical calculations \citep{Kochanek.2016.A} suggest that if $\dot{n}$ tracks the volume density of SMBHs, it will decline steeply with redshift.  We consider as an optimistic case $\dot{n}(z)=\dot{n}(0)$, and as a pessimistic case, $\dot{n}(z)=\dot{n}(0)/(1+10z)$, which crudely approximates the results of \citet{Kochanek.2016.A}.

The results of our approximate rate calculations are presented in Figure \ref{fig:cumulTDE}, which shows the cumulative number of TDEs found below a certain redshift.
For the most pessimistic combination of assumptions ($T_{\rm BB}=2\times 10^4~{\rm K}$; $\dot{n}(z)$ declines with $z$), the high-cadence survey will detect $N\approx 100$ TDEs per year and the low-cadence survey $N \approx 1500$.  For the most optimistic combination of assumptions ($T_{\rm BB} \approx 4\times 10^4~{\rm K}$; $\dot{n}(z)$ independent of $z$), the high-cadence survey will detect $N\approx 1400$ TDEs per year and the low-cadence survey $N \approx 14000$.

These detection rates are enormous, and dwarf the current sample of a few tens of TDEs \citep{vanVelzen2020}. Of course, the number of events that can be classified as TDEs via photometric and spectroscopic followup will be smaller, perhaps drastically so. This number depends on the limiting magnitude of available followup facilities. With commonly available followup resources, we will be able to spectroscopically follow dozens of TDEs at or brighter than 19th magnitude.  A smaller number of spectroscopic instruments will be able to follow targets peaking at $m\le 20$.  We repeat our calculations using these two limiting magnitudes (19 and 20) for both the low- and high-cadence fields.  The results are shown in Figure \ref{fig:followupTDE}.  When $m\le 19$ is used as the relevant threshold, the low-cadence survey will find $N\approx 32-88$ suitable TDEs per year across our range of temperature and redshift assumptions, while the high-cadence survey will struggle to find a single one ($N \approx 0.8 - 2.2$).  When $m \le 20$ is used, the low-cadence survey will find $N\approx 103-383$ suitable TDEs per year, and the low-cadence survey will find $N\approx 2.6-9.6$.

{\it ULTRASAT} will not be the only active wide-field survey searching for TDEs.  At present, the leader in TDE discovery is the ZTF optical survey, which detects $\approx 20$ new flares per year \citep{van_Velzen.2021.A}, although the {\it eROSITA} X-ray instrument, which has so far found a comparable number \citep{Sazonov.2021.A}, may ultimately be more productive once a longer temporal baseline is established for this low-cadence survey \citep{Khabibullin.2014.A, Jonker.2020.A}.  In the near future, the LSST optical survey will {\it detect} thousands of new TDEs every year \citep{van_Velzen.2011.A}, but only a small minority of these will actually be {\it identified} as TDEs from LSST photometry alone, perhaps as few as $5-10\%$ \citep{Bricman.2020.A}.  Both the cadence limitations of {\it eROSITA} and the wavelength limitations of LSST can potentially be ameliorated by contemporaneous {\it ULTRASAT} coverage of the same fields.  Likewise, the multiwavelength photometry from these other surveys may help to secure a TDE identification for the more marginal {\it ULTRASAT} detections ($m\approx 22.5$) which are inaccessible to spectroscopic confirmation yet nevertheless dominate the raw detection rate.  Joint {\it ULTRASAT}-optical detections will also be highly valuable for identifying nuclear transients, a task that will be challenging for {\it ULTRASAT}-only detections given resolution limitations.


\subsection{AGN}
\label{subsec:AGN}

The radiation emitted from the central engines of active galactic Nuclei (AGN), powered by accretion onto supermassive black holes (SMBHs), is known to vary on all timescales probed, and to peak in the UV regime. 
Assuming the canonical geometrically-thin, optically-thick accretion disk framework, ULTRASAT will be probing the inner parts of AGN accretion disks (10s-100s gravitational radii) down to dynamical timescales ($\sim$10 hours or less; see \citealt{Stern18,Trakhtenbrot19_CLAGN}, and references therein for scaling relations). 
It will also allow to better understand phenomena that arises from the reprocessing of this ``seed'' UV radiation in the circumnuclear gas regions gravitationallly bound to the accreting SMBH.

ULTRASAT is expected to probe significantly large samples of AGNs, thanks to its wide FoV and high sensitivity, combined with the steep redshift evolution of the AGN luminosity function.
Specifically, at the typical ULTRASAT depth of $NUV{\simeq}21.5$ AB mag in a single $3\times300\,{\rm s}$ visit (for $S/N=10$; see Fig.~\ref{fig:LimMag_SNR}), the sky density of unobscured (broad-line) AGNs is expected to reach ${\gtrsim}70\,{\rm deg}^{-2}$. 
This (somewhat conservative) estimate is based on the $g$-band number counts of confirmed broad-line AGNs out to $z\approx2$ \cite[e.g.,][]{Richards05_2SLAQ} and assuming a canonical SED shape of $f_{\nu} \propto \nu^{-1/2}$ \citep{VdB01}, which yields $NUV \simeq g + 0.32$ (neglecting additional foreground extinction).
Thus, every ULTRASAT pointing/FoV is expected to include several thousands of UV-detectable AGNs. Many of these, and particularly the brighter ones, will be previously known spectroscopically confirmed quasars (i.e., from the various SDSS projects, including SDSS-V in the southern hemisphere; \citealt{Kollmeier17}).
%
%

Below we briefly describe some of the main AGN-related science cases for ULTRASAT.

To enable some of these research efforts we plan to pursue follow-up observations in the optical regime, using several facilities. 
This includes 
(robotic) spectroscopy, using the Las Cumbres Observatory 2m telescopes \citep{Sand2011}, 
and narrow/medium-band imaging, using a new telescope under construction at Wise Observatory ($>$15 filters with widths ${\sim}100-400$\,\AA) 
as well as the Pan chromatic Array for Survey Telescopes (PAST; Ofek et al. in prep).

\subsubsection{Stochastic variability of persistent AGNs}

Multi-epoch imaging surveys of large AGN samples provide a phenomenological description of their variability, most commonly expressed through the Structure Function (SF) or the complementary Power Spectrum Distribution (PSD). 
Persistent AGN generally vary by ${\sim}$10\% over a year, and show greater (lesser) variability on longer (shorter) timescales. 
Moreover, the variability amplitude is anti-correlated with luminosity (at a given timescale), and there is ambiguous evidence for trends with other properties \cite[e.g.,][]{VdB04,Wilhite08,Caplar17_PTF}. 

UV variability is known to be higher compared to the optical regime \cite[e.g.,][]{Meusinger2011,Hung2016,Caplar17_PTF}, and it is key to probing the inner parts of the accretion disks that power AGNs (particularly on short timescales). 
Our current understanding of short-timescales UV variability in AGN is, however rather limited.
ULTRASAT will provide almost continuous observations,
with time scales of minutes to months (4.5 orders of magnitude in time scale) of a large sample of quasars with high $S/N$ (see above).
This will allow us to pursue the following studies, among others:

\paragraph{Quantifying AGN UV variability on short timescales}

ULTRASAT will survey the (nearly) uncharted territory of short timescale variability of ``normal'', persistent AGNs (minutes to days). 
This would allow to quantify the full distribution of AGN variability in this crucial regime, and to search for links with basic SMBH properties, such as BH mass, accretion rate (\Lagn\ or \lledd), radio jet activity, etc., to yield insights for accretion flow models.

Specifically, we will be able to construct the ensemble SF (and PSD) of various groups of AGNs, on all accessible timescales and drawing from all ULTRASAT surveys.
The shortest timescales (minutes-to-hours) are of particular interest, as explained below.
We will be able to compare the resulting SF to what is known from the optical regime, and specifically test whether the phenomenological (damped) random walk model, which was suggested to describe AGN variability in the optical regime, is applicable to the UV regime.
The ensemble SF and PSD can be also constructed in bins of \Lagn, \mbh, and \lledd, to search for correlations with any of these properties. This, in turn, can be directly used to test various models for accretion disk instabilities \cite[see, e.g.,][]{AraveloUttley06,Ruan14,Caplar17_PTF}.
Understanding the full distribution of variability is also important for our ability to identify and quantify ``extreme'' variability and/or AGN-related transients (see Section~\ref{subsubsec:AGN_transients} below).

\paragraph{Black hole mass from break in power spectrum}%

Observations suggests that AGN variability is well described by a PSD with a steep power-law index  \cite[e.g.,][and references therein]{Smith18,Caplar17_PTF}.
In principle, this power-law should have both high  frequency (``inner'') and low frequency (``outer'') break.
Since timescales in the disk scale with the distance from the central BH, the inner break in the PSD may reflect the innermost radius of the accretion disk, which in turn scales linearly with BH mass. 
%
Thus, if a high-frequency break in the PSD of AGNs, and its correlation with \mbh\ are established, this may provide a new tool to estimate BH masses in distant systems (and perhaps even probe BH spin).

Measuring this inner break likely requires high precision continuous UV observations, as the UV radiation is emitted from inner-most parts of the disk \cite[see][]{SpringerOfek21a}.
Even in the event that an inner break is found, but it is not correlated with BH mass, it may provide a new information on accretion physics in AGNs.
Unlike previous studies that focused on the PSD of the  corona-reprocessed X-ray emission \citep{Kelly13_PSD}, ULTRASAT would focus directly on inner disk itself, exploring a much broader range of frequencies and possibilities.

\subsubsection{Agile Reverberation Mapping}

The basic premise of reverberation mapping (RM) in AGN is to map the physical structure of and around the central engine by measuring time lags between variations in the ``seed'' (UV-dominated) radiation coming from the inner accretion disk, and the reprocessed radiation arising from various circumnuclear gaseous regions. 
These mainly include the (outer) disk itself and the broad line region (BLR), but also the dusty ``torus'' (via NIR RM; e.g., \citealt{Koshida14_NIR_RM}), or the X-ray emitting corona \cite[e.g.,][]{Kara16_X_RM}. 

All RM efforts face the same two key practical challenges. 
First, most campaigns focus on the optical regime, while the seed disk radiation is UV-dominated. This practical choice complicates the measurement with (persistent) host emission, and perhaps other reprocessing components \cite[e.g.,][]{Chelouche19}.
Second, monitoring a sizable AGN sample for a long period and a sufficiently high cadence is extremely challenging, although it is clearly required to ensure that significant variability in both the seed (disk) and responsive emission components is properly recorded.
Photometric RM alleviates some of these practical challenges, as it allows to robustly measure AGN time lags through photometric monitoring with various band widths instead of spectroscopy, 
as demonstrated by a growing number of studies \cite[see, e.g.,][and additional references below]{CheloucheZucker2013,Chelouche2014,Ramolla2018,Kim2019}.

ULTRASAT will allow to make great progress in AGN RM studies. It will continuously monitor the NUV emission of hundreds-to-thousands of luminous AGN, which is completely dominated by accretion disk radiation. 
The nearly real-time data processing would ensure that follow-up observations can be efficiently carried out, i.e. only when significant changes to the seed UV emission are detected. 
Similarly, NIR or X-ray imaging can be triggered to study the dusty torus or the corona (respectively).

\paragraph{Broad line region RM}

RM of broad emission lines in (unobscured) AGNs allows to study the structure and physics of the broad line region (BLR; \citealt{Peterson93}). 
The time lag, and thus size measurements of this dense circumnuclear gas \cite[\RBLR][]{Kaspi00,Bentz13}, combined with measurements of its virialized, SMBH-governed kinematics, provide a way to estimate \mbh\ for huge samples of distant AGNs (out to $z>6$; e.g., \citealt{TN12}). 
To date, reliable BLR lags were measured for several tens of AGNs \cite[e.g.,][and references therein]{Bentz15_DB,Homayouni20,Yu22}. 
The most recent optical studies demonstrate the difficulty in expensive monitoring of sources that do not exhibit sufficient variability; and the significant scatter in the optical $L-\RBLR$ relation, which may be alleviated if the UV continuum is monitored, instead.

The typical BLR sizes of several light-days to light-months mean that the few-days cadence, wider-field ULTRASAT survey(s) can be used for BLR RM, maximizing the pool of potential targets. Smaller samples of low-luminosity (low-mass) sources can be better studied within the nearly-continuous ULTRASAT fields. 
The variability-selected AGNs should then be dynamically targeted with ground-based spectroscopy and/or narrow/medium-band imaging (see above).
Tailoring the narrow/medium bands for each target will allow to decompose and measure line and continuum emission, per epoch. 
Spectroscopy for bright enough targets may facilitate velocity-resolved RM, and shed light on the BLR structure and kinematics.

\paragraph{Continuum \& accretion disk RM}

Recent RM and microlensing studies suggest that accretion disks in quasars are a factor of a few larger than predicted by simple thin disk models \cite[e.g.,][]{Blackburne11,Fausnaugh16}, with potential implications for understanding accretion in general, and SMBH fueling in particular.
However, some of those findings have been recently challenged by high-fidelity {\it continuum} RM campaigns, which suggest non-negligible continuum emission from the BLR in quasars may be contaminating the longer lag RM signal \cite[e.g.,][]{Chelouche19,Netzer22}. 
This generally overlooked BLR emission component could alleviate the tension between disk theory and observations, and also provide a novel window into BLR physics.
High-cadence and precise UV monitoring is crucial for disk (and continuum) RM studies, since this inner disk (seed) emission allows to anchor the entire observed lag--wavelength relation (see, e.g., Fig.~5 in \citealt{Fausnaugh16}). Such UV monitoring is obviously very challenging to pursue with existing facilities, even for single targets.

ULTRASAT will allow to measure disk and diffuse (BLR driven) continuum lags in a sizeable sample of $z\lesssim0.5$ AGNs. 
In this case, the typical thin disk sizes of ${\sim}1-5$ light-days mean that targets will be selected from the nearly-continuous, staring-mode ULTRASAT field(s).
As for the simultaneous optical follow-up observations, the narrow/medium-band imaging should include several line-free spectral regions (but see below for an alternative approach). 
Adding a few line-focused bands would enable broad line RM to be done in parallel.

\paragraph{Single band RM}%

Several studies have suggested that it is possible to measure BLR reverberation timescales using single band observations \cite{Zu2016_RM,SpringerOfek21a,SpringerOfek21b}.
Generally, the idea is that the observed quasar variability distribution would be comprised of a dominant driving signal (i.e., a power law or damped-random-walk shaped PSD or SF; see above) and a superimposed (weak) reverberating signal. 

The recent work by \cite{SpringerOfek21a,SpringerOfek21b} suggests that identifying this latter reverberating signal may be possible if the emission line is sufficiently strong, i.e. if the equivalent width is $\gtrsim10$\% of the observed band. 
For $z\sim1$ quasars, the Ly$\alpha$ line shifted to the ULTRASAT band and its equivalent width may be of the order of 30\% of the band. 
This novel method requires a large number of targets and a high cadence, which ULTRASAT may provide in its nearly-continuous, staring-mode fields if we maximize the number of accessible AGNs in them. Alternatively, this methods may be best applied for a future wider-area and somewhat slower cadence ULTRASAT survey (i.e. a possible medium tier survey).

We expect that ULTRASAT’s main advantage would arise by comparing the (effective) time delays of large subsets of quasars at given redshift intervals. This would allow us to better understand the scatter in the relations linking BLR size and AGN luminosity, and this scatter is influenced by additional AGN properties, such as the mass accretion rate (expanding on pioneering studies of small samples; see, e.g., \citealt{Du2018} and references therein).

\subsubsection{Extreme AGN variability and SMBH-related transients}
\label{subsubsec:AGN_transients}

Time-domain surveys have recently started to reveal new types of extreme variability and of transient phenomena associated with accretion onto SMBHs. 
Observationally, these include 
(1) persistent AGNs that vary by more than $\approx$1 mag on timescales of a few years \cite[or less; e.g.,][]{Lawrence16,Rumbaugh18}; 
(2) ``changing look'' AGN (CLAGN)---systems where the blue continuum and/or broad line emission typical of (unobscured) AGNs (dis-)appearances \cite[e.g.,][]{LaMassa15,Runnoe16,MacLeod19,Green22}, sometimes on timescales of weeks-to-months \cite[e.g.,][]{Gezari17,Trakhtenbrot19_CLAGN,Zeltyn22}; 
and (3) a growing number of nuclear transients in previously known AGN, exhibiting peculiar properties in their lightcurves, spectra, and/or multi-wavelength spectral energy distributions 
\cite[e.g.,][]{Drake11,Blanchard17,Kankare17,Trakhtenbrot19_BFF,Frederick21}.
These phenomena challenge decades-old models for SMBH accretion flows \citep{Lawrence18_NA}, and provide an unprecedented opportunity to explore accretion physics and SMBH fueling.

Most AGN-related transients are characterized by blue continuum emission, strongly rising towards the UV, and hints for intense ionizing radiation (based on certain spectral features). 
In many cases the UV flares are much more pronounced than the optical ones. 
These events are rare, comparable to (and perhaps rarer than) TDEs (Section~\ref{subsec:TDEs}). 
Practically, we are currently identifying roughly one transient of interest per month, based solely on public alerts based on optical imaging.
A sufficiently wide NUV survey (1000s of deg$^2$) with few-days cadence could yield 100s of events, per year.

ULTRASAT will allow to detect and survey AGN- and SMBH-related transients in the regime which most closely probes the inner workings of the accretion flows. 
The nearly real-time data analysis will allow us to trigger key follow-up multiwavelength observations (practically focusing on $z\lesssim1$), and to probe the light-curve peak, which is often missed in current efforts.
AGN transients of particular interest include 
(1) events that combine a sharp rise followed by a plateau near the Eddington luminosity--thus probing super-Eddington accretion; 
and (2) flaring AGNs driven by TDEs that perturb \emph{pre-existing} thin accretion disks \cite[e.g.][]{Merloni15,Chan19}.
The well-designed nature of the ULTRASAT survey(s) will allow us to determine the occurrence rates of various (classes of) AGN-related transients, which is important to understand their driving mechanisms and role in SMBH growth.

\subsubsection{Strongly lensed quasars}

Identifying strongly lensed quasars enable 
studies of 
dark matter halos \cite[e.g.,][]{MaozRix1993}, galaxy  evolution \cite[e.g.,][]{Ofek2003}, interiors \cite[][]{Kochanek2004}, and cosmography (e.g., \citealt{TreuMarshall16_rev}, but see also \citealt{Blum2020} and \citealt{Kochanek2020}).
While the low spatial resolution of ULTRASAT is insufficient to resolve the multiple images of strongly lensed quasars, the high-cadence ULTRASAT lightcurves will allow to identify such systems based on the combined variability of their flux and center-of-light \cite[following][]{SpringerOfek21a,SpringerOfek21b}.

\subsubsection{Other AGN-related science}

We briefly mention a few other AGN-related science projects enabled by ULTRASAT data:
    
\paragraph{Identifying AGN signatures in compact radio sources} to test models of jet launching and AGN re-ignition \cite[e.g.,][]{Mooley16}.
    
\paragraph{Studying the SEDs of jetted AGNs (blazars)} - to distinguish between disk- and jet-dominated emission mechanisms \cite[e.g.,][]{Rodrigues21}.
    
\paragraph{Selecting ``background'' quasars, based on variability} to enable studies of galaxies and their circumgalactic medium \citep{Tumlinson17}.
ULTRASAT will greatly expand on previous similar efforts (e.g., with GALEX; see \citealt{Wasleske2022}).


\subsection{Galaxies}
\label{subsec:galaxies}

ULTRASAT will produce the most sensitive all-sky near-UV galaxy survey yet.  Moreover, the combined depth achieved in its high-cadence field(s) will yield sensitivities greatly exceeding the deepest GALEX fields, and competitive with medium deep {\it Hubble Space Telescope} UV observations, albeit limited by crowding in the faintest fluxes.  

We base our galaxy count estimates on the point source sensitivity shown in figure~\ref{fig:LimMag_SNR}, since a majority of galaxies near the ULTRASAT detection threshold will be compact compared to the PSF.  
We adopt a near-UV number-magnitude relation based on a combination of {\it GALEX} NUV data \citep{Xu.2005.A}, {\it Swift} UVOT uvw1 data \citep{Hoversten.2009.A}, and {\it Hubble Space Telescope} F275W data \citep{Teplitz.2013.A}.  Both the {\it Swift} and {\it HST} data are well matched in central wavelength to the ULTRASAT band, while the {\it GALEX} NUV channel is slightly bluer.   Together, these sources provide estimates of the surface density for $15 \la m_{AB} \la 27$.  

\begin{figure}
\centerline{\includegraphics[width=0.5\textwidth]{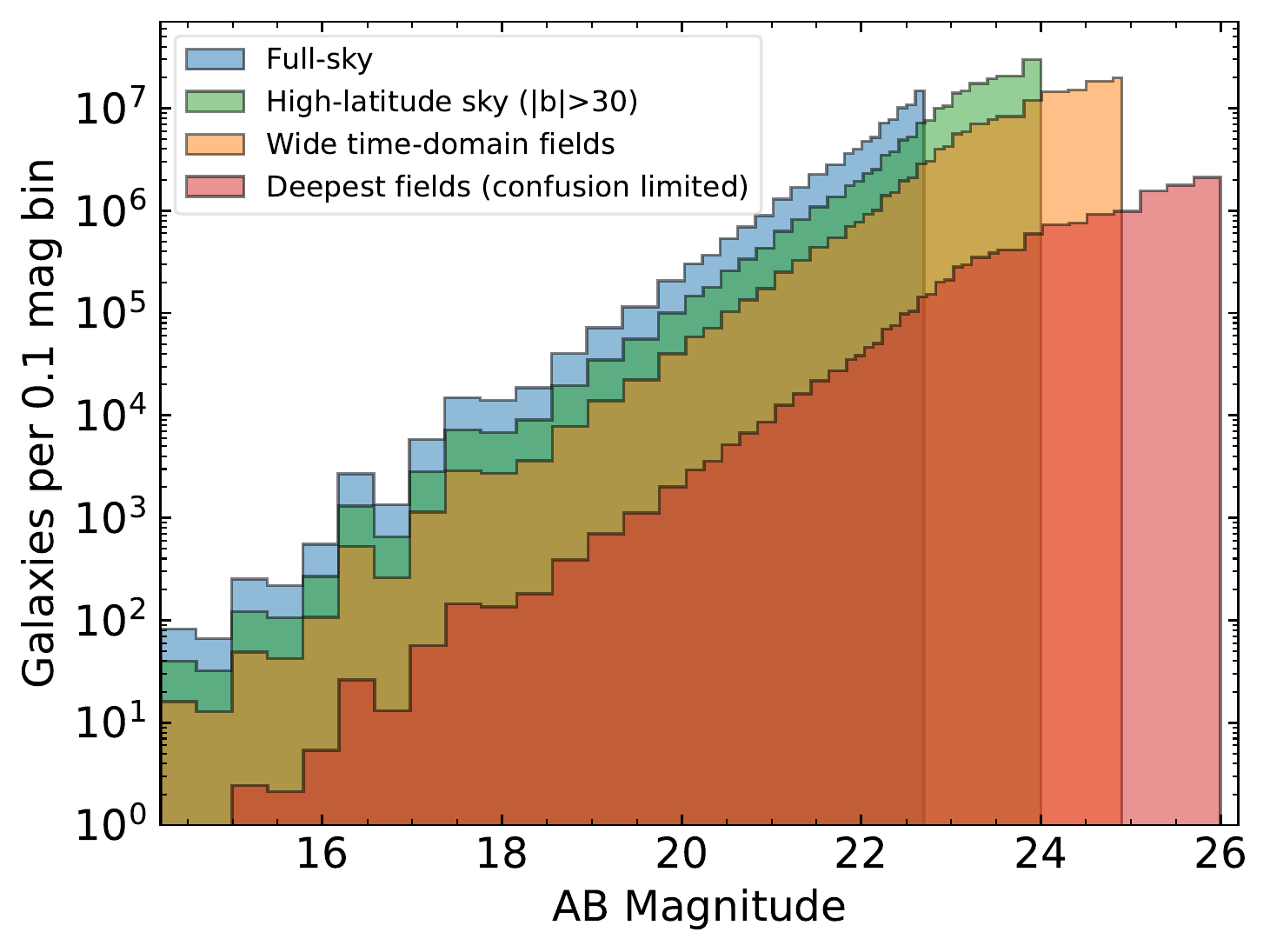}}
\caption{Expected number-magnitude distribution for galaxies in ULTRASAT surveys. Different colors represent different tiers of ULTRASAT survey.  The all-sky survey is subdivided into the high latitude portion (green) and the shallower observations over the full sky including low galactic latitudes (blue). The time domain fields are similarly divided into the low-cadence, wide area portion (yellow) and the deepest fields (red).  Each colored histogram is cut off at the magnitude limit appropriate to that survey's anticipated integration time and the 10--$11''$ PSF expected over most of the field, except for the deep field histogram, which terminates at the $5\sigma$ confusion limit for a $6''$ PSF. See text for discussion of samples upon which the plot is based.  
\label{fig:gal_sample}}
\end{figure}

Combining these number-flux relations with the survey modes outlined in section~\ref{sec:survey_mode}, we have estimated the total galaxy samples expected in ULTRASAT data.  The result is plotted in figure~\ref{fig:gal_sample}.  In total, we anticipate that ULTRASAT will detect (at $\ge 5\sigma$) a sample of $\sim 3\times 10^8$ galaxies, dominated by the high latitude sky $|b| > 30^\circ$ where the deeper portion of the all sky survey will typically achieve magnitude limits near $m_{AB}=24$.

\begin{figure}
\centerline{\includegraphics[width=0.5\textwidth]{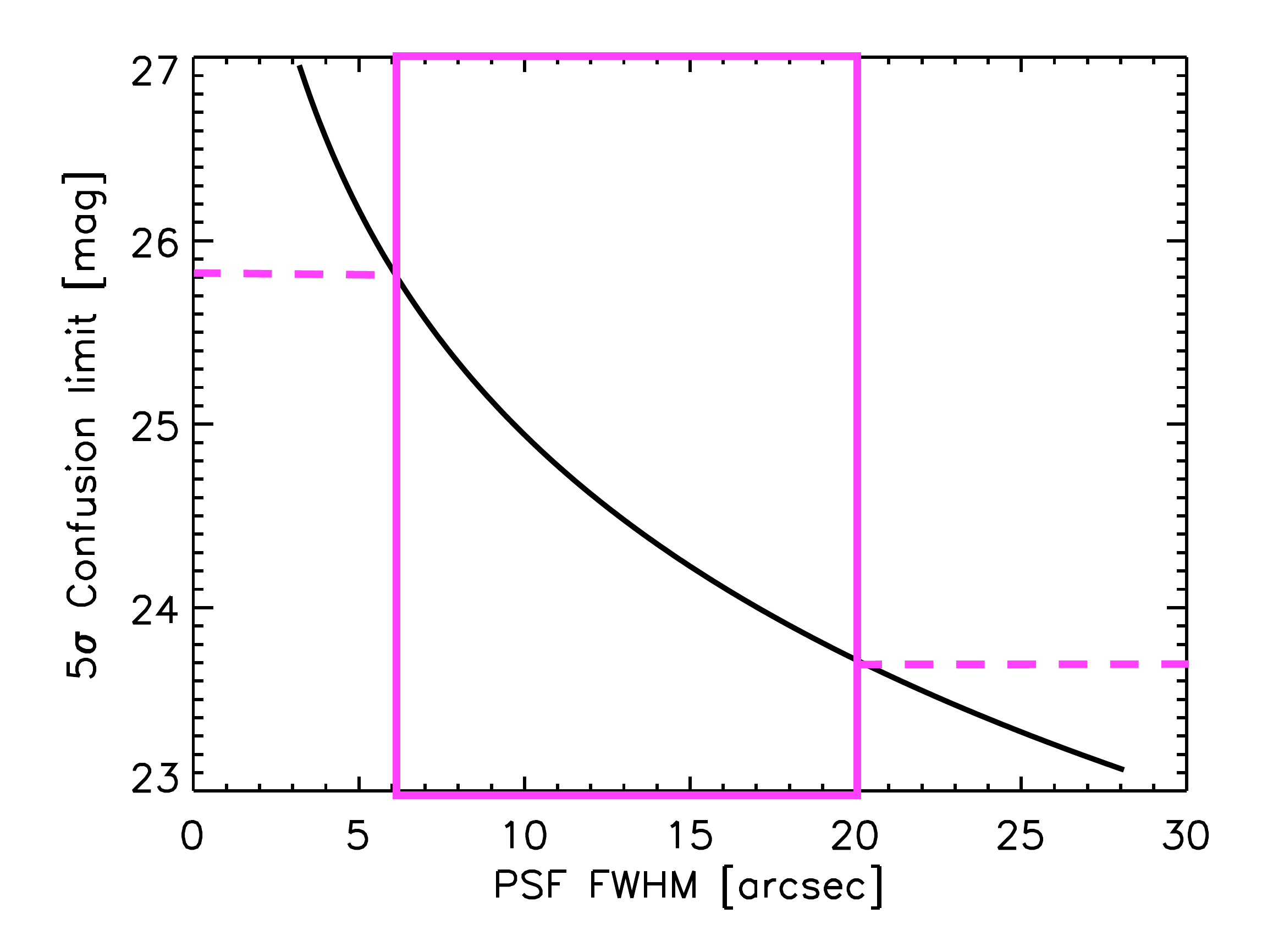}}
\caption{The 5-$\sigma$ confusion limit for near-UV imaging is plotted as a function of PSF size.  The key range for ULTRASAT is approximately $6'' \le \hbox{\rm FWHM} \la 20''$ (cf. figure~\ref{fig:Eff_PSF}).  Within its annulus of best focus ($4^\circ < \theta < 6^\circ$ off axis), ULTRASAT can be used to study galaxies as faint as $m_{AB} \approx 26$ with integration times $\sim 50$ hours.
\label{fig:confusion}}
\end{figure}

Because the total ULTRASAT exposures in the time domain fields may exceed $10^4$ hours, crowding is a critical consideration.  We have calculated the crowding limit expected by applying the analytical treatment from \citet{Condon.1974.A}, using a power law fit to the number-flux relation from \citet{Hoversten.2009.A}.  Because the PSF FWHM varies substantially with field angle for ULTRASAT, we kept FWHM as a free parameter. The resulting limits are shown in figure~\ref{fig:confusion}.
We have neglected the stellar contribution to the confusion noise.  This is justified at high galactic latitudes $b$ because galaxy counts exceed star counts for $m_{AB} \ga 21$ for $|b| \ga 30^\circ$ \citep{Xu.2005.A}.  (Near the plane of the galaxy, a more thorough treatment that includes confusion by Galactic stars would be appropriate.)

It will be possible to measure near-UV fluxes of objects somewhat fainter than the nominal crowding limit in cases where good priors can be derived from higher resolution imaging at other wavelengths.  Such approaches have been applied in past to analysis of {\it Spitzer Space Telescope} data, and are under active consideration for the analysis of Vera Rubin Observatory data in combination with higher resolution data from {\it Euclid} or the {\it Nancy Grace Roman Space Telescope}.

As an example of ULTRASAT’s impact on galaxy surveys, we estimate the benefit of having sensitive NUV data when computing photometric redshifts.  The accuracy of photometric reshifts can be greatly improved by sampling wavelengths that span prominent spectral features like the Lyman-break at 912\AA.  ULTRASAT’s bandpass covering 230--290 nm will be used to confirm this spectral break in $z\sim 2$ galaxies.  

To quantify our ability to recover photometric redshifts with ULTRASAT, we take the COSMOS2020 catalog \citep{Weaver2022} and degrade the ugrizy flux errors to the 5-year Rubin Observatory depths (ugrizy = 25.7,27.0,27.1,26.4,25.7,24.5 ($5\sigma$), respectively), which will be available over 18,000 deg$^2$ by the end of ULTRASAT’s prime mission.  The COSMOS2020 catalog also contains GALEX NUV data with a $3\sigma$ depth of 26 AB mag, comparable to the deepest ULTRASAT fields.  We produce two ugrizy catalogs, one with NUV data and one without, to simulate a survey with and without ULTRASAT.  We use the code EAZY \citep{Brammer2008} to obtain photometric redshifts and compare our results to known spectroscopic redshifts from the SDSS, zCOSMOS, and DEEP3 surveys \citep{Ahumada2020,Lilly2009,Cooper2012}.  These spectroscopic redshifts range from $z=0$ to 3.5 with an average of $z=0.5$. We find that including the NUV photometry primarily improves the fraction of catastrophic photo-z failures ($\Delta(z)/(1+z)>0.15$), which drops from 25\% with optical data alone to 17\% with the inclusion of the NUV band, over the full redshift range tested (which is dominated by galaxies at $0<z<2$). 

We anticipate a stronger impact on the false-positive rate for Lyman break galaxies (LBGs) at $2.2<z<2.8$, where the Lyman break results in no flux in the ULTRASAT NUV band, while all optical passbands accessible to ground-based telescopes will show strong detections for blue, actively star-forming galaxies.  At this redshift, the characteristic 1500\AA\ magnitude $M^* = -20.97$ \citep{Bouwens2015}, corresponding to $m^*_{AB} \approx 24.3$.  The deepest ULTRASAT photometry will go 2 magnitudes deeper than this, and combined with Rubin data will yield complete samples of LBGs down to $M^*$ for hundreds or thousands of square degrees.


\subsection{Solar system}
\label{subsec:solar_system}

ULTRASAT’s all-sky UV map observations  will be sufficient to measure the NUV color of more than $10^4$ asteroids, at different sections of the main belt of asteroids. Studies examining the NUV range of minerals and meteorites show that the NUV spectral region is sensitive to different mineral properties, thus it might provide complementary capabilities and opportunities for asteroid classification ($\it e.g.$, \citealt{Cloutis2008}, \citealt{HendrixVilas2006}). Color differences between the main asteroid types (S-type - silicate-based minerals; C-type – carbonaceous-based minerals) were measured by the GALEX space telescope (\citealt{GALEX_asteroids}) and the International Ultraviolet Explorer (IUE; \citealt{IUE_asteroids}).

\begin{figure}
\centerline{\includegraphics[width=0.45\textwidth]{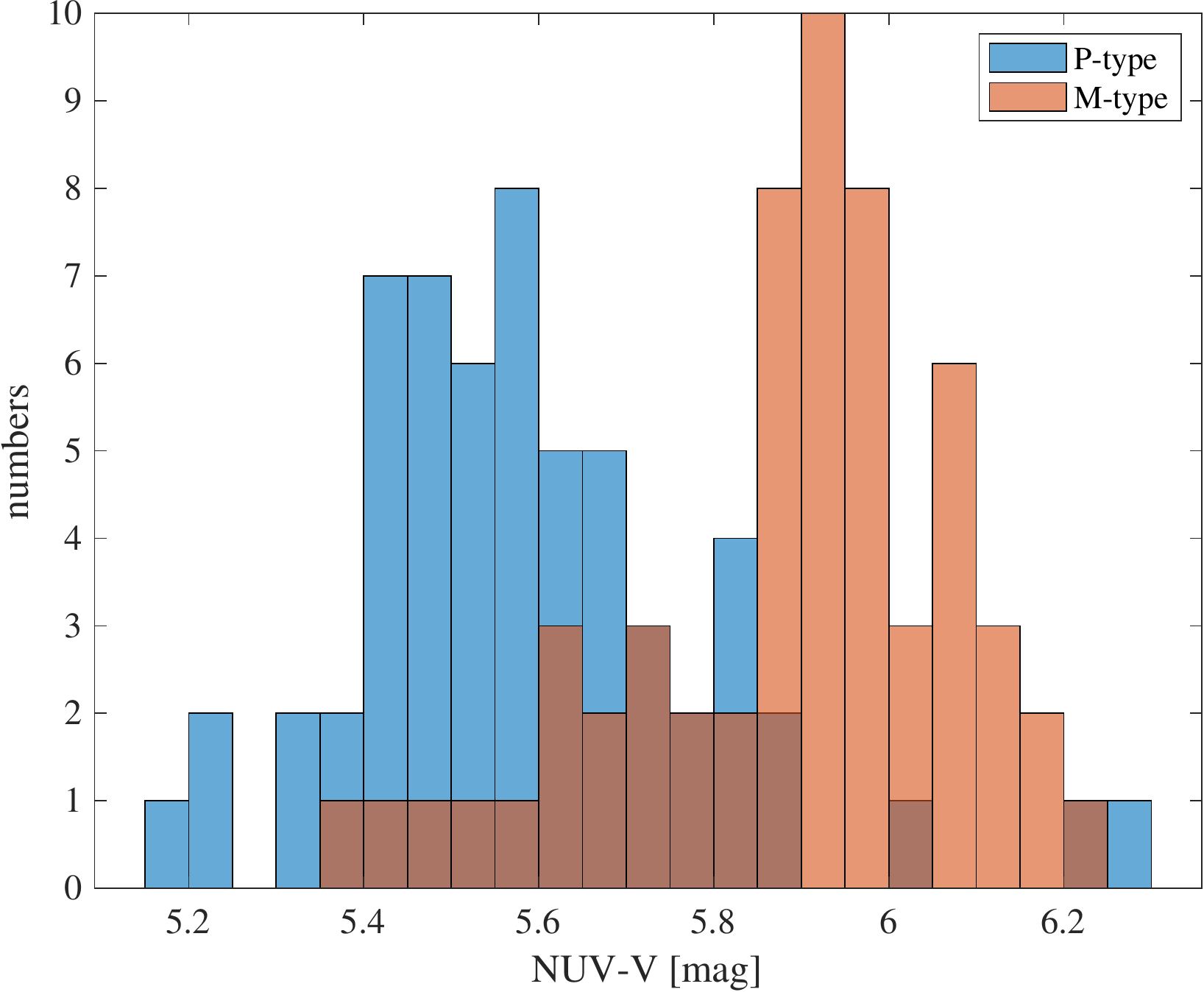}}
\caption{Distribution of NUV-V colors of P- (bluish) and M-type (reddish) asteroids measured by the GALEX space telescope (\citealt{GALEX_asteroids}). The Vmag is from the Minor Planet Center (MPC). Mean values are separated by about 0.3 mag, a value which is larger value than ULTRASAT photometric precision.
\label{fig:Asteroids_NUV}}
\end{figure}

Since the NUV region is a sensitive indicator of the presence of even trace amounts ($<0.01 wt\%$) of Ferrum (\citealt{Cloutis2008}), it has the potential to break the degeneracy at both the visible and near-IR range (\citealt{DeMeoTax2009}) between the spectral signature of metal asteroids (M-type) and the primitive/organic asteroids (P-type). NUV measurements from GALEX show a mean difference of about 0.3 mag between these two types (Fig. \ref{fig:Asteroids_NUV}). This difference is well above the photometric uncertainty of ULTRASAT, estimated as $1\%$. This difference can also be seen on the surface of the large asteroid 1 Ceres (which belongs to the C-type classification), that was found to present a deep absorption band at around 250nm \citep{Li_CeresUV}, while metal meteorites have fixed slope at the NUV range with no light diminution \citep{Cloutis2008}.

\begin{figure}
\vspace{5mm}
\centerline{\includegraphics[width=0.5\textwidth]{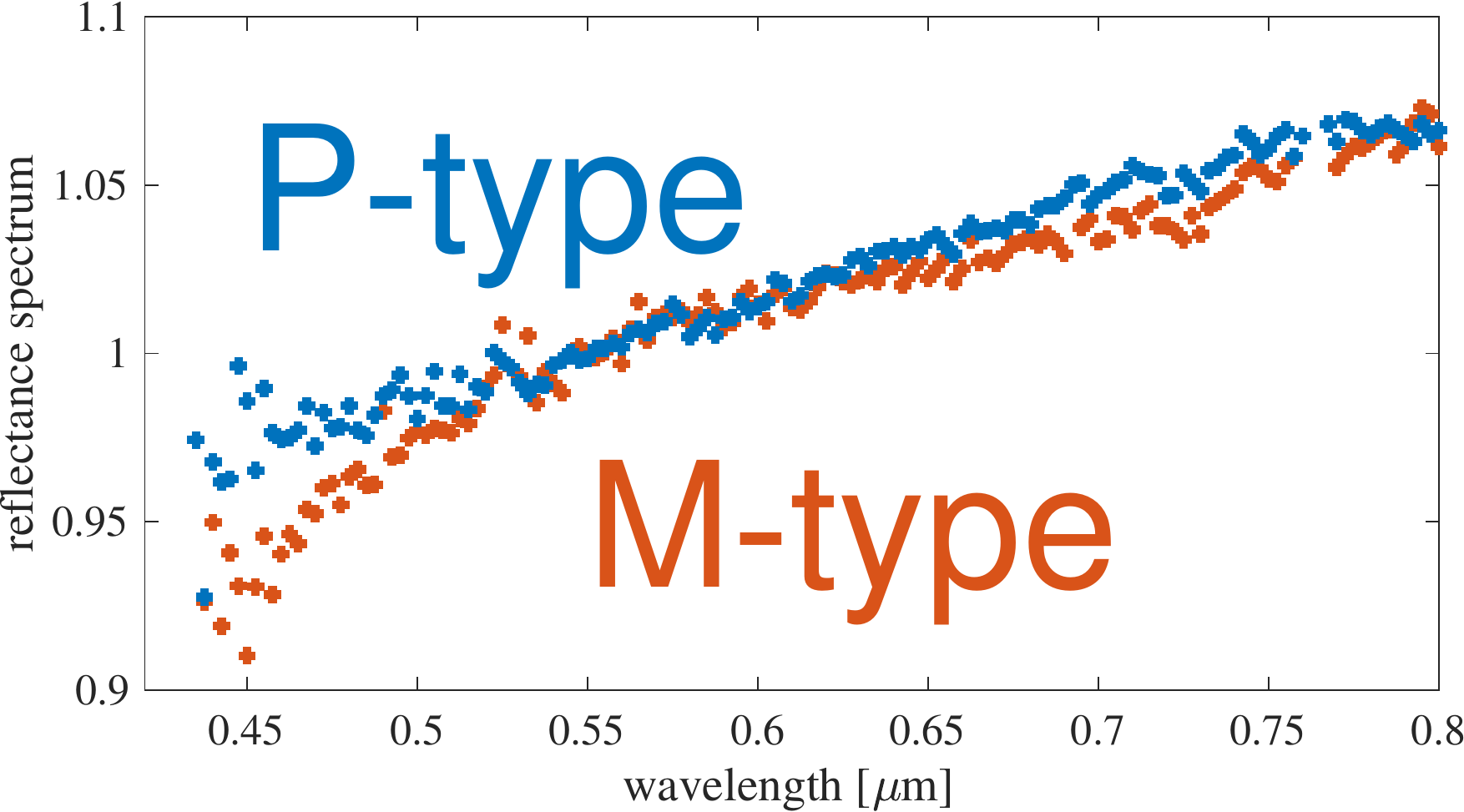}}
\caption{Visible reflectance spectra of asteroids 125 Liberatrix (M-type, reddish) and 46 Hestia (P-type, bluish). Data from the SMASS library (http://smass.mit.edu/), observation and classification conducted by \citealt{BusBinzel2002TaxA}.
\label{fig:Asteroids_SpectraPM}}
\end{figure}

Both M-type and P-type asteroids are defined by a flat, almost feature-less, reflectance spectra (Fig. \ref{fig:Asteroids_SpectraPM}), that usually hides well within the error range of measured spectra. They are only distinguished from one another when an albedo value is available, with P-types have low values and M-types have mid-values. However, a secure albedo value with low uncertainty is hardly available, which prevents their correct identification, and makes the ULTRASAT measurements valuable for asteroid studies.

Disentangling between the M-types and the P-types might solve the “missing mantle problem” (\citealt{Burbine1996}) – the inconsistency between the numbers of M-types, that supposedly originate from the inner cores of differentiated planetesimals later destroyed by catastrophic collisions, to the number of asteroids originated from the mantle layers of the same destroyed differentiated planetesimals (dubbed as A-type asteroids). A possible answer lays within a wrong identification of P-type asteroids and defining them as M-type asteroids, due to their reflectance spectra resemblance in both the visible and near-IR range. Therefore, ULTRASAT’s NUV color of asteroids, have the potential to solve this inconsistency, resulting in fine-tuned models of planetesimals formation.

\section{Summary}
\label{sec:summary}

We have described ULTRASAT, a wide-field time-domain UV space telescope that is expected to be launched in 2026 (see table~\ref{tab:properties} for ULTRASAT's key properties). ULTRASAT's design (\S~\ref{sec:design}) and operation modes (\S~\ref{sec:modes}) are optimized for studying the transient and variable UV sky. With a grasp (i.e. instantaneous volume of the universe accessible for transients' search) much larger than that of current surveys and comparable to that of the planned VRO LSST (see fig.~\ref{fig:Grasp}), ULTRASAT will be at the forefront of time domain astronomy both as a transient discovery machine driving vigorous ground- and space- based follow-up campaigns, as well as an optimal and fast response follow-up facility. ULTARSAT's location at GEO will enable it to publicly distribute transient alerts within 15-min from imaging, and its agility will enable it to access any point in the observable sky ($>50\%$ of the sky at any time) within 15-min from an external transient alert received at the science operation center (SOC, \S~\ref{sec:SOC}). 

The large enhancement of the discovery rate of transients and the continuous min-mon cadence UV light cruves, that will be provided by ULTRASAT, will have a significant impact on a wide range astrophysics research areas (see table~\ref{tab:highlights} and \S~\ref{sec:Science}), ranging from high-energy extragalactic sources such as BNS and NS-BH mergers, SNe, TDEs, AGN, and GRBs, through active and flaring stars and exoplanet host stars, up to solar system objects. 

The key science goals, that defined the required technical capabilities of ULTRASAT, are the discovery and observation of electromagnetic emission from BNS mergers and SNe. 
\\ - With a large fraction ($> 50\%$) of the sky instantaneously accessible, fast (minutes) slewing capability and a field-of-view that covers the error ellipses expected from gravitational wave (GW) detectors beyond 2025 for events at distances of $\lesssim300$~Mpc, ULTRASAT is expected to be the best observatory for detecting EM counterparts of GW events produced by mergers involving neutron stars (see fig.~\ref{fig:Rates}, \S~\ref{subsec:GW}). Measuring the EM emission following BNS/NS-BH mergers will (i) provide direct constraints on the structure and composition of the ejected material, thus providing unique diagnostics of the properties of matter at nuclear density and of the merger dynamics, and (ii) enable to determine whether mergers are the sources of r-process elements and gamma-ray bursts, and (iii) allow to determine the location in, and properties of, the host galaxy, thus revealing the stellar antecedents of the binary systems. The early UV light curves will provide unique constraints and will be highly useful for discriminating between different models for the structure and composition of the ejecta (\S~\ref{subsubsec:GW-UV-importance}).
\\ - ULTRASAT will detect hundreds of SNe within the first day from explosion, and tens within the first hour- a discovery rate which is an order of magnitude larger than that of any other survey (see fig.~\ref{fig:Rates}, \S~\ref{subsec:SNe}). This will allow, for the first time, to systematically detect shock breakouts and to construct early ($<1$d) continuous high (minutes) cadence UV light curves for hundreds of core-collapse SNe, including for rarer BSG and WR SN progenitor types. Measuring this early shock breakout/cooling part of SN light curves will provide unique information on the progenitor stars and their pre-explosion evolution, in particular mapping the different types of SNe to the different stellar progenitors, and hence also constraints on the explosion mechanisms, which are not fully understood.

\begin{acknowledgments}
EW's research is partially supported by ISF, GIF and IMOS grants.
AGY’s research is supported by the EU via ERC grant No. 725161, the ISF GW excellence center, an IMOS space infrastructure grant and a GIF grant, as well as the André Deloro Institute for Advanced Research in Space and Optics, The Helen Kimmel Center for Planetary Science, the Schwartz/Reisman Collaborative Science Program and the Norman E Alexander Family M Foundation ULTRASAT Data Center Fund, Minerva and Yeda-Sela;  AGY is the incumbent of the The Arlyn Imberman Professorial Chair.
E.O.O. is grateful for the support of grants from the Willner Family Leadership Institute, André Deloro Institute, Paul and Tina Gardner, The Norman E Alexander Family M Foundation ULTRASAT Data Center Fund, Israel Science Foundation, Israeli Ministry of Science, Minerva, BSF, BSF-transformative, NSF-BSF, Israel Council for Higher Education (VATAT), Sagol Weizmann-MIT, Yeda-Sela, and Weizmann-UK.
S.B.A is grateful for support from the Willner family foundation, Israel Science Foundation, Israel Ministry of Science, Minerva and the Azrieli Foundation. 
N.C.S is supported by the Israel Science Foundation (Individual
Research Grant 2565/19) and the Binational Science foundation (grant Nos. 2019772 and 2020397).
B.T. acknowledges support from the European Research Council (ERC) under the European Union's Horizon 2020 research and innovation program (grant agreement 950533) and from the Israel Science Foundation (grant 1849/19).
IA is a CIFAR Azrieli Global Scholar in the Gravity and the Extreme Universe Program and acknowledges support from that program, from the European Research Council (ERC) under the European Union’s Horizon 2020 research and innovation program (grant agreement number 852097), from the Israel Science Foundation (grant number 2752/19), from the United States - Israel Binational Science Foundation (BSF), and from the Israeli Council for Higher Education Alon Fellowship.

\end{acknowledgments}

\bibliography{yossi_articles,BT_references, cosmo_references}
\bibliographystyle{aasjournal}

\end{document}